\documentclass[a4paper,11pt]{article}
\pdfoutput=1

\usepackage{jcappub}
\usepackage[T1]{fontenc}
\usepackage[dvipsnames]{xcolor}

\newcommand{\be}{\begin{equation}}
\newcommand{\ee}{\end{equation}}
\newcommand{\bea}{\begin{eqnarray}}
\newcommand{\eea}{\end{eqnarray}}

\newcommand{\RE}{\mathfrak{Re}}

\newcommand{\PartI}{\textit{Part I}~\cite{Antusch:2021aiw}}

\newcommand{\CL}{{\tt ${\mathcal C}$osmo${\mathcal L}$attice}~}

\title{\boldmath Characterizing the post-inflationary reheating history, Part II: \newline Multiple interacting daughter fields}

\author{Stefan Antusch, Kenneth Marschall, Francisco Torrenti}
\affiliation{Department of Physics, University of Basel, Klingelbergstr.\ 82, CH-4056 Basel, Switzerland.}

\emailAdd{stefan.antusch@unibas.ch}
\emailAdd{kenneth.marschall@unibas.ch}
\emailAdd{f.torrenti@unibas.ch}

\abstract{We characterize the post-inflationary dynamics of an inflaton $\phi$ coupled to multiple interacting daughter fields $X_n$ ($n=1,\dots N_d$) through quadratic-quadratic interactions $g_n^ 2\phi^2 X_n^2$. We assume a monomial inflaton potential $V(\phi) \propto |\phi|^p$ ($p \geq 2$) around the minimum. By simulating the system in 2+1-dimensional lattices, we study the post-inflationary evolution of the energy distribution and equation of state, from the end of inflation until a stationary regime is achieved. We show that in this scenario, the energy transferred to the daughter field sector can be larger than 50\%, surpassing this way the upper bound found previously for single daughter field models. In particular, for $p \geq 4$ the energy at very late times is equally distributed between all fields, and only $100/(N_d + 1) \%$ of the energy remains in the inflaton. We also consider scenarios in which the daughter fields have scale-free interactions $\lambda_{nm} X_n^2 X_m^2$, including the case of quartic daughter field self-interactions (for $n=m$). We show that these interactions trigger a resonance process during the non-linear regime, which in the single daughter field case already allows to deplete more than 50\% of the energy from the inflaton for $p\geq 4$.
}

\begin{document}

\maketitle

\section{Introduction}

The inflationary paradigm describes an early phase of accelerated expansion of the universe \cite{Starobinsky:1980te,Guth:1980zm,Linde:1981mu,Albrecht:1982wi}. Inflation provides a solution to the initial condition problems of classical cosmology, and generates a spectrum of quantum fluctuations that seeds the later structure formation. In the simplest model realizations, the accelerated expansion is sourced by the vacuum energy of a scalar field $\phi$ (the inflaton) in a slow-roll regime. Cosmic Microwave Background experiments such as Planck \cite{Akrami:2018jri} or BICEP \cite{BICEP:2021xfz} have been able to significantly constrain the range of observationally-viable scalar potentials. 

Inflation must be followed by a reheating stage, during which the universe must transition to a radiation-dominated thermal state before the onset of Big Bang Nucleosynthesis at $T\:{\sim}\:1\;{\rm MeV}$ \cite{Kawasaki:1999na,Kawasaki:2000en,Hannestad:2004px,Hasegawa:2019jsa}. Relevant questions concerning this phase are how the energy stored in the inflaton is transferred to other light fields and eventually to the Standard Model particles, or the exact post-inflationary evolution of the equation of state. In fact, a complete characterization of the expansion history after inflation is crucial for reducing the theoretical uncertainty in the predictions of CMB observables for specific inflationary models \cite{Dai:2014jja,Martin:2014nya,Munoz:2014eqa,Gong:2015qha,Cook:2015vqa}. Details of reheating depend strongly on the physics model under consideration, but its early stage is typically characterized by a non-perturbative, out-of-equilibrium excitation of field fluctuations called `preheating' \cite{Traschen:1990sw,Kofman:1994rk,Shtanov:1994ce,Kaiser:1995fb,Khlebnikov:1996mc,Prokopec:1996rr,Khlebnikov:1996wr,Kaiser:1997mp,Kofman:1997yn,Greene:1997fu,Khlebnikov:1996zt,Kaiser:1997hg}. Although these fluctuations can be treated with a linearized approximation immediately after inflation, non-linearities become relevant at later times, so one must resort to real-time lattice simulations in order to fully capture the dynamics of this stage  \cite{Khlebnikov:1996mc,Prokopec:1996rr,Khlebnikov:1996zt}. For reviews on (p)reheating see e.g.~\cite{Bassett:2005xm,Allahverdi:2010xz,Amin:2014eta,Lozanov:2019jxc,Allahverdi:2020bys}, while for a review on lattice techniques for the simulation of the early universe field dynamics see \cite{Figueroa:2020rrl}. 

In this work we consider inflaton potentials that can be approximated as a monomial function $V(\phi) \propto |\phi|^p$ ($p \geq 2$) after inflation. In this case, the post-inflationary oscillations of the inflaton homogeneous mode generate an effective equation of state $\bar w_{\rm hom} \equiv (p-2)/(p+2)$ during the linear regime \cite{Turner:1983he}. If the inflaton is decoupled from other fields, it fragments through a process of self-resonance for $p>2$, which gives rise to a radiation-dominated universe ${\sim} 10$ e-folds of expansion after the end of inflation \cite{Lozanov:2016hid,Lozanov:2017hjm}. If $p=2$, the inflaton fragments instead due to gravitational effects at much longer time scales \cite{Musoke:2019ima}.

In any case, a successful reheating process requires an efficient depleting of energy from the inflaton, which can be naturally achieved by coupling it to other fields (which we refer to as `daughter' fields). One possibility is to couple the inflaton to a daughter scalar field $X$ through a quadratic-quadratic interaction $g^2 \phi^2 X^2$, with $g^2$ a dimensionless coupling strength. This is e.g.~the leading term in the interaction between charged scalars and gauge fields \cite{Figueroa:2015rqa, Lozanov:2016pac}, and has the advantage of not introducing new dimensional scales in the theory (which facilitates lattice simulations). In this case, the field fluctuations get excited through a process of broad parametric resonance, and the inflaton fragments instead ${\sim}3{-}4$ e-folds after the end of inflation. The evolution of the energy distribution and equation of state during preheating in these models has been studied in detail with lattice simulations, see \cite{Podolsky:2005bw,Figueroa:2016wxr,Maity:2018qhi,Saha:2020bis}. The case of an inflaton coupled to a daughter field through a trilinear interactions has also been studied on the lattice, see \cite{Dufaux:2006ee}. Other works have studied the role of non-minimal kinetic terms during preheating in different scenarios: in the context of DBI inflation in \cite{Child:2013ria}, in the context of $\alpha$-attractor scenarios in \cite{Krajewski:2018moi}  (see \cite{Iarygina:2018kee} for a semi-analytical study), and in the context of multi-field models with non-minimal couplings to gravity (which appear as non-minimal kinetic terms in the Einstein frame) in \cite{Nguyen:2019kbm,vandeVis:2020qcp} (see also~\cite{DeCross:2015uza, DeCross:2016fdz, DeCross:2016cbs} for semi-analytical studies). The fields eventually achieve a turbulent regime at late times, see \cite{Micha:2002ey,Micha:2004bv}.

While most lattice studies of (p)reheating have focused on the linear and early non-linear stages, in the \textit{Letter} \cite{Antusch:2020iyq} we instead characterized the entire evolution of the energy distribution and equation of state after inflation, from the end of inflation until the achievement of a stationary regime. We considered an observationally-viable inflaton potential that behaves as $V(|\phi|) \sim |\phi|^p$ around the minimum, and coupled the inflaton to one (effectively massless) daughter field through a $g^2 \phi^2 X^2$ term. By simulating the post-inflationary dynamics in 2+1 dimensions, we were able to parametrize how the energy density distributes between its components at late times as a function of $p$ and $g^2$, as well as the final values for the equation of state. We found that the fraction of energy transferred to the daughter field is always negligible for $p < 4$, while it is at most $\sim$ 50 \% for $p \geq 4$. We continued this work in Ref.~\cite{Antusch:2021aiw} (which we refer to as \textit{Part I} from now on), in which we expanded the results of our \textit{Letter} \cite{Antusch:2020iyq}, as well as generalised our analysis to a class of inflaton potentials that have a `displaced' minimum $V(|\phi|) \sim |\phi-v|^p$ (with $v \geq 0$). Moreover, by using our information of the equation of state evolution, we were able to obtain exact predictions for the inflationary observables $n_s$ and $r$ in the model under consideration, for those cases in which the universe ends up in a radiation-dominated state.

\subsection*{~~\textit{Part II: Multiple interacting daughter fields}}

Notably, most (p)reheating studies (including our \textit{Letter} \cite{Antusch:2020iyq} and \PartI) have considered scenarios in which the inflaton is coupled to at most one daughter field. However, the correct physics model at high energies may well contain many scalar fields. For example, one could consider models with multiple inflaton fields. In this work we consider instead a different scenario, in which the stage of inflation is still generated by a single inflaton $\phi$, but the inflaton is coupled to multiple daughter fields during the subsequent phase of reheating. The existence of multiple daughter fields does doubtlessly change the post-inflationary evolution of the energy distribution and equation of state with respect to single daughter field scenarios, as they give rise to additional channels through which the energy of the inflaton can be extracted. Knowing the exact evolution of the equation of state after inflation is essential, for example, to make accurate predictions for the CMB observables $n_s$ and $r$, as shown in \textit{Letter} \cite{Antusch:2020iyq} and \PartI~for the single daughter field case. The post-inflationary equation of state is also important for e.g.~determining the exact redshift of a GW signal produced during inflation or preheating until today. Moreover, the energy transfer to daughter fields can have a significant influence on the produced baryon asymmetry of the universe from non-thermal leptogenesis, see e.g.\ \cite{Antusch:2018zvu}. In addition, systems of multiple daughter fields can have further interesting phenomenological consequences. For example, during (p)reheating a daughter field gets excited during the linear regime at a specific momentum scale, set by the strength of its coupling to the inflaton \cite{Figueroa:2017vfa}. As shown in \cite{Figueroa:2022iho}, in systems with multiple daughter fields these different scales can be imprinted in the produced spectrum of gravitational waves, which features a `stairway' pattern that potentially allows for particle coupling spectroscopy (see also \cite{Giblin:2010sp} for a previous study with multiple daughter fields with equal couplings).

The aim of this paper (which we refer to as \textit{Part II}) is, therefore, to study the post-inflationary dynamics of models with an arbitrary number of daughter fields. This work constitutes a direct continuation of the research carried out in \textit{Letter} \cite{Antusch:2020iyq} and \PartI, which considered single-daughter field scenarios. As in these past works, we will characterize the post-inflationary evolution of the energy distribution and equation of state, from the end of inflation until the establishment of a stationary regime. Special emphasis will be put on describing how the energy gets distributed at very late times. Our analysis will be based both on a numerical analysis of the linearized field equations under a Hartree approximation, as well as on 2+1-dimensional lattice simulations of the system carried out with the code \CL  \cite{Figueroa:2021yhd}.\footnote{Note that in \textit{Letter}~\cite{Antusch:2020iyq}, we explicitly compared the output from lattice simulations in 2+1 dimensions with the one from (3+1)-dimensional ones. This way, we showed that (2+1)-D simulations mimic very well the dynamics of (3+1)-D ones for the kind of models under consideration, both qualitatively and quantitatively.} Regarding our lattice-based analysis, we will start by studying the post-inflationary dynamics of an inflaton coupled to multiple daughter fields $X_n$ ($n=1,\dots,N_d$) through quadratic-quadratic interactions $g_n^2 \phi^2 X_n^2$, where the coupling strength of each daughter field can be different. After that, we study again the case of one daughter field $X$, but include now a quartic self-interaction $\lambda X^4$ into our analysis, which was not taken into account in Refs.~\cite{Antusch:2020iyq,Antusch:2021aiw}. Finally, we will consider different examples of multi-daughter field theories with scale-free interactions of the type $\lambda_{nm} X_n^2 X_m^2$, which include both quartic self-interactions (when $n=m$) and quadratic-quadratic interactions between different daughter fields (when $n \neq m$).

The structure of this work is as follows. In Sect.~\ref{sec:LinearAnalysis} we present the details of our model set-up. In Sect.~\ref{Sec:Linear} we investigate the early stage of preheating by means of a linearized analysis of the field equations in the Hartree approximation. In Sect.~\ref{sec:LatSims} we deploy lattice simulations to study the non-linear regime of the field dynamics in the different scenarios explained above. In Sect.~\ref{sec:Summary} we summarize and discuss our results.

\section{Our set-up} \label{sec:LinearAnalysis}

\subsection{Inflaton potential}

Let us consider an inflaton field $\phi$ with monomial potential around the minimum,
\be V_{\rm m} (\phi) \simeq \frac{1}{p} \lambda \mu^{4-p}   |\phi |^p \ ,  \label{eq:powlaw-pot}\ee
where $\lambda$ is a dimensionless parameter, $\mu$ is an energy scale, and $p$ is an arbitrary coefficient obeying $p \geq 2$. Cosmological observations rule out the case of potential  (\ref{eq:powlaw-pot}) sustaining inflation at all amplitudes \cite{Akrami:2018jri}. However, we can still consider observationally-viable potentials that behave like (\ref{eq:powlaw-pot}) around the minimum but flatten at larger amplitudes. One example is the $\alpha$-attractor T-model \cite{Kallosh:2013hoa},
\be V_{\rm t} (\phi) = \frac{1}{p} \Lambda^4{\rm tanh}^{p} \left( \frac{|\phi|}{M} \right) \ , \hspace{0.5cm} \frac{\Lambda^4}{M^p} \equiv \lambda \mu^{4-p}  \label{eq:inflaton-potential} \ , \ee
where $\Lambda$ and $M$ have dimensions of energy. The ratio $\Lambda^4 / M^p$ can be fixed so that we recover the monomial potential (\ref{eq:powlaw-pot}) in the limit $M \rightarrow \infty$. The observed value of the tensor-to-scalar ratio ($r <0.036$ at $95\%$ confidence level \cite{BICEP:2021xfz}) imposes the upper bound $M \lesssim (8.5-9.5) m_{\rm pl}$ when the number of e-folds from the horizon crossing of the pivot scale until the end of inflation is fixed to $N_k = 60$ (the exact bound for $M$ depends on $p$). Moreover, by fitting the theoretical prediction for the  scalar amplitude to the observed value $A_s=2.1\times10^{-9}$, one can determine a relation between the model parameters as $\Lambda = \Lambda (p,M,N_k)$ (the explicit expression is written in Eqs.~(A5) and (A6) of \PartI).

Slow-roll inflation takes place at large field values, and ends approximately when the condition  $\epsilon_{V}\equiv m_{\rm pl}^2 V_{,\phi}^2/(2V^2) = 1$ holds, at the amplitude
\be \phi_* = \frac{1}{2} M {\rm arcsinh} \left( \frac{\sqrt{2} p m_{\rm pl}}{M}\right) \ . \ee 
The inflaton then starts oscillating around the minimum of the potential. Note that if $M \gtrsim 1.65 m_{\rm pl}$, we have $\phi_* < \phi_{\rm i}$ for all values of $p\geq2$, where $\phi_{\rm i} \equiv  M \,{\rm arcsinh} \left( \sqrt{(p-1)/2} \right)$ is the inflection point of the potential. For these values of $M$, the oscillations always take place in the positive-curvature region of the potential, and can be approximately parametrized during the initial linear regime as \cite{Turner:1983he}
\be \label{eq:ApproxVarPhi} \phi (t) \simeq  \mathcal{A}_{\phi} (t) \mathcal{F} (t) \ , \hspace{0.5cm} \mathcal{A}_{\phi}(t) \equiv \phi_{\star} \left(\frac{t}{t_{\rm {\star}}}\right)^{-2/p} \ ,  \ee
where $\phi_{\star} \equiv \phi (t_{\star}) \simeq \phi_*$ is the field amplitude at a time scale $t = t_{\star}$ close to the end of inflation, and $\mathcal{F}(t)$ is an oscillatory function with oscillation period
\be \Omega_{T_ {\bar \varphi}} \equiv \lambda^{\frac{1}{2}}  \mu^{\frac{4-p}{2}}   \mathcal{A}_{\phi} ^{\frac{p-2}{2}} = \omega_{\star} \left(\frac{t}{t_{\star}}\right)^{\frac{2}{p} - 1} \ , \hspace{0.5cm} \omega_{\star} \equiv  \lambda^{\frac{1}{2}} \mu^{\frac{4-p}{2}} \phi_{\star}^{\frac{p-2}{2}} \ . \label{eq:OmegaOsc}\ee
The period is constant for $p=2$, but time-dependent for $p \neq 2$. The \textit{effective} equation of state, i.e.~the ratio between the oscillation-averaged pressure and energy densities, is approximately
\be \bar w_{\rm hom} \equiv \frac{ \langle {p}_\phi \rangle_{T_ {\bar \varphi}}}{ \langle {\rho}_\phi \rangle_{T_ {\bar \varphi}}} = \frac{p-2}{p+2} \ . \label{eq:EoSoscillations}
\ee

\subsection{Daughter field sector}

The objective of this work is to study the post-inflationary dynamics of an inflaton $\phi$ coupled to multiple \textit{daughter} scalar fields $X_n$, with $n=1,2,\dots N_d$ and $N_d \geq 1$. We will consider different particularizations of the following potential,
\be
V(\phi, \{ X_n \}) 
=  V_{\rm inf} (\phi) + \frac{1}{2} \phi^2 \sum_{n=1}^{N_d}  g_n^2 X_n^2  + \frac{1}{4} \sum_{\substack{n,m=1}}^{N_d} \lambda_{n m} X_n^2 X_m^2\  , \label{eq:potential} \ee
where $V_{\rm inf} (\phi)$ is the potential sustaining inflation (either (\ref{eq:powlaw-pot}) in Sect.~\ref{Sec:Linear} or (\ref{eq:inflaton-potential}) in Sect.~\ref{sec:LatSims}). Each daughter field is coupled to the inflaton and  other daughter fields via quadratic-quadratic interactions, with $g_n^2$ and $\lambda_{nm} (= \lambda_{mn})$ denoting the corresponding dimensionless coupling strengths. The last term also contains quartic self-interactions of the daughter fields when $n=m$: in this case we will use the short notation $\lambda_n\equiv\lambda_{nn}$.

Mimicking the procedure in \PartI, it is convenient to work in \textit{natural variables}, defined for field amplitudes and spacetime coordinates as
\be \varphi \equiv \frac{1}{\phi_*} a^{\frac{6}{p+2}} \phi \ , \hspace{0.4cm} \chi_n \equiv \frac{1}{\phi_*} a^{\frac{6}{p+2}} X_n \ , \label{eq:newvars2} \ee
\be t \rightarrow u \equiv \omega_{*} \int_{t_{*}}^t { \, a(t')^{\frac{3 (2-p)}{2+p}}dt'} \ , \hspace{0.4cm} \vec{x} \rightarrow \vec{y} \equiv \omega_{*} \vec{x} \ . \label{eq:newvars1}\ee
This way, the amplitude and oscillation period of the `natural' inflaton $\varphi$ during the linear regime are constant in `natural' time $u$. The equations of motion can then be written as
\begin{align}
& \varphi'' - a^{\frac{-(16 - 4p)}{2+p}} \nabla^2_{\vec{y}} \, \varphi +  \left( |\varphi|^{p-2} + \sum_{n=1}^{N_d} \tilde{q}^{(n)} (a) \chi_n^2  + F(u) \right) \varphi = 0 \ , \label{eq:fullEOMs1} \hspace{0.7cm}  \\
& \chi_n ''- a^{\frac{-(16 - 4p)}{2+p}} \nabla^2_{\vec{y}} \, \chi_n + \left( \tilde{q}^{(n)} (a)  \varphi^2 + \sum_{\substack{m=1}}^{N_d} \tilde{\sigma}^{(nm)}(a) \chi_m^2 + F(u) \right) \chi_n = 0 \ , \label{eq:fullEOMs2}
\end{align}
where we have fixed $a(t_*) = 1$ at the end of inflation, $F(u) \equiv  \frac{6(p-4)}{(p+2)^2} \left( a' / a \right)^2 - \frac{6}{2+p} (a'' / a)  \sim u^{-2}$ is a function that becomes subdominant after a few inflaton oscillations, and we have defined the following time-dependent functions,
\begin{align}
   \tilde{q}^{ (n)} (a) &\equiv q_{*}^{(n)} a^{\frac{6 (p-4)}{p+2}} \ , \hspace{1.4cm} q_*^{(n)} \equiv g^2_n \frac{\phi_{*}^2}{ \omega_{*}^2}  \ ,  \label{eq:respar}\\
    \tilde{\sigma}^{(nm)} (a) &\equiv  \sigma_*^{(nm)} a^{\frac{6 (p-4)}{p+2}} \ , \hspace{0.8cm} \sigma_*^{(nm)} \equiv \lambda_{nm} \frac{ \phi_*^2}{\omega_*^2}  \ .  \label{eq:resint}
\end{align}  
Here, $q_*^{(n)}$ is the (initial) \textit{resonance parameter} of the daughter field $X_n$, and $\tilde{q}^{ (n)}$ the corresponding time-dependent (effective) one. We have defined $\sigma_*^{(nm)}$ and $ \tilde{\sigma}^{(nm)} $ for the daughter-daughter interactions in an analogous way. At the end of inflation we have $\tilde q^{(n)} = q_*^{(n)}$ and $\tilde \sigma^{(nm)} = \sigma_*^{(nm)}$, but both functions evolve in different ways depending on $p$: they decrease for $p<4$, grow for $p>4$, and remain constant for $p=4$.

\section{Linearized analysis of the field equations} \label{Sec:Linear}

\begin{figure}
    \centering
    \includegraphics[width=7.5cm]{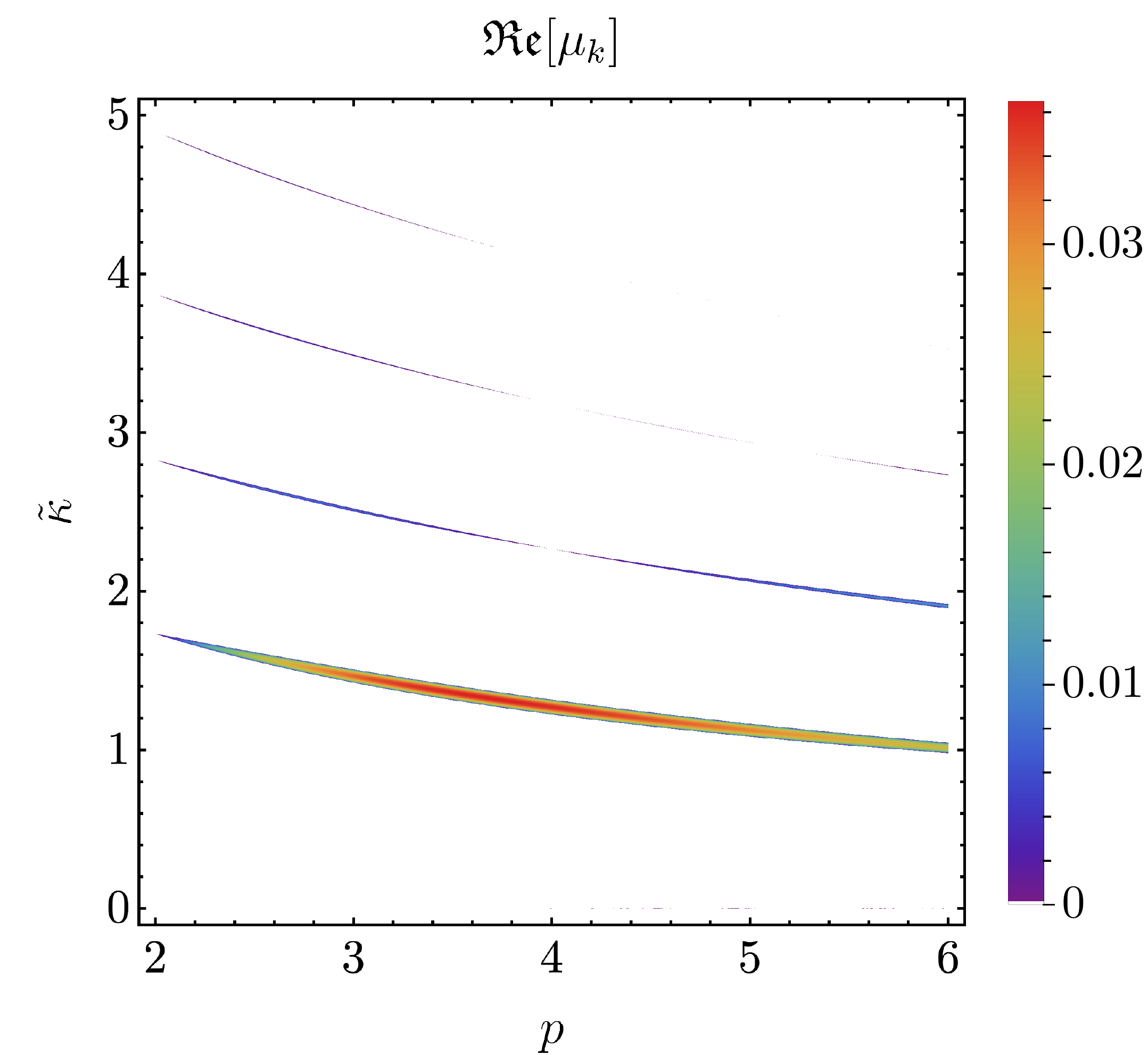} \,  
    \includegraphics[width=7.5cm]{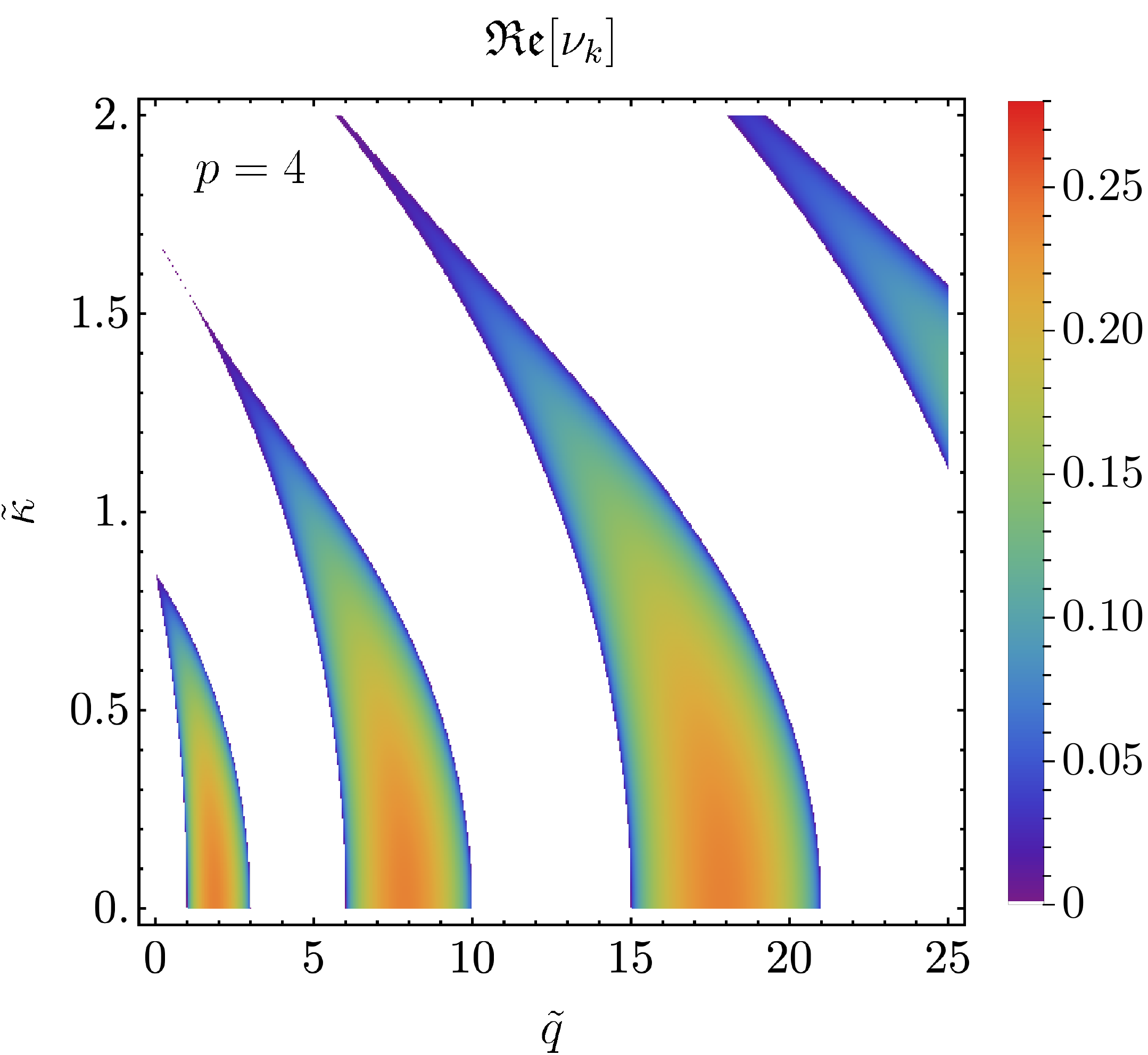}
    \caption{Floquet charts for inflaton self-resonance (left panel) and parametric resonance of the daughter field for $p=4$ (right panel).} 
    \label{fig:stabilitychart}
\end{figure}

The inflaton oscillations may lead to a strong growth of fluctuations for either the inflaton (in a process of \textit{self-resonance}), the daughter field (in a process of \textit{parametric resonance}), or both. In order to illustrate this, let us expand the fields as
\begin{align}
\varphi (\vec{y}, u)  &\equiv  \bar{\varphi} (u) + \delta \varphi (\vec{y},u) \ , \label{eq:exp-phi} \\
\chi_n (\vec{y}, u) &\equiv   \delta \chi_n (\vec{y},u) \ , \label{eq:exp-daug} 
\end{align}
where $\bar{\varphi}$ is the homogeneous component of the inflaton (note that we have $\bar{\chi}_n\simeq0$ at the end of inflation). Under the approximation $F=0$, the homogeneous part of the inflaton obeys the equation ${\bar \varphi'' +  |\bar \varphi |^{p-2} \bar \varphi \simeq 0}$, whose solution is $\bar \varphi = \cos (u)$ for $p=2$ and $\bar \varphi \simeq \cos (\beta_{\bar{\varphi}} u)$ with $\beta_{\bar{\varphi}} \sim 1$ for $p > 2$. The fluctuations can be described by their mode equations in Fourier space,
\begin{align}
\delta \varphi^{''}_k + \tilde \omega_{k,\varphi}^2 \delta \varphi_k \simeq 0& \ , \hspace{1cm} \tilde \omega_{k,\varphi} \equiv \sqrt{ \tilde \kappa^2 (a) + (p-1) |\bar{\varphi}|^{p-2} }   \ , \label{eq:modephi}  \\
\delta \chi^{''}_{n,k} + \tilde \omega_{k,\chi_n}^2  \delta \chi_{n,k} \simeq 0&   \ ,  \hspace{1cm} \tilde \omega_{k,\chi_n} \equiv \sqrt{ \tilde \kappa^2 (a) + \tilde{q}^{(n)} (a) \bar{\varphi}^2} \ , \label{eq:modechi} \end{align}
where $\tilde \kappa (a) \equiv  (k/\omega_{*}) a^{\frac{-(8 - 2p)}{2+p}}$ is the `natural' resonance momentum. The effective frequencies $\tilde \omega_{k,\varphi}$ and $\tilde \omega_{k,\chi_n}$ may vary non-adiabatically when the inflaton homogeneous mode crosses the minimum of the potential, which leads to a strong growth of the field fluctuations. More specifically, the mode equations (\ref{eq:modephi}) and (\ref{eq:modechi}) allow for solutions of the form $\delta\varphi_k \sim e^{\mu_k u}$ and $\delta \chi_{n,k} \sim e^{\nu_{k} u}$, with $\mu_k\equiv\mu_k(p)$ and $\nu_{k}\equiv\nu_{k}(p,\tilde{q})$ the corresponding Floquet indices. For certain values of $p$ and $\tilde{q}
^{(n)}$, the real parts of the Floquet indices are positive, thus leading to the following two resonance phenomena:

\begin{itemize}
    \item \textbf{Inflaton self-resonance:} The structure of (narrow) resonance bands is shown as a function of $p$ in the left panel of Fig.~\ref{fig:stabilitychart}. For a given choice of $p$, the dominant band is the one of lowest momenta, with $\RE[\mu_k] \lesssim 0.036$. Note that there is no self-resonance for $p=2$.

    \item \textbf{Parametric resonance of the daughter field:} The structure of resonance bands is shown in the right panel of Fig.~\ref{fig:stabilitychart} (we show the case $p=4$, but very similar charts can be depicted for other values of $p$, see Fig.~7 of \PartI). The regime of strongest resonance corresponds to $\tilde{q} \gtrsim 1$  (broad resonance), while for $\tilde{q} \lesssim 1$ the resonance is very weak (narrow resonance). The type of resonance can change as the universe expands: for $p<4$ an initially broad resonance becomes narrow at later times, while if $p>4$ an initially narrow resonance becomes broad at later times. If $p=4$, the type of resonance never changes. The maximum Floquet index for broad parametric resonance is $\RE[\nu_k] \simeq 0.26$. Thus, parametric resonance is much stronger than inflaton self-resonance.
    
\end{itemize}

The linear regime of resonant excitation ends when the energy of the fluctuations becomes comparable with the one of the inflaton homogeneous mode. In \PartI~we computed an analytical estimate of the \textit{backreaction time} $u_{\rm br}$ when the equality $\rho_{|\bar \varphi |} \simeq \rho_{\delta f}$ holds exactly, for both fields $f = \varphi, \chi$.

\subsection{Termination of resonant growth} \label{sec:Hartree}

Let us perform a Hartree or mean-field approximation \cite{Boyanovsky:1994me, Kofman:1994rk,Kofman:1997yn, Prokopec:1996rr, Khlebnikov:1996wr} to the field equations, which allows to partially capture the early backreaction effects of the $\lambda_{nm} X_n^2 X_m^2$ interactions during the linear regime. Following the procedure of Refs.~\cite{Boyanovsky:1994me, Bassett:2005xm}, we substitute $\delta f^2\rightarrow\langle\delta f^2\rangle$ and $\delta f^3\rightarrow3\langle\delta f^2\rangle\delta f$ in the fluctuation equations in position space, where $f=\varphi,\chi$ labels the fields and $\langle\delta f^2\rangle \equiv (2\pi^2)^{-1} \int {\rm d}k k^2 |\delta f_k|^2$. This approximation neglects the couplings between modes of different momenta, but captures sufficiently well the field dynamics during the linear regime as we shall see. We obtain the following equations for the field modes
\begin{align}
    \delta \varphi^{''}_k + \biggr( \tilde{\kappa}^2 + (p-1)|\bar{\varphi}|^{p-2} + \sum_{n=1}^{N_d}\tilde{q}^{(n)} \langle \delta\chi_n^2 \rangle + \frac{1}{3!}\partial^4_{\bar{\varphi}} V(\bar{\varphi})\langle\delta\varphi^2\rangle\biggr) \delta \varphi_k &\simeq 0 \ ,  \label{eq:InflFluct} \\
    \delta \chi^{''}_{n,k} + \biggr( \tilde{ \kappa}^2 + \tilde{q}^{(n)} (\bar{\varphi}^2 + \langle\delta\varphi^2\rangle) + 3\tilde{\sigma}^{(nn)} \langle \delta \chi_n^2 \rangle + \sum_{\substack{m=1\\ (n\neq m)}}^{N_d}\tilde{\sigma}^{(nm)}\langle \delta \chi_m^2 \rangle \biggr) \delta \chi_{n,k} &\simeq 0 \ ,  \label{eq:DaugFluct} 
\end{align} 
which incorporate the corrections generated by the $\lambda_{nm} X_n^2 X_m^2$ interaction terms to Eqs.~(\ref{eq:modephi}) and (\ref{eq:modechi}). Note that, if we set $k=0$ and $\delta \varphi_0 \equiv \bar{\varphi}$ in Eq.~(\ref{eq:InflFluct}), we obtain the corrected equation for the inflaton homogeneous mode. 

The field variances contribute to the effective masses of both fields, and grow exponentially during the initial inflaton oscillations. If they become large enough, they trigger the decay of the inflaton homogeneous mode due to backreaction effects, which terminates the resonant growth. However, they can also block the growth of the field modes before that happens. As an example, consider the case of one daughter field in broad parametric resonance: the first situation happens when $q_* \gtrsim \sigma_*$ (as long as $\tilde{q}>1$) via the coupling $\sim\tilde{q}\langle\delta \chi^2\rangle$ in Eq.~(\ref{eq:InflFluct}),\footnote{In the following we will use the shorter notation of $\chi\equiv\chi_1$, $q_*\equiv q_*^{(1)}$ and $\sigma_*\equiv \sigma_*^{(1)}$ for single daughter field scenarios.} while the second situation happens when $q_* \lesssim \sigma_* $ and the effective mass becomes of the order of the term responsible for the resonance, $3\tilde{\sigma}\langle\delta \chi^2\rangle \approx \tilde{q}\bar{\varphi}^2$. The maximum variance attained in this case is $\langle \delta \chi^2\rangle_{\rm max} \approx q_*/(6\sigma_*)$ (this result can be obtained by setting $\bar \varphi^2 \simeq \langle \bar \varphi^2 \rangle_{T_ {\bar \varphi}} \approx 1/2$). In these cases, the inflaton homogeneous mode survives and continues to dominate the energy budget (at least as long as the Hartree approximation is valid).

\begin{figure}
    \centering
    \includegraphics[width=7.5cm]{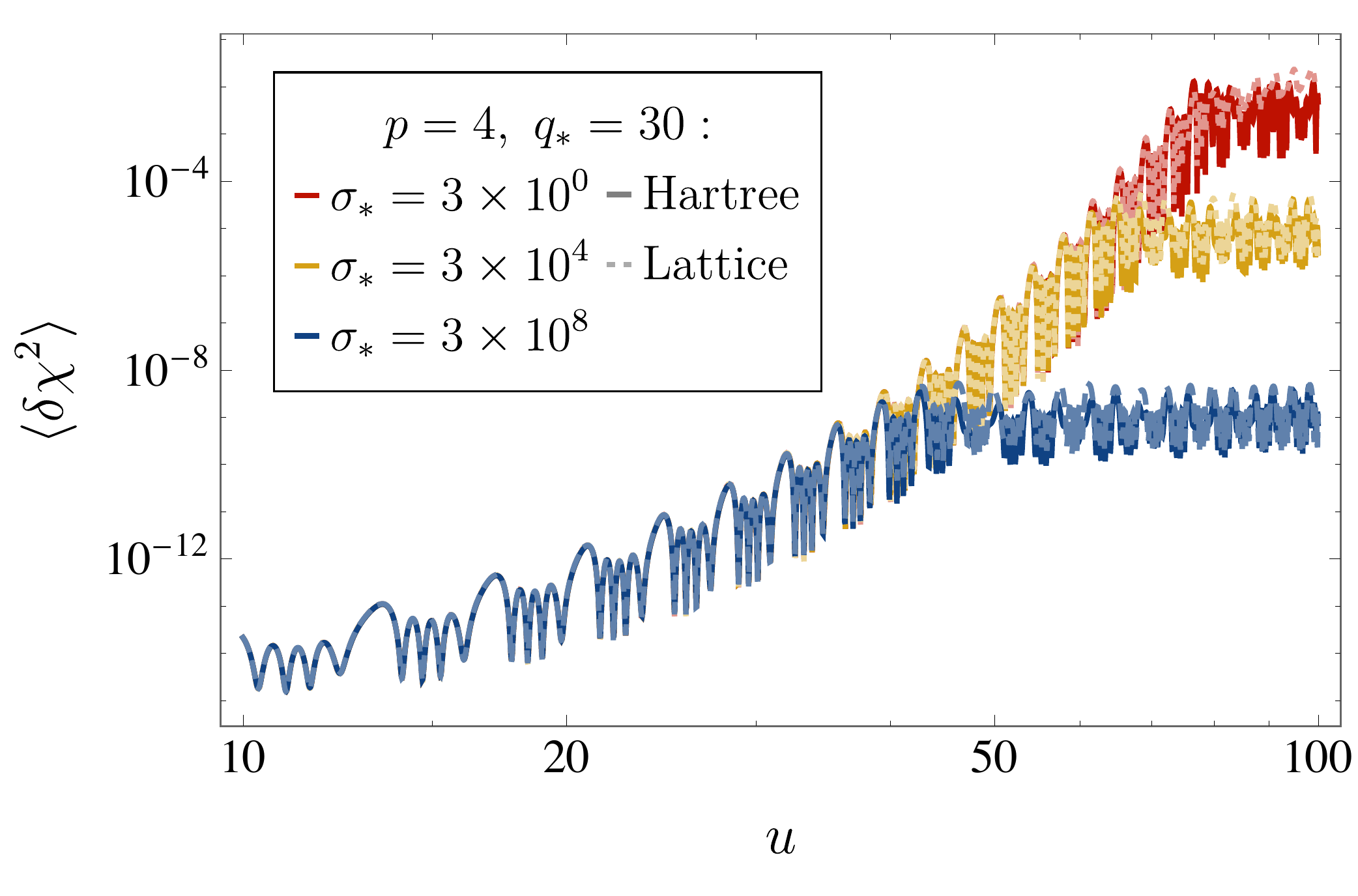} \,  
    \includegraphics[width=7.5cm]{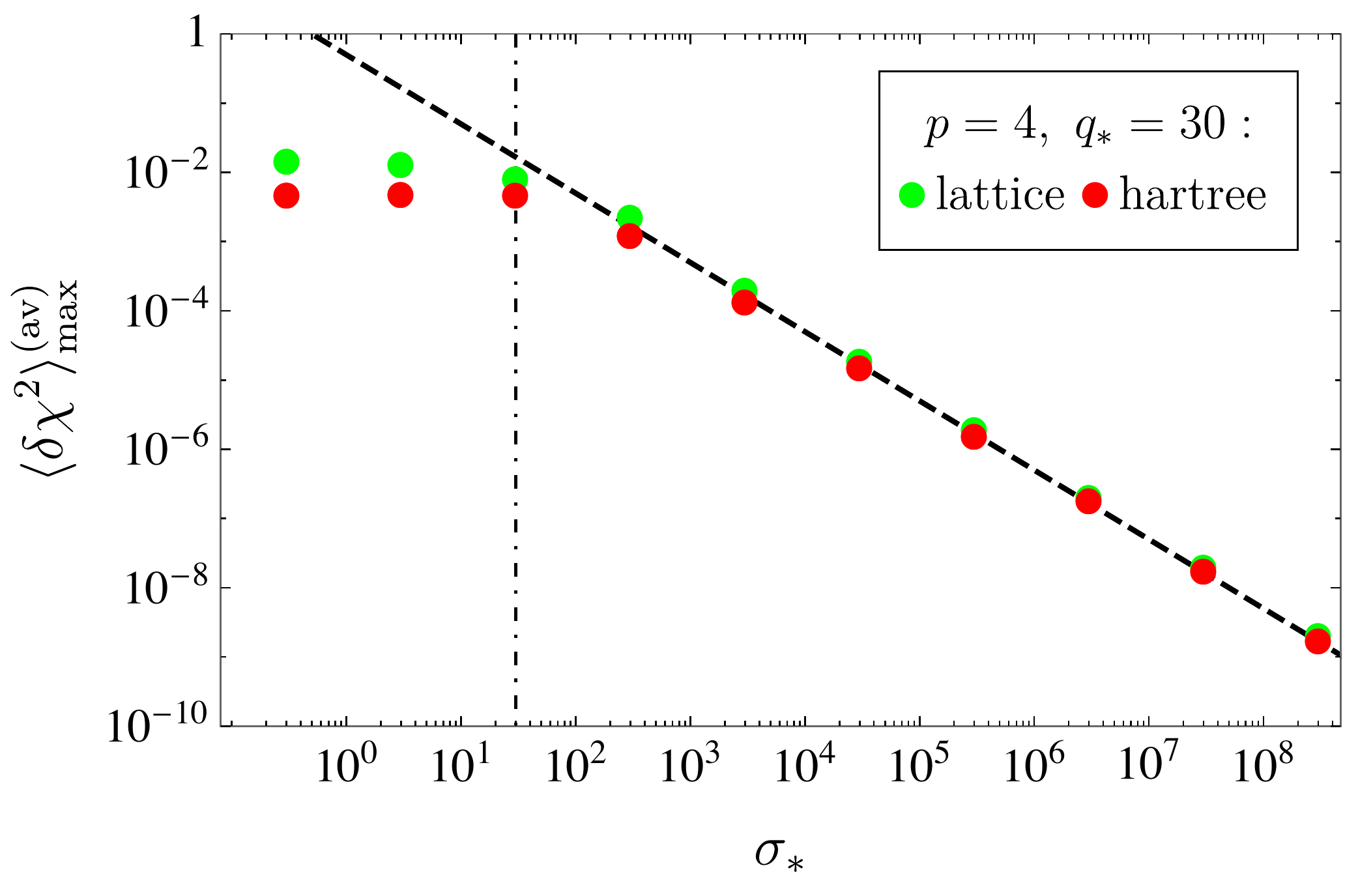}
    \caption{Left: Evolution of the daughter field variance for $N_d = 1$, $p=4$ and $q_* = 30$, obtained by 1) solving the linearized equations under the Hartree approximation (continuous lines), and 2) simulating the system in the lattice (dashed lines). We consider three different self-coupling parameters: $\sigma_* = 3 \cdot 10^0$, $3 \cdot 10^4$, and $3 \cdot 10^8$.  Right panel: Variance of the daughter field attained at late times for different choices of $\sigma_*$, computed with both the Hartree approximation and with lattice simulations. The black dashed line indicates the estimate $\langle \delta \chi^2\rangle_{\rm max} \sim  0.1\times q_*/ (6\sigma_*)$. The vertical line indicates $q_*=\sigma_*$.}
    \label{fig:hartree} \end{figure}

We illustrate this in the left panel of Fig.~\ref{fig:hartree}, where we show the evolution of the variance of one daughter field for $p=4$, $q_*=30$, and three different self-coupling parameters:  $\sigma_* = 3 \cdot 10^0\,  ( < q_*)$, $3 \cdot 10^4\,  (> q_*)$ and $3 \cdot 10^8\, (>q_*)$. The solution is obtained by solving numerically Eq.~(\ref{eq:DaugFluct}) together with (\ref{eq:InflFluct}) for the homogeneous mode $k=0$ (we have ignored the growth of the inflaton fluctuations $\delta \varphi_k$ with $k>0$, as inflaton self-resonance is a much weaker effect). We compare each solution with the result from a lattice simulation, which takes all non-linearities into account (see Sect.~\ref{sec:LatSims} for more details). For $\sigma_* = 3 \cdot 10^4$ and $\sigma_* = 3 \cdot 10^8$, the growth of the variance saturates at $\langle \delta \chi^2\rangle_{\rm max}  = \mathcal{O} (0.1)q_*/(6\sigma_*)$, where we have included a correction factor of order $\mathcal{O} (0.1)$ to our naive analytical estimation. For $\sigma_* = 3 \cdot 10^0$ we have $\sigma_* < q_*$ and the results obtained by the Hartree approximation and the lattice differ more strongly (see left panel). In this case, the inflaton homogeneous mode decays due to backreaction effects, which is not fully captured by the Hartree approximation. The dynamics at later times can only be properly studied with lattice simulations. Parametric resonance terminates before the growth of $\langle \delta \chi^2 \rangle$ saturates due to its own effective mass, so the estimation $\langle \delta \chi^2\rangle_{\rm max}  = \mathcal{O} (0.1)q_*/(6\sigma_*)$ does not work in this case.
 
The right panel of Fig.~\ref{fig:hartree} shows the saturated variance $\langle \delta \chi^2 \rangle_{\rm max}$ for $q_*=30$ and different choices of $\sigma_*$, obtained with both a Hartree approximation and lattice simulations. The vertical dash-dotted line indicates the case $\sigma_* = q_*$. For $\sigma_*/q_* > 1$ the saturated variance approximates quite well the estimation  $\langle \delta \chi^2\rangle=0.1\cdot q_* / (6\sigma_*)$ (in particular for very large ratios). For $\sigma_*/q_* \leq 1$ the resonance is terminated by backreaction effects from the daughter field modes onto the homogeneous inflaton condensate and the variance saturates roughly at $\langle \delta \chi^2\rangle\sim(p-1)|\bar{\varphi}|^{p-2}/q_*$. The results from the lattice and the Hartree approximation differ more strongly.

    \begin{figure}
    \includegraphics[width=7.5cm]{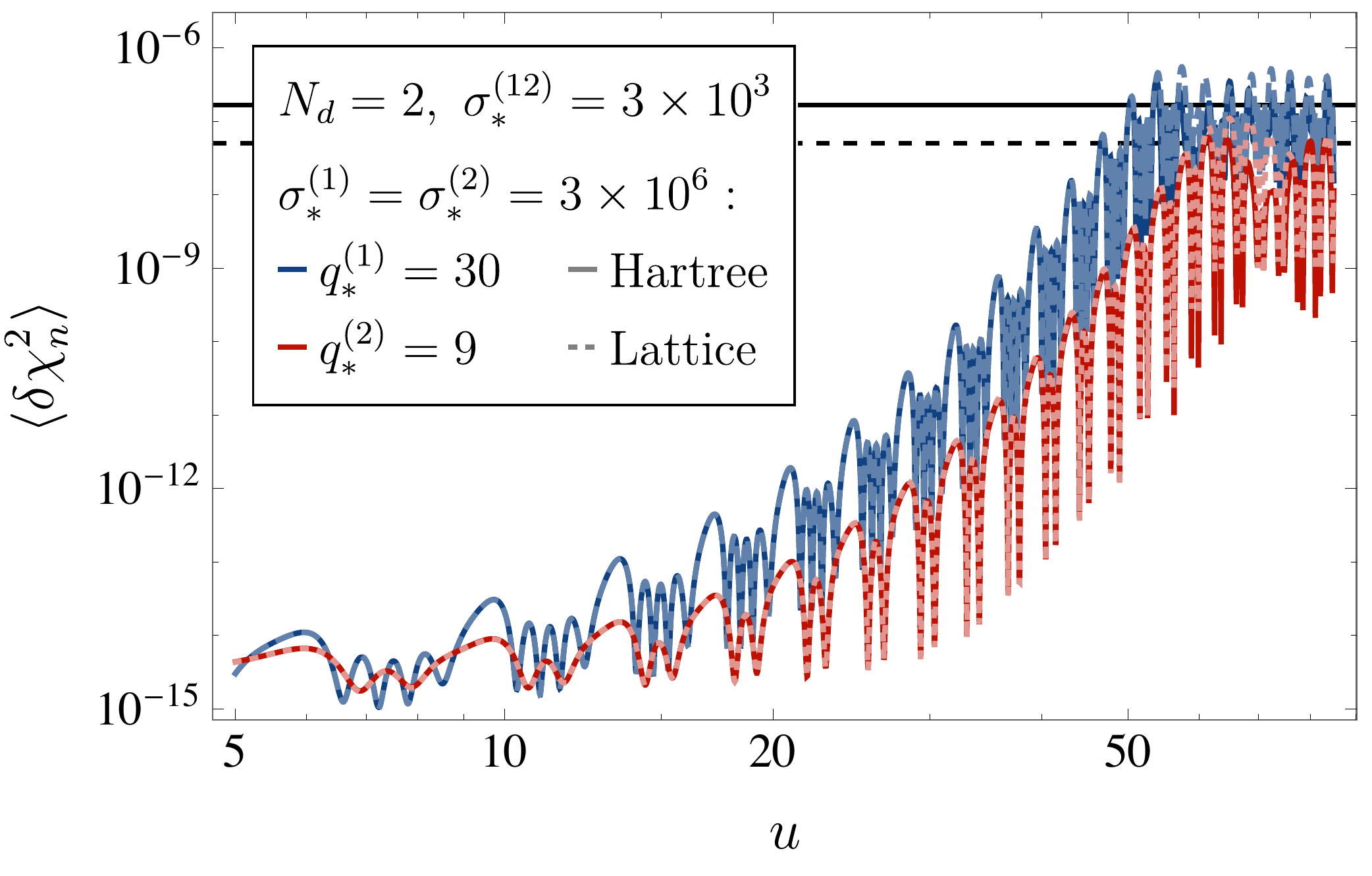}
    \includegraphics[width=7.5cm]{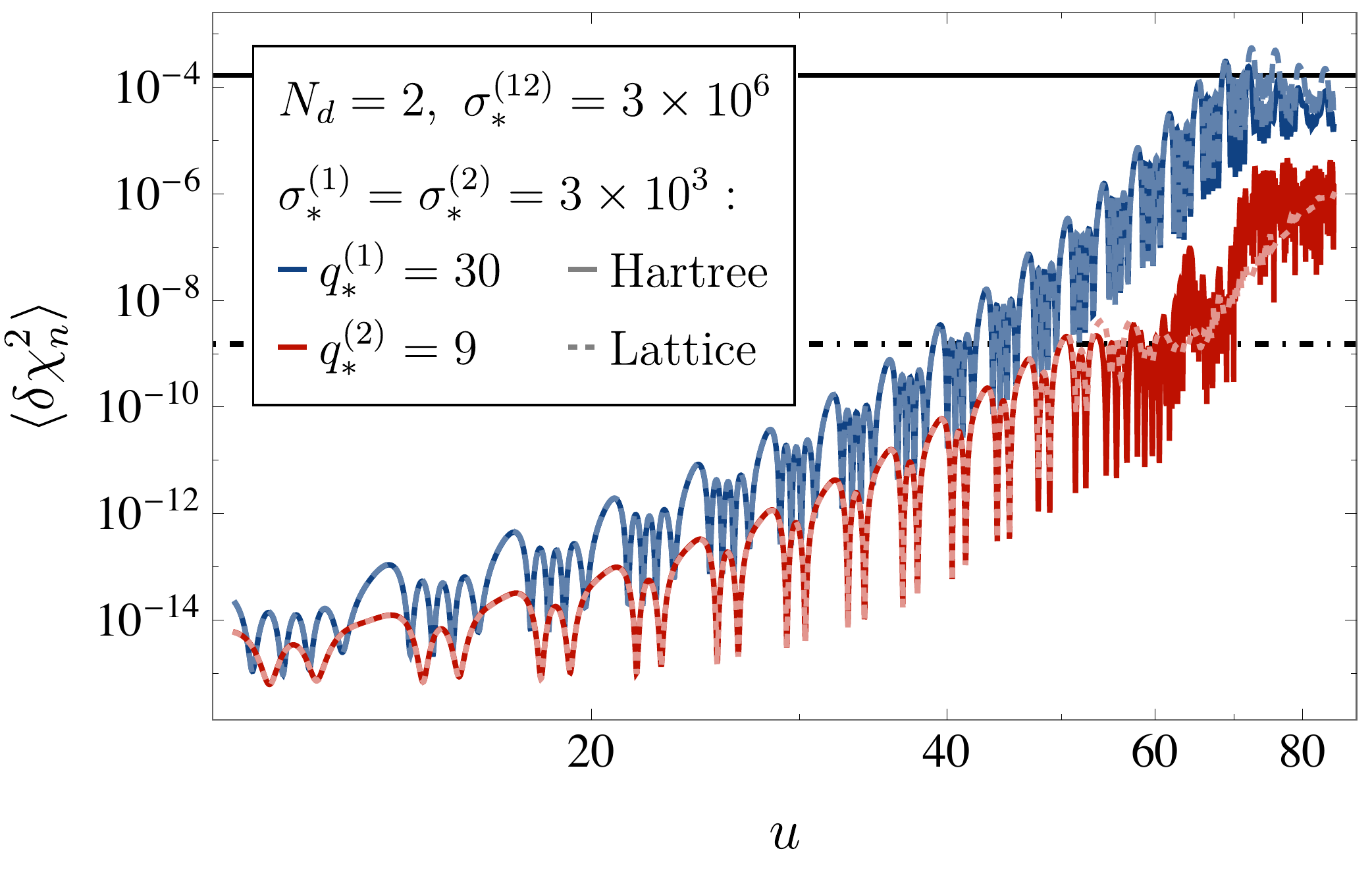}
    \caption{Evolution of the daughter field variances for $N_d = 2$ (two daughter field scenario), $p=4$, $q_*^{(1)}=30$ and $q_*^{(2)}=9$. Left: We show the case $\sigma_*^{(1)} = \sigma_*^{(2)}=3 \cdot 10^6$ and $\sigma_*^{(12)}=3 \cdot 10^3$. The horizontal solid and dashed lines show the estimates  $\langle \delta \chi^2_1 \rangle_{\rm max} = 0.1 \times q_*^{(1)}/(6\sigma_*^{(1)})$ and $\langle \delta \chi^2_2 \rangle_{\rm max} = 0.1 \times q_*^{(2)}/(6\sigma_*^{(2)})$ respectively. Right: We show the case $\sigma_*^{(1)} = \sigma_*^{(2)}=3 \cdot 10^3$ and $\sigma_*^{(12)}=3 \cdot 10^6$. The horizontal solid line indicates the variance $\langle \delta \chi_1^2 \rangle_{\rm max}$, while the dash-dotted line indicates the temporary plateau attained by $\langle \delta \chi_2^2 \rangle$, achieved when $\langle \delta \chi_1^2\rangle \sim q_*^{(2)}/(2\sigma_*^{(12)})$.}
    \label{fig:hartree2fields}
\end{figure}

Analytical estimations for $\langle \delta \chi^2_n\rangle_{\rm max}$ ($n=1,\dots N_d$) can also be obtained for two or more interacting fields, but expressions become significantly more complicated. For illustrative purposes, let us consider a two-daughter field scenario. Fig.~\ref{fig:hartree2fields} shows the evolution of the daughter field variances, obtained again by solving the field equations under the Hartree approximation and with lattice simulations. The left panel shows a scenario in which $\sigma_*^{\rm (1)}, \sigma_*^{\rm (2)} \gg \sigma_*^{\rm (12)} ( \gg q_*^{(1)}, q_*^{(2)})$. In this case, the role of the daughter-daughter interaction is negligible, so the evolution of each individual variance is similar to the single-daughter field case: they saturate at roughly the value $\langle \delta \chi_{n}^2\rangle_{\rm max} \sim 0.1 \times q_*^{(n)}/(6\sigma_*^{(n)})$. The right panel shows instead a scenario in which $\sigma_*^{\rm (12)} \gg \sigma_*^{\rm (1)}, \sigma_*^{\rm (2)}$ ($\gg q_*^{(1)}, q_*^{(2)})$. In this case, the variance of the first daughter field (the one with largest $q_*$) saturates roughly at the value $\langle \delta \chi_{1}^2\rangle_{\rm max} \sim  0.1 \times q_*^{(1)}/(6\sigma_*^{(1)})$, in agreement again with the single-daughter prediction. However, the evolution of the second daughter field gets affected by the fast growth of $\langle \delta \chi_{1}^2\rangle$ due to the strong coupling $\sim \sigma_*^{(12)} \chi_1^2 \chi_2^2$. In fact, we find that $\langle \delta \chi_{2}^2\rangle$ temporarily saturates when the variance of the \textit{first} daughter field attains the value $\langle \delta \chi_{1}^2\rangle\sim q_*^{(2)}/(2\sigma_*^{(12)})$ ($\ll \langle  \delta \chi_{1}^2 \rangle_{\rm max}$). However, we observe that $\langle \delta \chi_2^2 \rangle$ starts growing again after a while in both the Hartree approximation and the lattice.

\section{Energy distribution after inflation: lattice simulations} \label{sec:LatSims}

We now investigate the post-inflationary evolution of the energy distribution with lattice simulations, going beyond the linearized analysis of Sect.~\ref{Sec:Linear}. The simulations have been carried out with the publicly available code  ${\mathcal C}$osmo${\mathcal L}$attice \cite{Figueroa:2021yhd}. We have done simulations in 2+1 dimensions, which allow to properly capture the very late-time regime of the field evolution. In Appendix A of \textit{Letter}~\cite{Antusch:2020iyq} we carried out lattice simulations of single field scenarios (described by potential (\ref{Sec:Nd_1_lambda_0}) below) in both 2+1 and 3+1 dimensions and compared explicitly the output: this way, we showed that (2+1)-D simulations mimic the dynamics in (3+1)-D very well, at the level of both volume-averaged quantities and field spectra\footnote{We plan to publish a technical note explaining the implementation of 2+1-dimensional lattice simulations in \CL in \url{https://cosmolattice.net/technicalnotes/}, together with an update of the code.}. Furthermore, we checked that this holds true for simulations with multiple daughter fields as well. The discrete field equations have mainly been solved with the velocity-verlet integrator of 2nd order of accuracy implemented in the code, although for some model parameters we have required the 4th order one. Depending on the particular scenario, we have used lattices between $N^2=128^2$ and $1024^2$ points.

In the following subsections we consider different particularizations of potential (\ref{eq:potential}):
\begin{align}
    V(\phi,X) &= V_{\rm t} (\phi) + \frac{1}{2} g^2 X^2 \phi^2 \ ,  \hspace{0.2cm} &\text{[Sect.~\ref{Sec:Nd_1_lambda_0}]} , \label{eq:Pot1} \\
    V(\phi,\{X_n\}) &=  V_{\rm t} (\phi) + \frac{1}{2}\phi^2  \sum_{n=1}^{N_d}  g_n^2 X_n^2 \ , \label{eq:Pot2} \hspace{0.2cm} &\text{[Sect.~\ref{Sec:MultipleDaughterFields}]} ,\\ 
    V(\phi,X) &=   V_{\rm t} (\phi) +  \frac{1}{2}g^2\phi^2 X^2 + \frac{1}{4} \lambda X^4  \ ,  \label{eq:Pot3} \hspace{0.2cm} &\text{[Sect.~\ref{Sec:SelfInteract}]} ,\\
    V(\phi,\{X_n\}) &=   V_{\rm t} (\phi) + \frac{1}{2} \phi^2 \sum_{n=1}^{N_d}  g_n^2 X_n^2  +  \frac{1}{4} \sum_{n,m=1}^{N_d}  \lambda_{nm} X_n^2 X_m^2 \ , \label{eq:Pot4} \hspace{0.2cm} &\text{[Sect.~\ref{Sec:Ndg1-lambdag0}]} .
\end{align}  
Sect.~\ref{Sec:Nd_1_lambda_0} reviews the scenario of \textit{one} daughter field coupled to the inflaton through a quadratic-quadratic interaction, already studied in detail in \PartI. Sect.~\ref{Sec:MultipleDaughterFields} considers the case of \textit{multiple} daughter fields coupled to the inflaton with different strengths. Sect.~\ref{Sec:SelfInteract} analyzes the case of one daughter field with a quartic self-interaction ${\sim}\lambda X^4$. Finally, in Sect.~\ref{Sec:Ndg1-lambdag0} we consider multi-daughter field scenarios, but including now interactions between them of the type ${\sim}\lambda_{mn}X_n^2 X_m^2$ (which incorporates quartic self-interactions when $n=m$).

In all cases we take the $\alpha$-attractor T-model potential (\ref{eq:inflaton-potential}) for the inflaton, and consider different values of $p$. We have fixed $M=10m_{\rm pl}$, so that $\phi_{\rm i}>\phi_*$ and the inflaton oscillates in the positively curved region of the potential (note that this value of $M$ is in slight tension with the upper bound for the tensor-to-scalar ratio of the inflationary perturbations \cite{BICEP:2021xfz}, $M \lesssim (8.5 - 9.5) m_{\rm pl}$, but the dynamics is very similar for smaller values of $M$ as long as $\phi_{\rm i} \gtrsim \phi_*$). The amplitude of the plateau is fixed through the relation $\Lambda = \Lambda (p,M,N_k)$ for $N_k = 60$ (see comment after Eq.~\ref{eq:inflaton-potential}).\footnote{Note that this gives rise to a slight inconsistency, as the exact value of $N_k$ can only be determined with knowledge of the full post-inflationary evolution of the equation of state, which we obtain from a lattice simulation in which $N_k$ has a priori been fixed. This was solved in \PartI~with an iterative procedure, in which several lattice simulations allow to determine $N_k$ up to a factor $\mathcal{O} (10^{-2})$ (see Section 5.A of that paper for more details). However, we found that the post-inflationary dynamics remains basically unchanged between different simulations, so here we have fixed $N_k=60$ for simplicity.}

In the following analysis, we will characterize the evolution of the energy distribution in terms of `energy density ratios' (or simply `energy ratios') $\varepsilon_i \equiv \langle E_i \rangle / \langle \sum_i E_i \rangle$,  defined as the relative contribution of each (volume-averaged) energy density component $i$ to the total (volume-averaged) energy density. Different contributions include the kinetic and gradient energies of each field (defined as $\varepsilon_{\rm k}^{f}$ and $\varepsilon_{\rm g}^{f}$ for $f=\varphi$, $\chi$ respectively), and the different terms of the potential (defined as $\varepsilon_{\rm p}^t$ with $t$ labelling the corresponding term). Expressions for the energy ratios $\varepsilon_i$ and equation of state $w$ are given in Appendix \ref{App:EnergiesEos}. Bared quantities $\bar \varepsilon_i$ and $\bar w$ denote the corresponding oscillation-averaged expressions.\footnote{In the Figures presented in this paper, the oscillation averages of the energy ratios and equation of state have been obtained by means of a mean filter.}

\subsection{One daughter field without self-interaction: \texorpdfstring{$N_d=1$}{Nd=1}, \texorpdfstring{$\lambda=0$}{l=0}} \label{Sec:Nd_1_lambda_0}

The case of one daughter field without quartic self-coupling was studied extensively in \PartI, so here we simply summarize our main results. The values attained by the energy ratios and effective equation of state at late times are given in Table~\ref{Tab:Resultsn1ql0} for different choices of $p$ and $q_*$ (by setting $N_d = 1$). Let us briefly consider each case: \vspace{0.2cm}

\textbf{a)} $\boldsymbol{p=2}$: The inflaton does not get excited via self-resonance, but the daughter field does via broad parametric resonance as long as $\tilde{q} \equiv q_* a^{-3} \gtrsim 1$. If the stage of broad resonance takes long enough, backreaction effects trigger the decay of the inflaton homogeneous mode, and the equation of state deviates from $\bar w = \bar w_{\rm hom}\, (=0)$ towards $ \bar w =\bar w_{\rm max}\, ( \lesssim 1/3)$. In any case, the production of daughter field fluctuations terminates once $\tilde q \lesssim 1$. As fluctuations dilute as radiation, the inflaton homogeneous mode (which dilutes as matter in the present case) eventually dominates the energy budget again, and we get $\bar w \rightarrow 0$ and  $\bar \varepsilon_{\rm k}^\varphi \simeq  \bar \varepsilon_{\rm p}^{\varphi^2} \rightarrow 1/2$ at late times. \vspace{0.2cm}
 
\textbf{b)} $\boldsymbol{2<p<4}$: The daughter field gets excited through broad parametric resonance as long as $\tilde q \equiv q_* a^{\frac{6 (p-4)}{p+2}} \gtrsim 1$. However, unlike the $p=2$ case, now the inflaton also develops fluctuations via self-resonance. Although the excitation of the daughter field is initially much stronger than the one of the inflaton (see the typical Floquet indices in Fig.~\ref{fig:stabilitychart}), it eventually terminates when the daughter field resonance becomes narrow ($\tilde q \lesssim 1$), like in the $p=2$ case. On the other hand, the strength of the inflaton self-resonance remains constant. Therefore, inflaton fluctuations are continuously produced even during the non-linear regime, and we end up in a universe with $\bar \varepsilon_{\rm k}^{\varphi}$, $\bar \varepsilon_{\rm g}^{\varphi} \rightarrow 1/2$ and $\bar w \rightarrow 1/3$ at late times.\vspace{0.2cm}

\textbf{c)} $\boldsymbol{p \geq 4}$: The resonance parameter $\tilde q \equiv q_* a^{\frac{6 (p-4)}{p+2}}$ is either constant (for $p=4$) or grows with time (for $p>4$), so both  inflaton self-resonance and broad parametric resonance of the daughter field are present at late times. In this case we end up in an equilibrium regime in which both fields end up completely fragmented and have the same energy:  we get $\bar \varepsilon_{\rm k}^{\varphi}$, $\bar\varepsilon_{\rm g}^{\varphi} $, $\bar\varepsilon_{\rm k}^{\chi} $, $\bar \varepsilon_{\rm g}^{\chi} \rightarrow 1/4$ and $\bar w \rightarrow 1/3$ at late times.

\begin{table}
    \centering \textbf{Final energy ratios for $N_d \geq 1$, $\lambda_{nm} = 0$}
    \def\arraystretch{1.7}
    \begin{center}
        \begin{tabular}{ |p{3cm}|c|c|c|c|c|c|c| } 
            \hline
            \centering  $p$, $q_*$ & $\bar w$ & $\bar{\varepsilon}_{\rm k}^{\varphi}$ & $\bar{\varepsilon}_{\rm g}^{\varphi}$ & $\bar{\varepsilon}_{\rm k}^{\chi}$ & $\bar{\varepsilon}_{\rm g}^{\chi}$ & $\bar{\varepsilon}_{\rm p}^{\varphi^p}$ & $\bar{\varepsilon}_{\rm p}^{\varphi^2 \chi^2}$\\   \hline
            \centering $p=2, \forall\, q_*$ &0 & 1/2 &  0 & 0 & 0 & 1/2 & 0 \\   \hline
            \centering $2 < p < 4, \forall\, q_*$ & 1/3  & 1/2 & 1/2 & 0 & 0 & 0 & 0 \\   \hline
            \centering $p \geq 4$, $q_* = 0$ & 1/3 & 1/2 & 1/2 & 0 & 0 & 0 & 0 \\   \hline
            \centering $p \geq 4$, $q_*>0$ & 1/3 & $\frac{1}{2(1+N_d)}$ & $\frac{1}{2(1+N_d)}$ & $\frac{1}{2(1+N_d)}$ & $\frac{1}{2(1+N_d)}$ & 0 & 0 \\   \hline
        \end{tabular} 
    \end{center}
    \caption{[Potential (\ref{Sec:MultipleDaughterFields})] Values achieved by the equation of state and energy ratios at late times for different choices of $p$, $q_*$, and number of daughter fields $N_d$.} \label{Tab:Resultsn1ql0}
\end{table}

\subsection{Multiple daughter fields without (self-)interactions: \texorpdfstring{$N_d\geq 1$}{Nd>1}, \texorpdfstring{$\lambda_{nm}=0$}{lmn=0}} \label{Sec:MultipleDaughterFields}

Let us now consider the case of an inflaton coupled to multiple daughter fields via quadratic-quadratic interactions, described by potential (\ref{eq:Pot2}). We consider the power-law coefficients: a) $p \geq 4$, b) $p=2$, and c) $2 < p < 4$.\vspace{0.2cm}

\begin{figure}
    \centering
    \includegraphics[width=7.5cm]{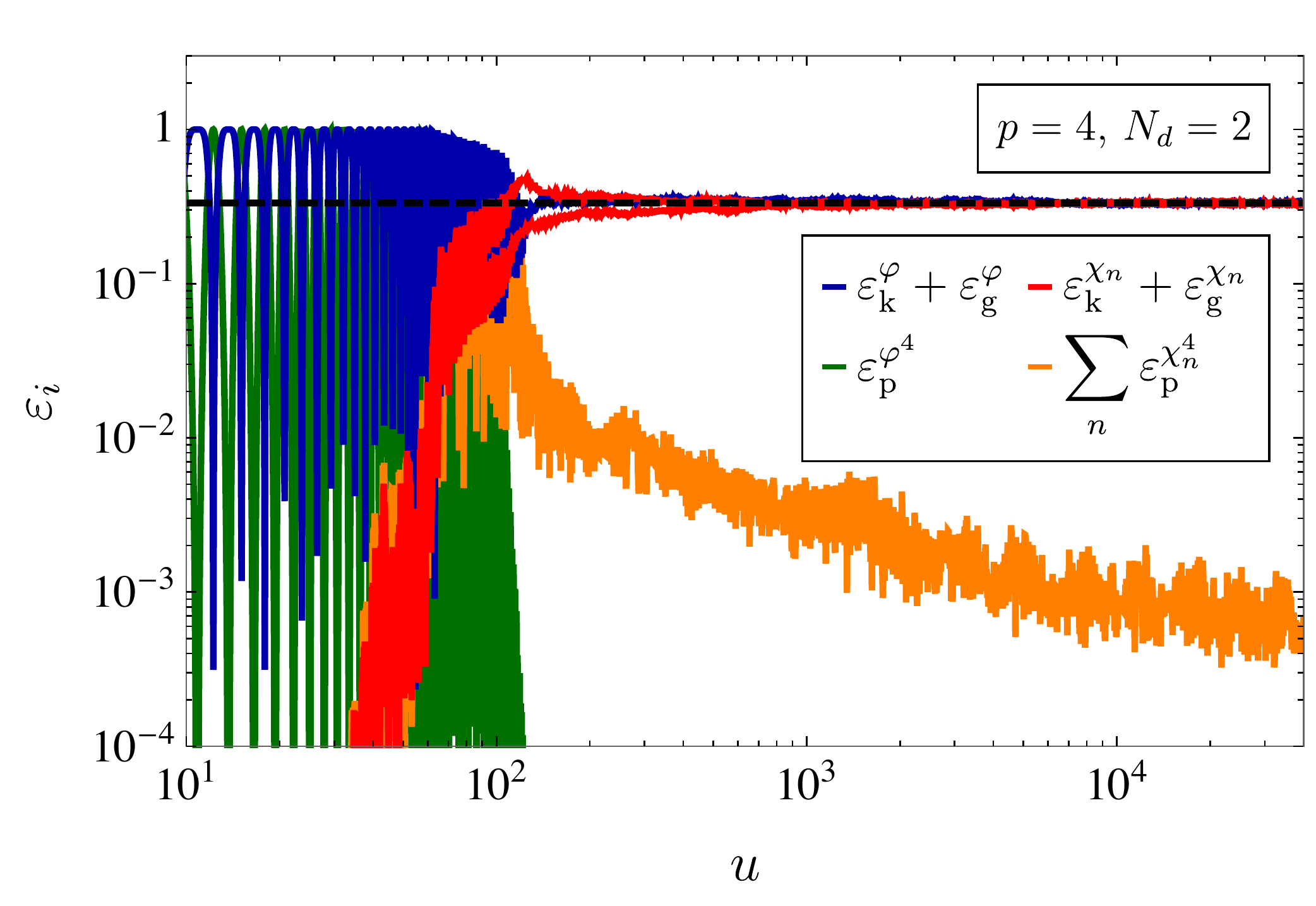}  
    \includegraphics[width=7.5cm]{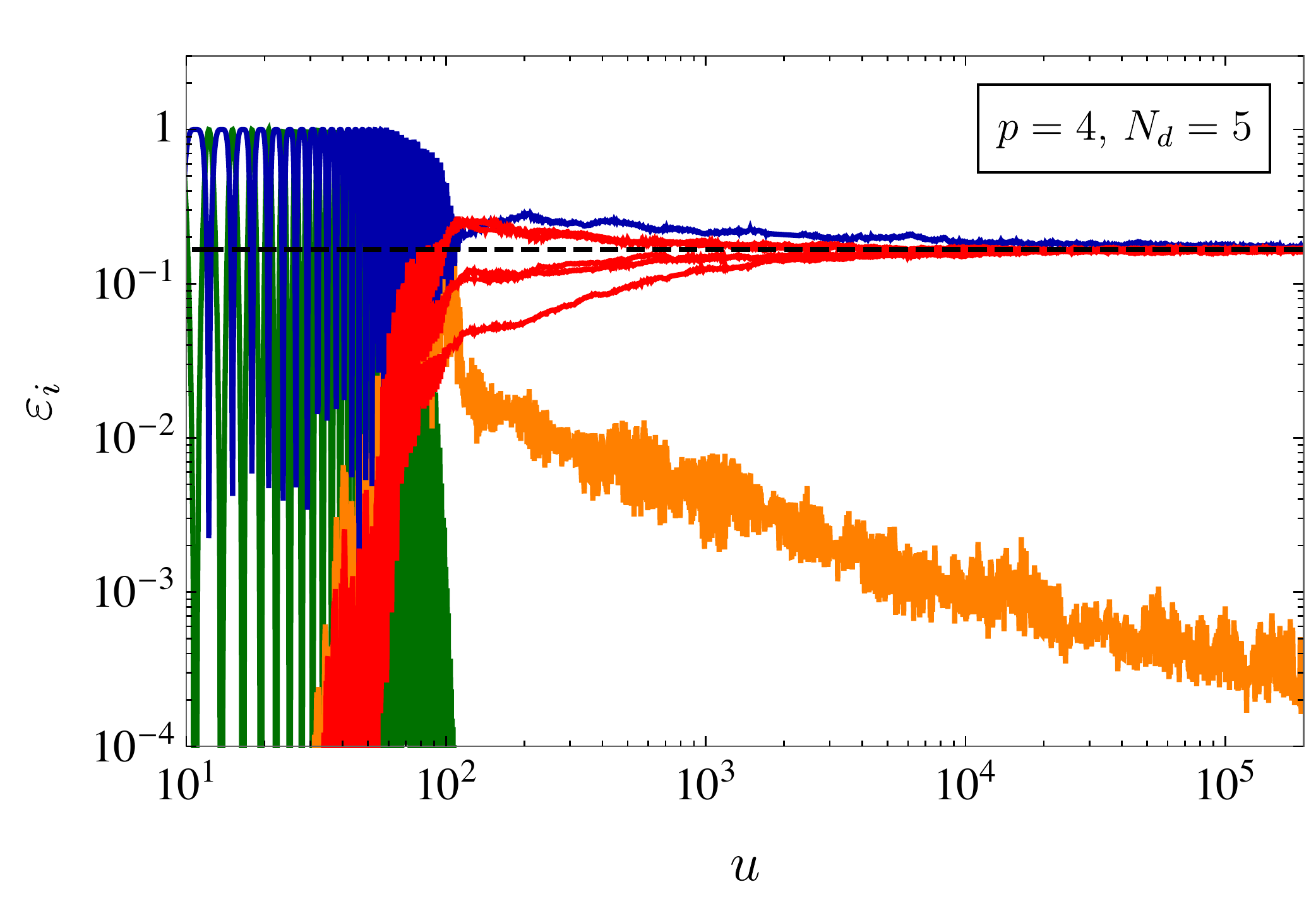}  
    \includegraphics[width=7.5cm]{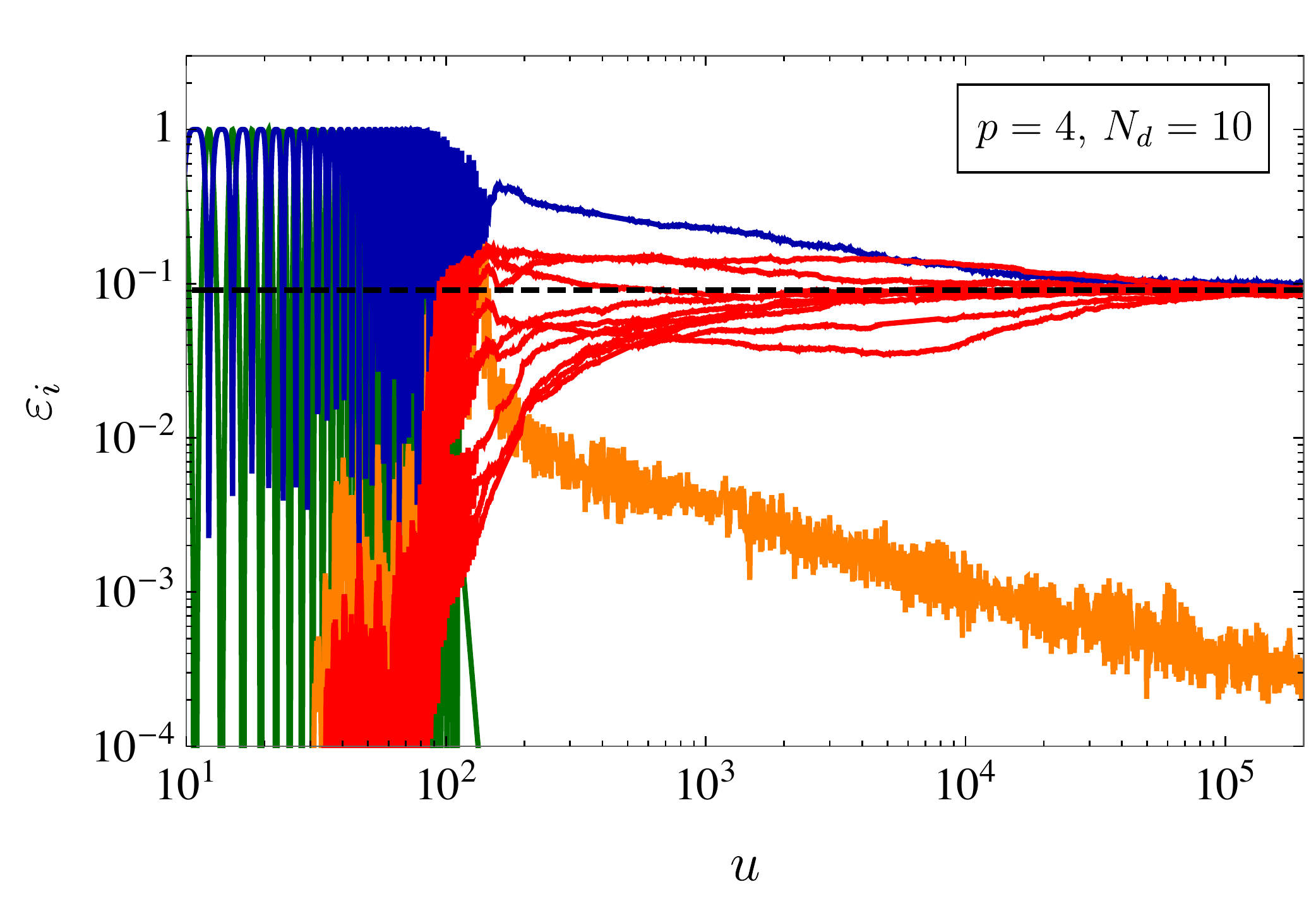}
    \includegraphics[width=7.5cm]{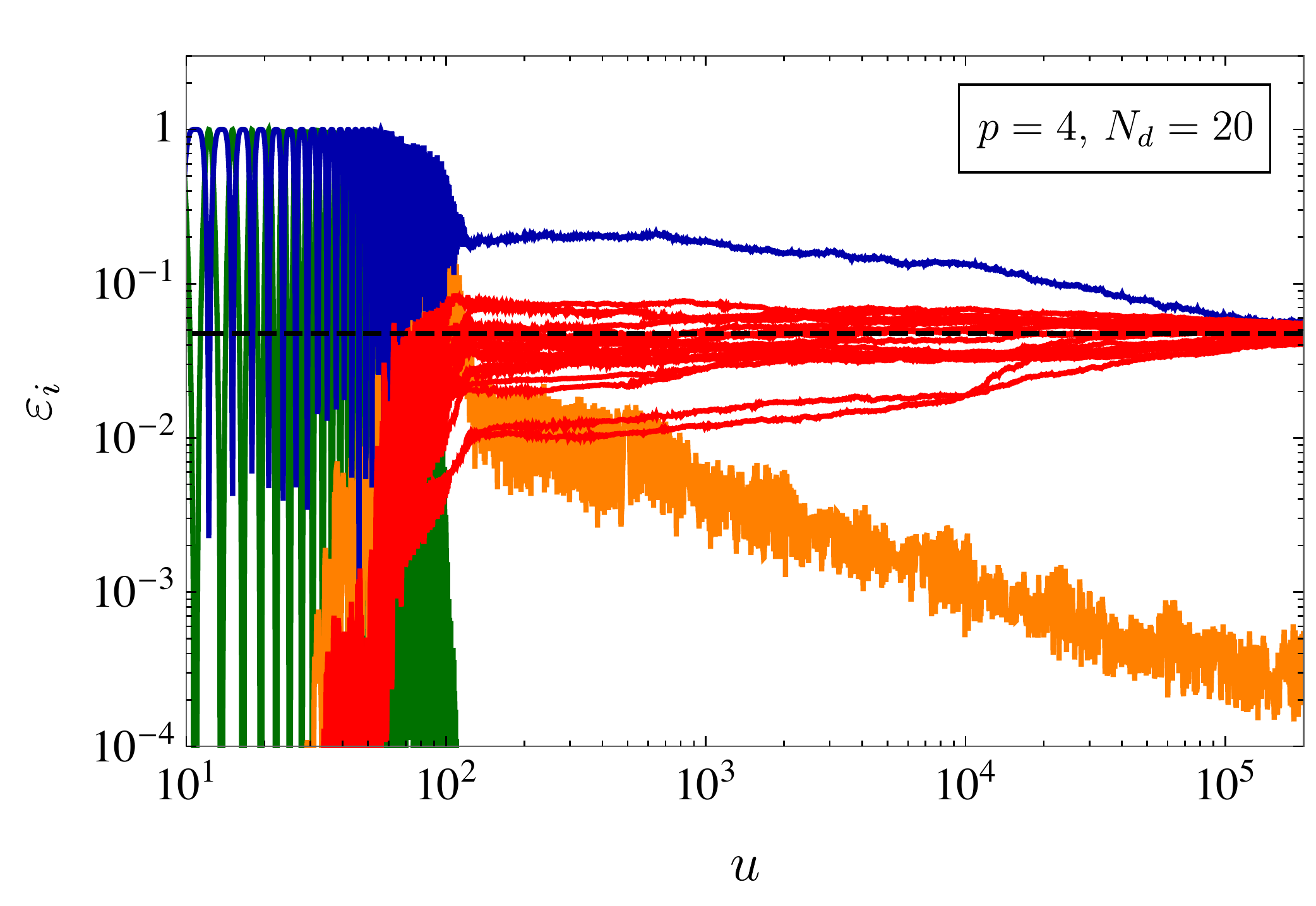}
    
     \caption{[Potential (\ref{eq:Pot2}), $p=4$] Evolution of the energy ratios for different number of daughter fields: $N_d=2,5,10$ and $20$. All of them are coupled to the inflaton with the same resonance parameter $q_*=10^4$. We have plotted the sum of the kinetic and gradient contributions of the inflaton (blue) and the different daughter fields (red lines). The potential contributions, which become negligible at late times, are depicted in green and orange. The dashed line indicates the predicted value (\ref{eq:EnergyRatios-Ndg1}) for the energy ratios at late times.} \label{fig:En-MultipleDaughter0} \vspace*{0.5cm}

     \includegraphics[width=7.5cm]{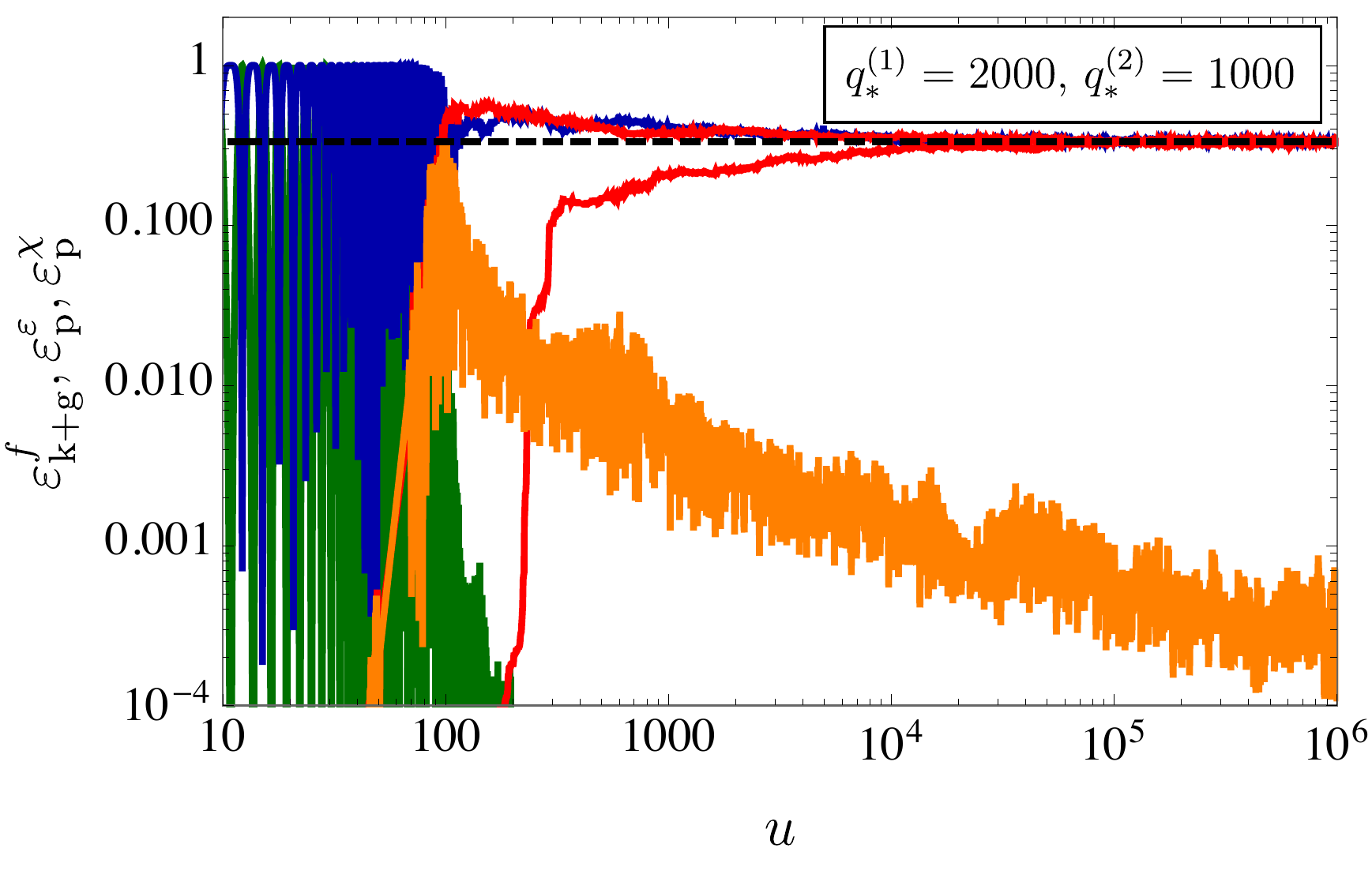}
      \includegraphics[width=7.5cm]{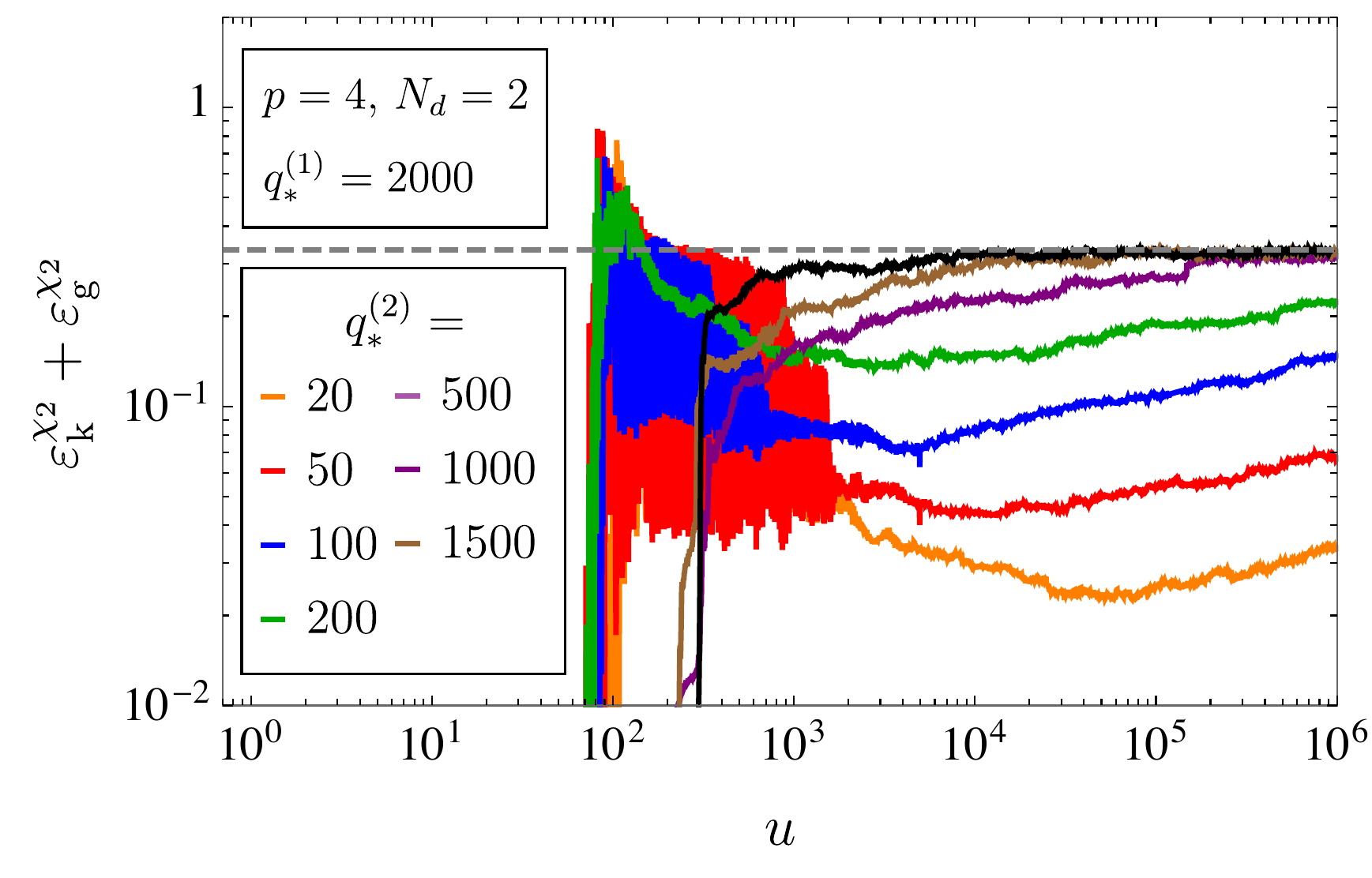}
        
      \caption{[Potential (\ref{eq:Pot2}), $p=4$] Left: Evolution of the energy ratios for a system of two daughter fields with $q_*^{(1)} > q_*^{(2)}$. Right: Evolution of the fraction of energy stored in $\chi_2$, for $q_*^{(1)} = 2000$ and different values of $q_*^{(2)} (< q_*^{(1)})$. The dashed line indicates the expected value $\varepsilon_k^{\chi_2} + \varepsilon_g^{\chi_2} \simeq 1/3$, see Eq.~(\ref{eq:EnergyRatios-Ndg1}).} 
     \label{fig:En-MultipleDaughter}
       
\end{figure}

\textbf{a)} $\boldsymbol{p \geq 4}$: In Fig.~\ref{fig:En-MultipleDaughter0} we show the post-inflationary evolution of the energy ratios for $p=4$ and different number of daughter fields ($N_d = 2, 5, 10$ and $20$). Each daughter field is coupled to the inflaton with the same resonance parameter, $q_*^{(n)}=10^4$. For each field we depict the sum of its kinetic and gradient energy ratios: at late times this represents the total energy fraction stored in the field, as the potential energy contributions become negligible.

During the linear regime (for times $u \sim 0 - 100$), the excitation strength for each daughter field is characterized by its resonance parameter $q_{*}^{(n)}$. However, although all daughter fields in Fig.~\ref{fig:En-MultipleDaughter0} have the same $q_{*}^{(n)}$, we observe that the amount of energy transferred during this regime is larger for some daughter fields than for others. This effect can be attributed to the randomness of the initial fluctuations: varying the seed of the random generator changes which particular fields receive more energy. 

However, in the deep non-linear stage we always get an equilibrium regime between the inflaton and the daughter fields, analogous to the one in the single daughter field case discussed in Sect.~\ref{Sec:Nd_1_lambda_0}. More specifically, the energy is equally distributed at very late times between the inflaton and all daughter fields (as long as they are coupled in broad resonance), with the energy ratios satisfying
\be \bar \varepsilon_{\rm k}^{\varphi} \simeq \bar \varepsilon_{\rm g}^{\varphi} \simeq \bar\varepsilon_{\rm k}^{\chi_n} \simeq \bar \varepsilon_{\rm g}^{\chi_n} \simeq \frac{1}{2(1+N_d)} \ , \hspace{0.5cm} 
\varepsilon_p \ll 1 \ ,  \hspace{0.8cm}  [p\geq4] \ . \label{eq:EnergyRatios-Ndg1} \ee
For example, for $N_d = 2, 5, 10$ and $20$, each of the $N_d + 1$ fields of the system get 33\%, 17\%, 9.1\% and 4.8\% of the total energy respectively.  Remarkably, this equilibrium regime between the different daughter fields is achieved without including explicit interactions between them, such as $X_i^2 X_j^2$.

A similar equilibration regime is achieved even if the daughter fields are coupled to the inflaton with different strengths. This can be seen in the left panel of Fig.~\ref{fig:En-MultipleDaughter}, where we depict the case $N_d=2$, $q_*^{(1)}=2000$ and $q_*^{(2)}=1000$. Both daughter fields end up with the same energy despite having different resonance parameters. However, the smaller the ratio $q_*^{(2)} / q_*^{(1)}$ is, the later the equilibration between all fields is achieved. We illustrate this in the right panel of  Fig.~\ref{fig:En-MultipleDaughter}, where we depict the fraction of energy transferred to $\chi_2$ as a function of time, for different ratios $q_*^{(2)} / q_*^{(1)}$. For small ratios $q_*^{(2)} / q_*^{(1)} \lesssim 1/10$, we observe that the fraction of energy stored in $\chi_2$ decreases for some time during the early non-linear regime, develops a local minimum and starts growing again. Eventually it attains the same energy as $\chi_1$ at late times. In reality, for these ratios we are unable to observe the complete achievement of equilibration on the lattice, but the long-term trend can be extrapolated in all cases.

These examples illustrate that a significant depletion of the inflaton energy can be achieved in multi-field scenarios like the ones considered here, without relying on perturbative decay channels like in combined preheating scenarios  \cite{Bezrukov:2008ut,Garcia-Bellido:2008ycs,Repond:2016sol,Fan:2021otj}.

\begin{figure}
    \centering
    \includegraphics[width=7.5cm]{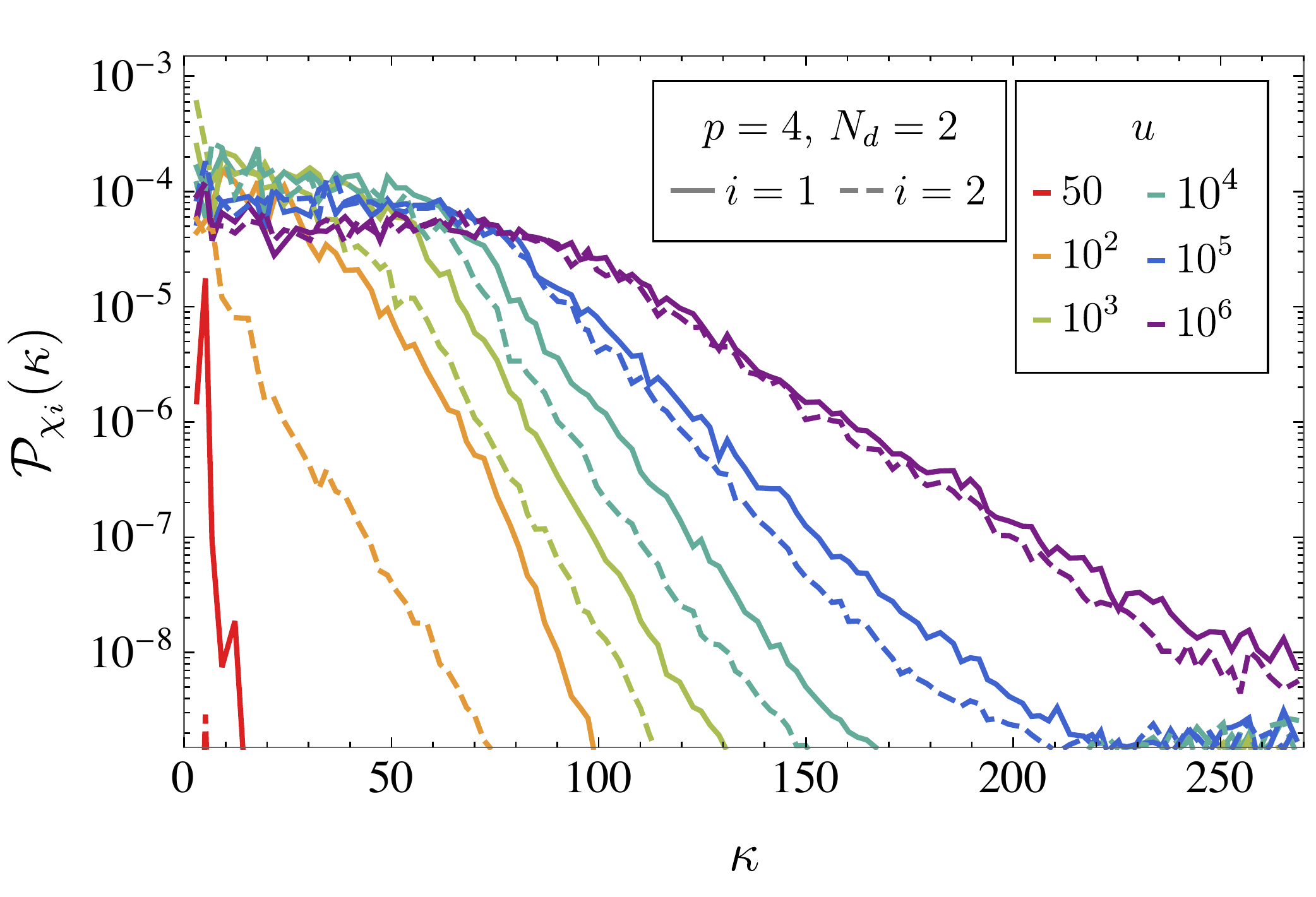}
    \includegraphics[width=7.5cm]{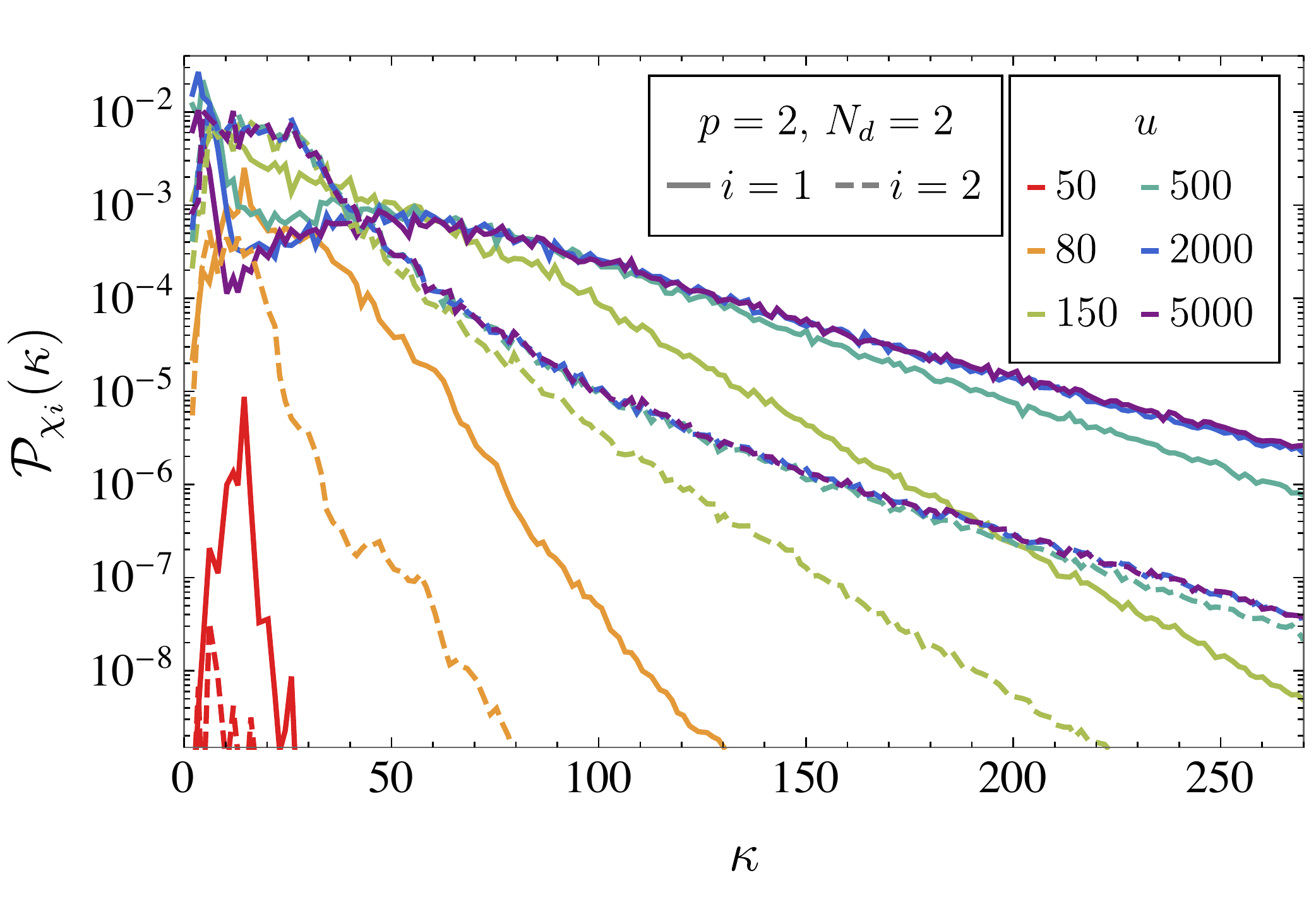}
    \caption{[Potential (\ref{eq:Pot2}), $p=2$] Evolution of the daughter field spectra for a system with two daughter fields ($N_d=2$). Solid and dashed lines correspond to $\chi_1$ and $\chi_2$ respectively, and we compare the spectra at the same times. The left panel shows the case:  $p=4$, $q_*^{(1)}=2\cdot10^3$ and $q_*^{(2)}=10^3$. The right panel shows the case: $p=2$, $q_*^{(1)}=10^5$ and $q_*^{(2)}=10^4$. } 
\label{fig:spectra-p2p4}
\end{figure}

It is also interesting to show how the equilibration regime between the different daughter fields is achieved in momentum space. For this purpose we inspect the power spectrum of the fields in natural variables, defined as $\langle f^2 \rangle \equiv \int d\, \rm{log}\,\kappa\, {\mathcal{P}}_f(\kappa)$ for $f=\{\varphi,\chi_i\}$. In the left panel of Fig.~\ref{fig:spectra-p2p4} we show the time-evolution of the daughter field spectra for $N_d=2$ and $p=4$ (i.e.~the same case as in the left panel of Fig.~\ref{fig:En-MultipleDaughter}). We observe that different ranges of momenta are populated during the linear regime for both fields, which are characterized by $q_*^{(1)}$ and $q_*^{(2)}$ for $\chi_1$ and $\chi_2$ respectively. However, the spectra of both fields converge at later times.

Finally, let us mention that we have simulated multi daughter field systems for values $p > 4$ (such as $p=4.5$ or $5$), and observed that a similar equilibration regime between the field emerges, i.e.~the energy ratios at late times are also given by Eq.~(\ref{eq:EnergyRatios-Ndg1}).\vspace{0.2cm}

\begin{figure}
    \centering
    \includegraphics[width=7.5cm]{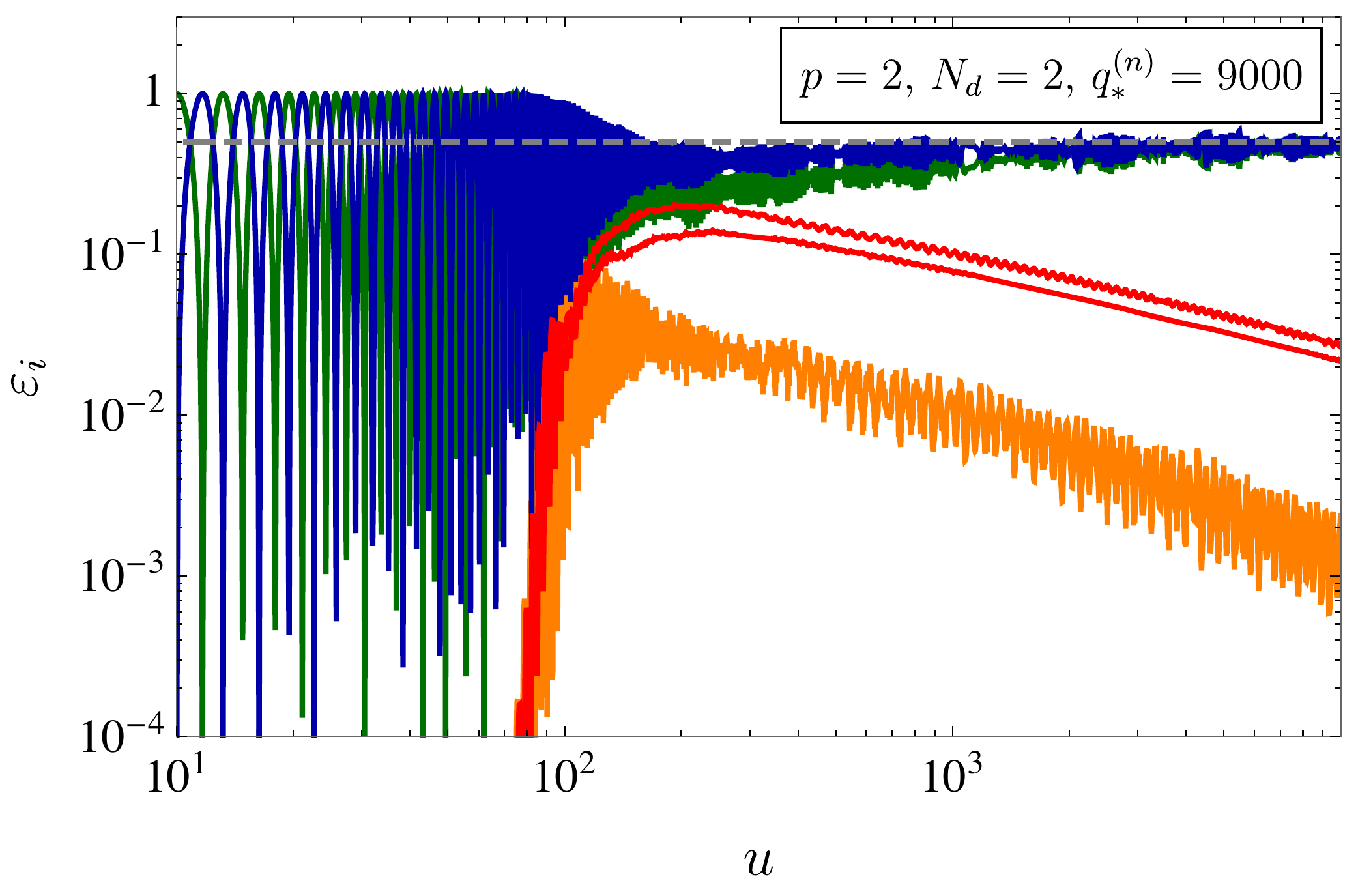} \,  
    \includegraphics[width=7.5cm]{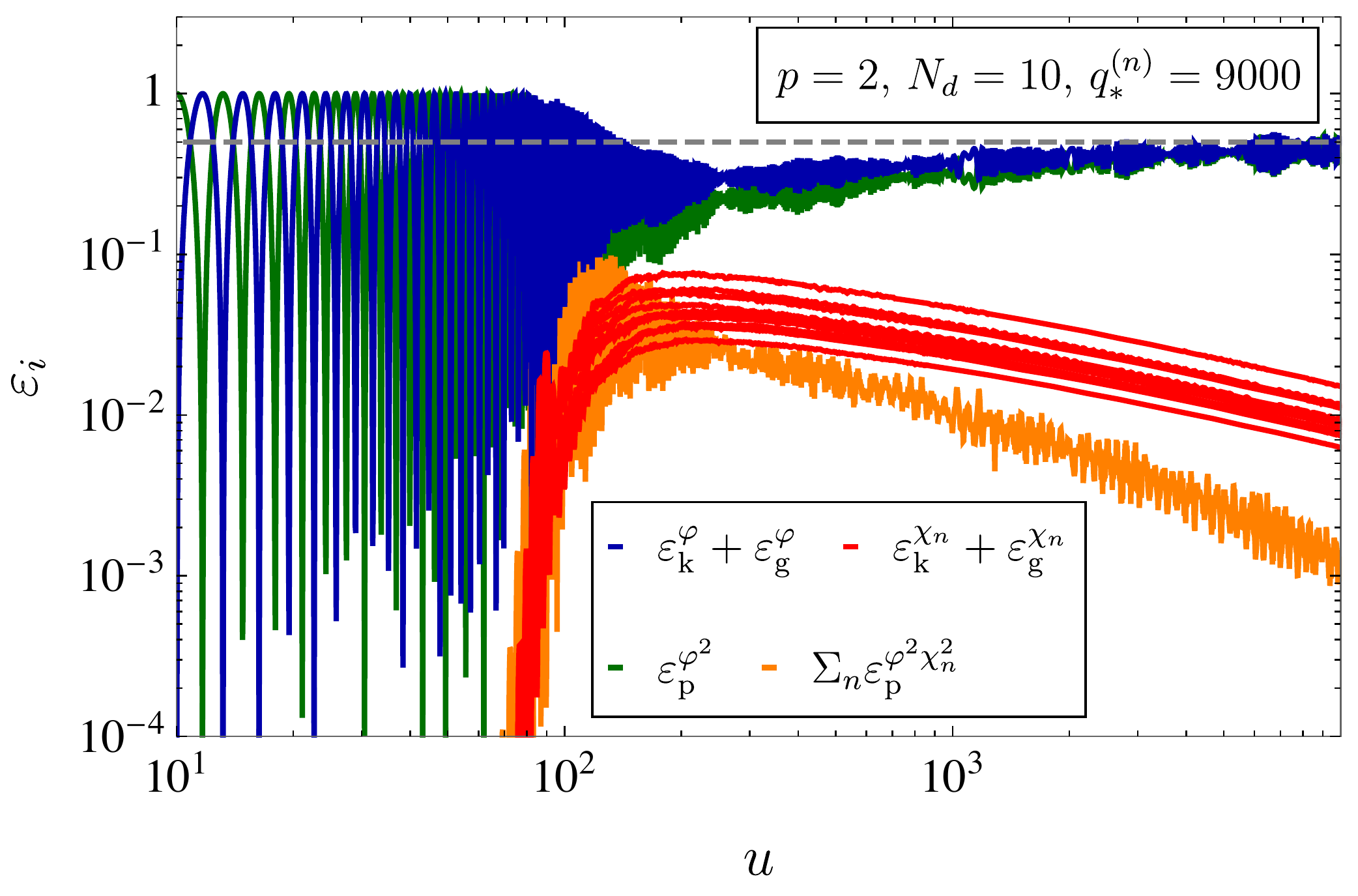} 
    \caption{[Potential (\ref{eq:Pot2}), $p=2$] Evolution of the energy distribution for $N_d =2$ (left) and $N_d =10$ (right). The resonance parameters of all daughter fields are set to the same value $q_{*}^{(n)}=9\cdot 10^3$.}\label{fig:En-Ndp2} \vspace{0.6cm}
    
    \includegraphics[width=7.5cm]{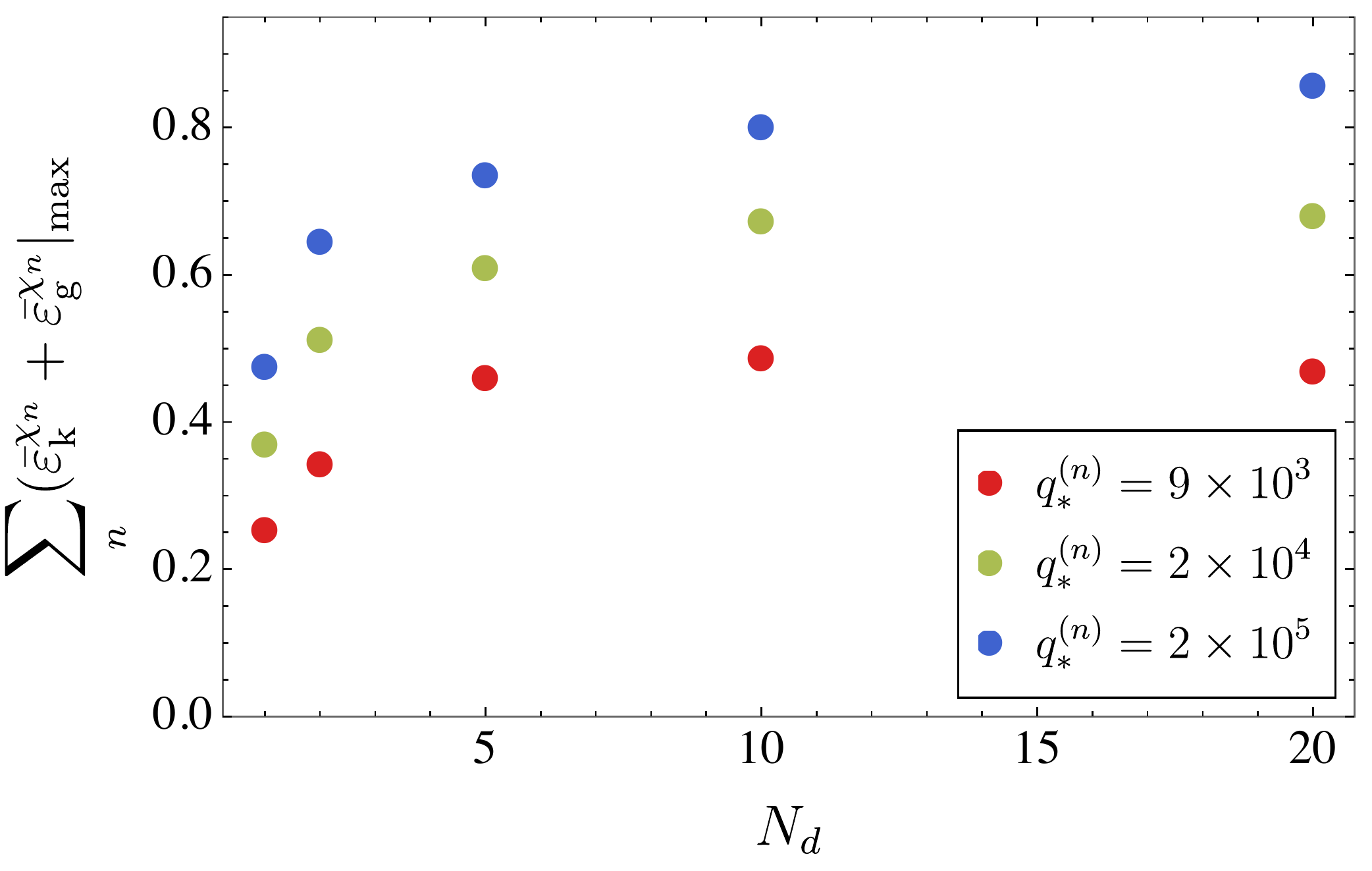} \,
    \includegraphics[width=7.5cm]{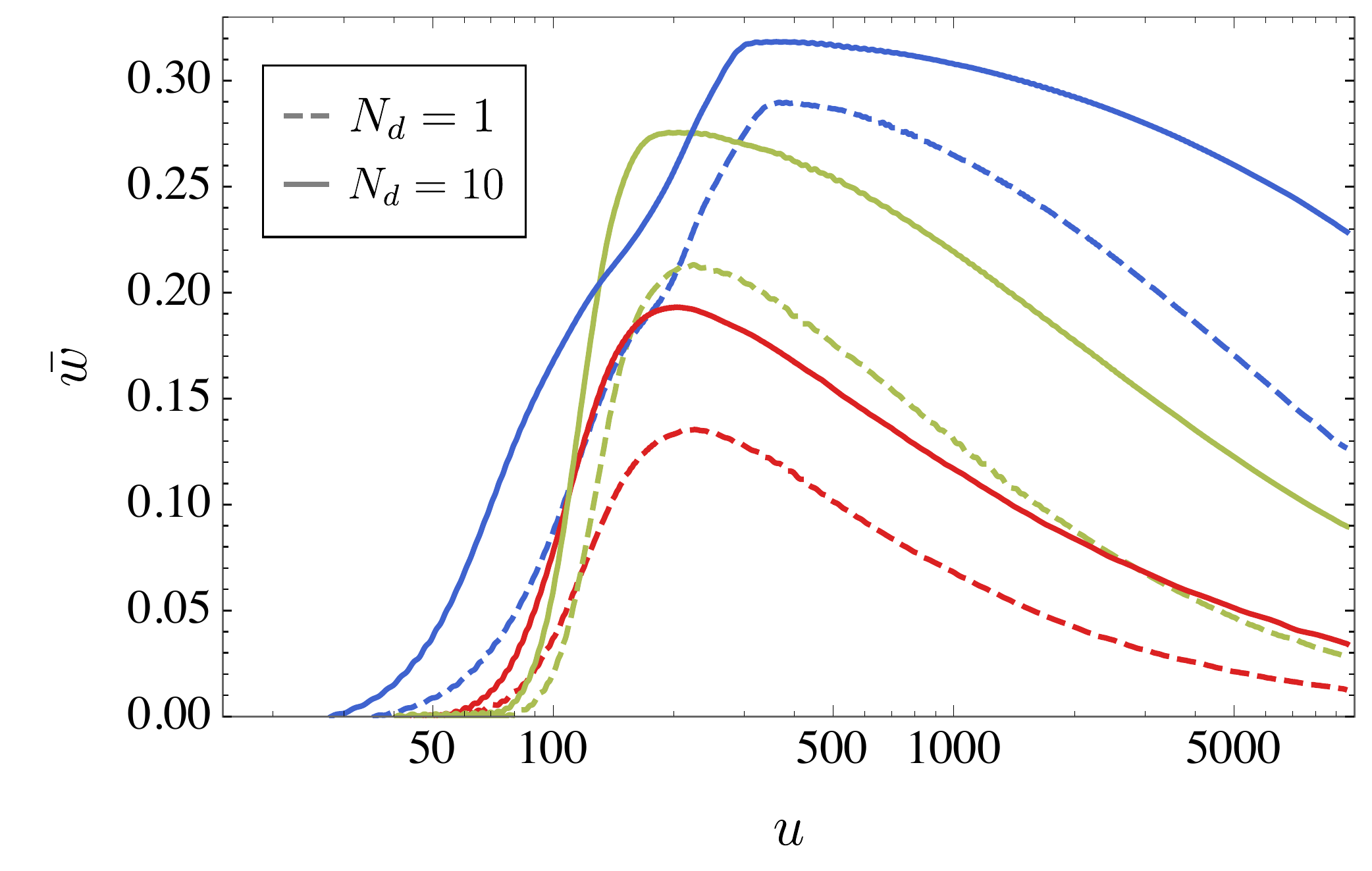}
    \caption{[Potential (\ref{eq:Pot2}), $p=2$] Left: Maximum fraction of energy attained by the daughter field sector during the simulation. We consider three choices of $q_{*}^{(n)}$ (the same one for all daughter fields), and plot the sum $\sum_n  ( \bar{\varepsilon}_{\rm k}^{\chi_n}+\bar{\varepsilon}_{\rm g}^{\chi_n} )|_{\rm max}$ as a function of $N_d$. Right: Evolution of the effective equation of state for the same values of $q_{*}^{(n)}$, and in each case for $N_d = 1$ (dashed) and $N_d = 10$ (solid).
    } \label{fig:En-Ndp2B}
\end{figure}

 \textbf{b)} $\boldsymbol{p=2}$: We depict in Fig.~\ref{fig:En-Ndp2} the evolution of the energy ratios for $p=2$ and $N_d=2$ (left panel), $N_d =10$ (right panel), where the resonance parameters of all daughter fields are the same $q_*^{(n)}=9000$. Similarly to the single daughter field case, during the initial stage of broad parametric resonance, the energy ratios of all daughter fields increase exponentially, until backreaction effects trigger the decay of the inflaton homogeneous mode and the non-linear regime starts. Remarkably, the amount of energy transferred to each daughter field during this stage is different even if they have the same $q_*^{(n)}$. As in the $p=4$ case, this effect is induced by the initial random fluctuations, as different realizations change which particular fields get more energy. In any case, the broad resonance terminates once $\tilde q \lesssim 1$, so the inflaton homogeneous mode eventually recovers 100\% of the total energy, and the energy transferred to the daughter field sector becomes subdominant. Similarly, the fluctuations of the inflaton dilute faster than its homogeneous mode, so the equation of state goes again to $\bar w \rightarrow \bar w_{\rm hom} = 0$ at late times. Remarkably, we do not observe an equipartition regime between the different daughter fields during the non-linear regime: the field that has (randomly) received more energy during the linear stage will dominate the energy budget of the daughter field sector later on.

Although  adding more daughter fields does not change the final energy distribution and equation of state for $p=2$, it can have relevant effects at intermediate times. In order to illustrate this, we consider systems where the $N_d$ daughter fields have all the same $q_*^{(n)}$. In the left panel of Fig.~\ref{fig:En-Ndp2} we show the maximum fraction of energy attained by the daughter field sector during the simulation as a function of $N_d$ and three different choices of $q_*$ (more specifically, we depict the sum $\sum_n  ( \bar{\varepsilon}_{\rm k}^{\chi_n}+\bar{\varepsilon}_{\rm g}^{\chi_n} )|_{\rm max}$). For fixed $q_*$, the larger the number of daughter fields the greater the energy transfer is (though the average energy transferred to each individual daughter field decreases for larger $N_d$).

As the energy transferred to field gradients increases with $N_d$, so does the transitory behaviour of the equation of state. In the right panel of Fig.~\ref{fig:En-Ndp2B} we compare the evolution of the (effective) equation of state for $N_d = 1$ and 10, for the same three choices of $q_*$. As expected, the deviation from $\bar w = \bar w_{\rm hom} (= 0)$ towards $\bar w = \bar w_{\rm max} (\lesssim 1/3)$ is stronger in the $N_d = 10$ case than in the $N_d = 1$ one, although at late times we always recover the matter-dominated state $\bar w \rightarrow 0$.

Finally, we show the evolution of the daughter field spectra in the right panel of Fig.~\ref{fig:spectra-p2p4}, for $N_d=2$, $q_*^{(1)}=10^5$ and $q_*^{(2)}=10^4$.  Initially, a narrow infrared band of modes is strongly amplified for each daughter field, characterized by $q_*^{(1)}$ and $q_*^{(2)}$ respectively. Once backreaction effects become relevant, the daughter fields populate a wider range of momenta. However, once $\tilde q^{(1)}$, $\tilde q^{(2)} \lesssim 1$,  the exchange of energy ceases and the spectra freeze. Therefore, the daughter field spectra do not converge and end up with different shapes, unlike in the $p=4$ case.\vspace{0.2cm}

\textbf{c)} $\boldsymbol{2<p<4}$: The dynamics of the daughter field sector are, for these power-law coefficients, similar to the $p=2$ case discussed above. In particular, as their resonance also becomes narrow at late times, the fraction of energy stored in the daughter field sector eventually becomes negligible, $\bar \varepsilon_{\chi} \rightarrow 0$. However, the inflaton now fragments due to its self-resonance, so we have $\bar \varepsilon_{\rm k}^{\varphi}$, $\bar \varepsilon_{\rm g}^{\varphi} \rightarrow 1/2$ and $\bar w \rightarrow 1/3$ at late times.

\subsection{One daughter field with quartic self-interaction: \texorpdfstring{$N_d=1$}{Nd=1}, \texorpdfstring{$\lambda > 0$}{l>0}} \label{Sec:SelfInteract}

Let us consider the case of one daughter field with a quartic self-interaction, described by potential (\ref{eq:Pot3}). We now need to fix  two free parameters: $q_* \equiv g^2 (\phi_* / \omega_*)^2 $ and $\sigma_* \equiv \lambda (\phi_* / \omega_*)^2$. If $\sigma_* / q_* = \lambda /g^2 \gtrsim 1$, the self-interaction can significantly affect the field dynamics during both the linear and non-linear regimes (this was already noted and investigated for the early preheating phase in \cite{Prokopec:1996rr}). During the linear regime, the self-interaction gives an effective mass to the daughter field and suppresses its resonant growth, as discussed in Sect.~\ref{sec:Hartree}. However, during the non-linear regime the self-interaction also triggers a self-resonant excitation, as we shall see. We will consider the power-law coefficients: a) $p=2$,  a) $2 <p<4$, and c) $p \geq 4$. 

\vspace{0.2cm}

\begin{figure*}
    \centering
    \includegraphics[width=7.5cm]{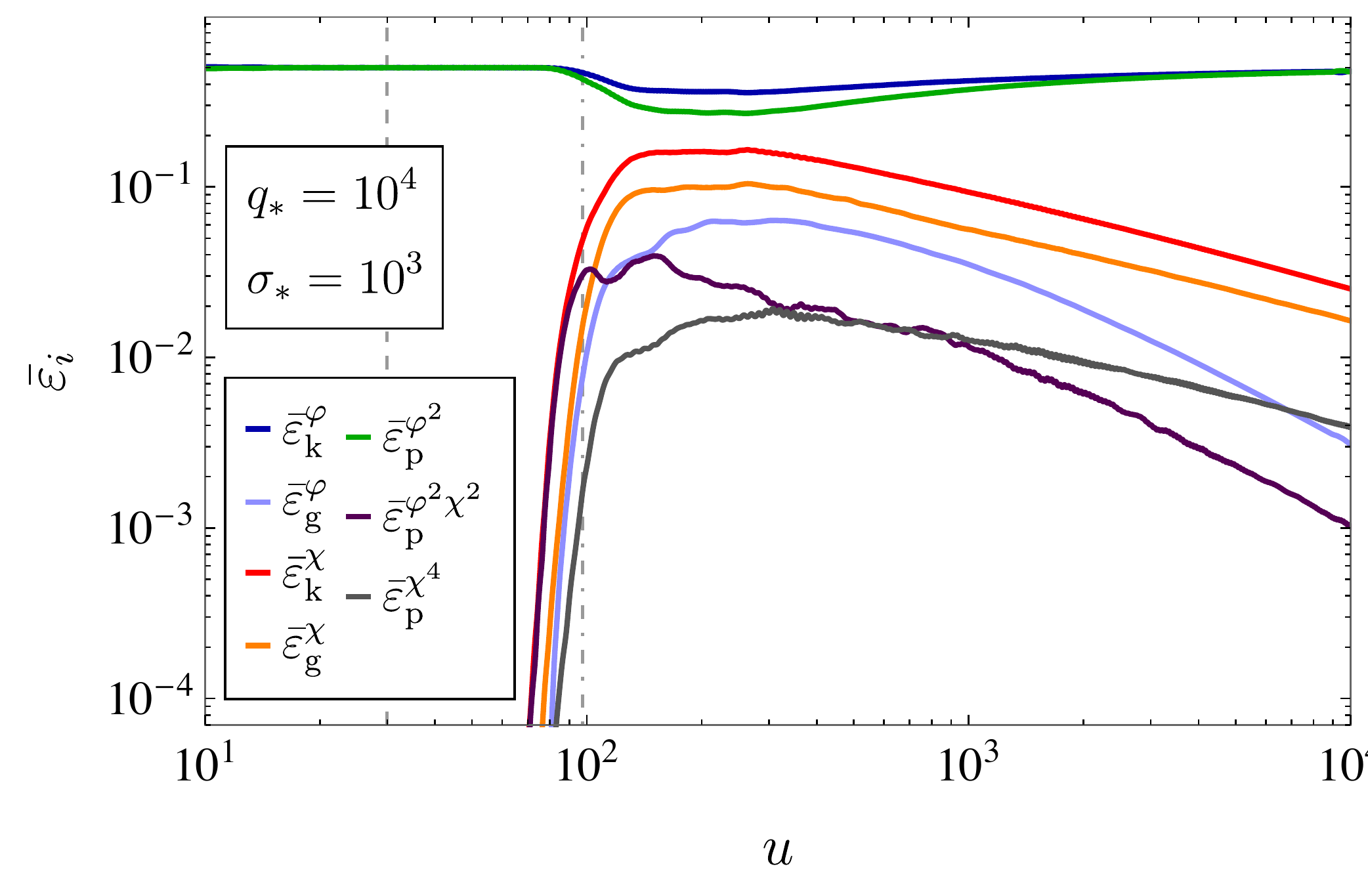} \,  
    \includegraphics[width=7.5cm]{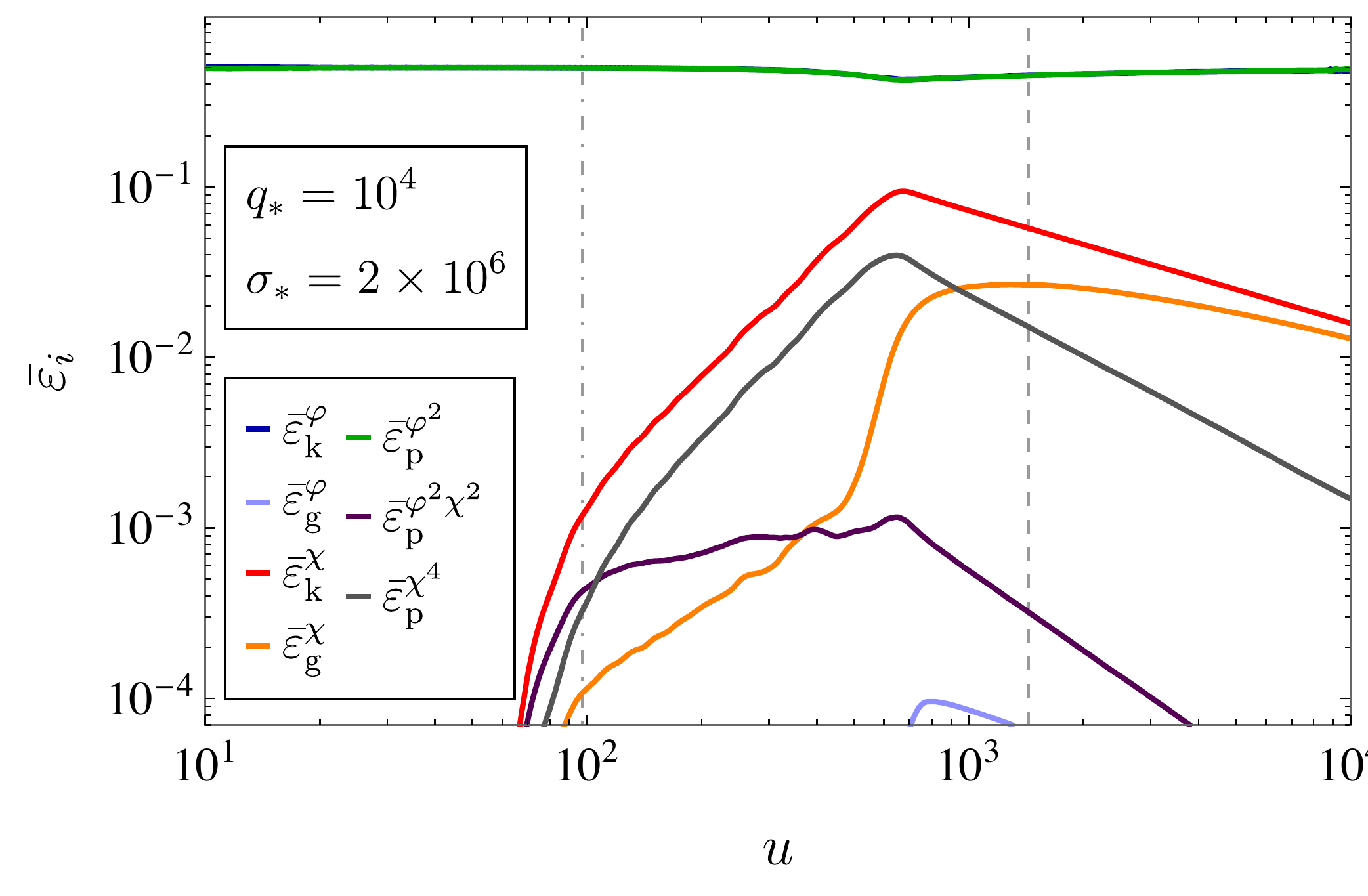} \,
    \includegraphics[width=7.5cm]{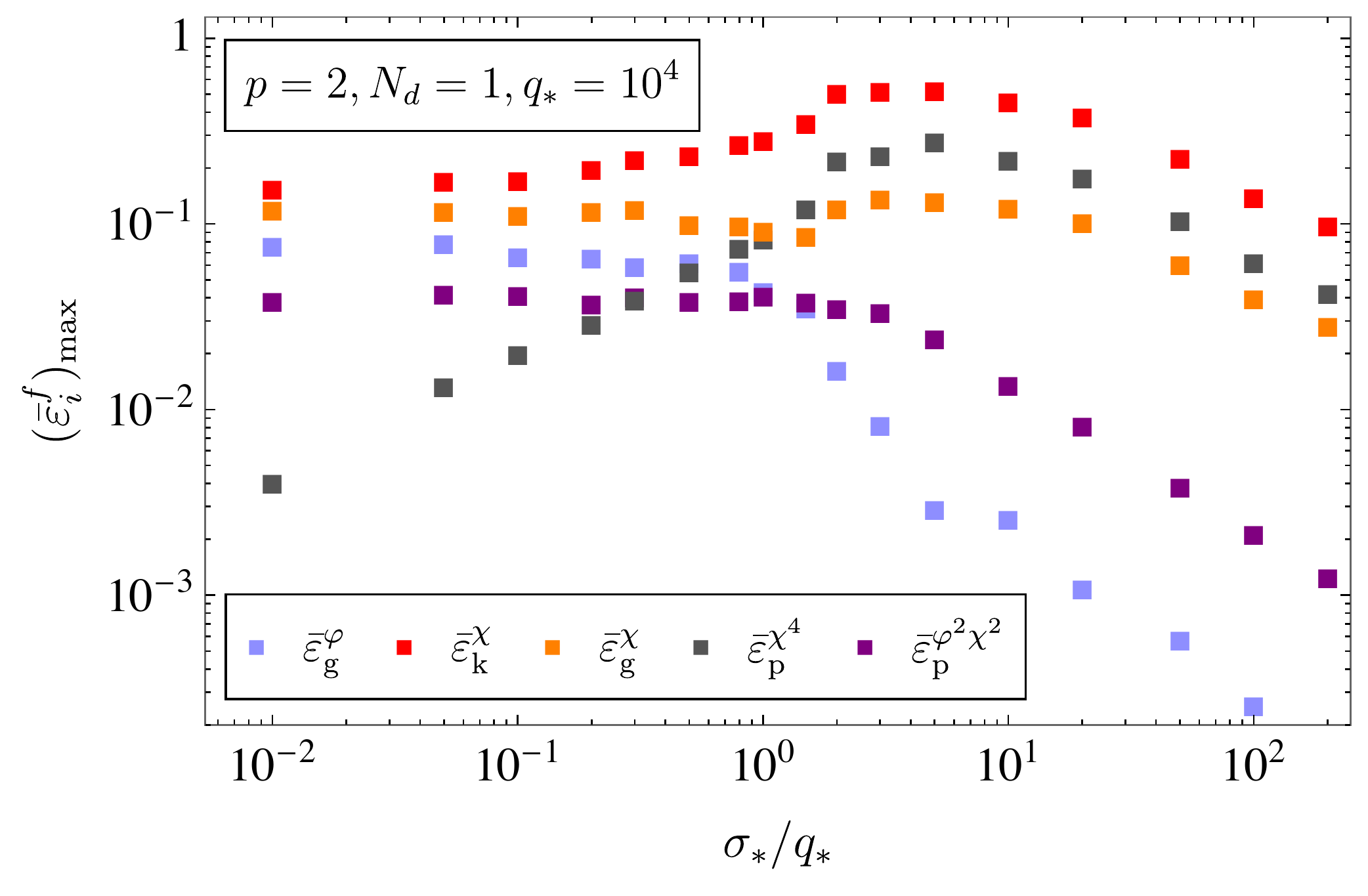} \, 
    \includegraphics[width=7.5cm]{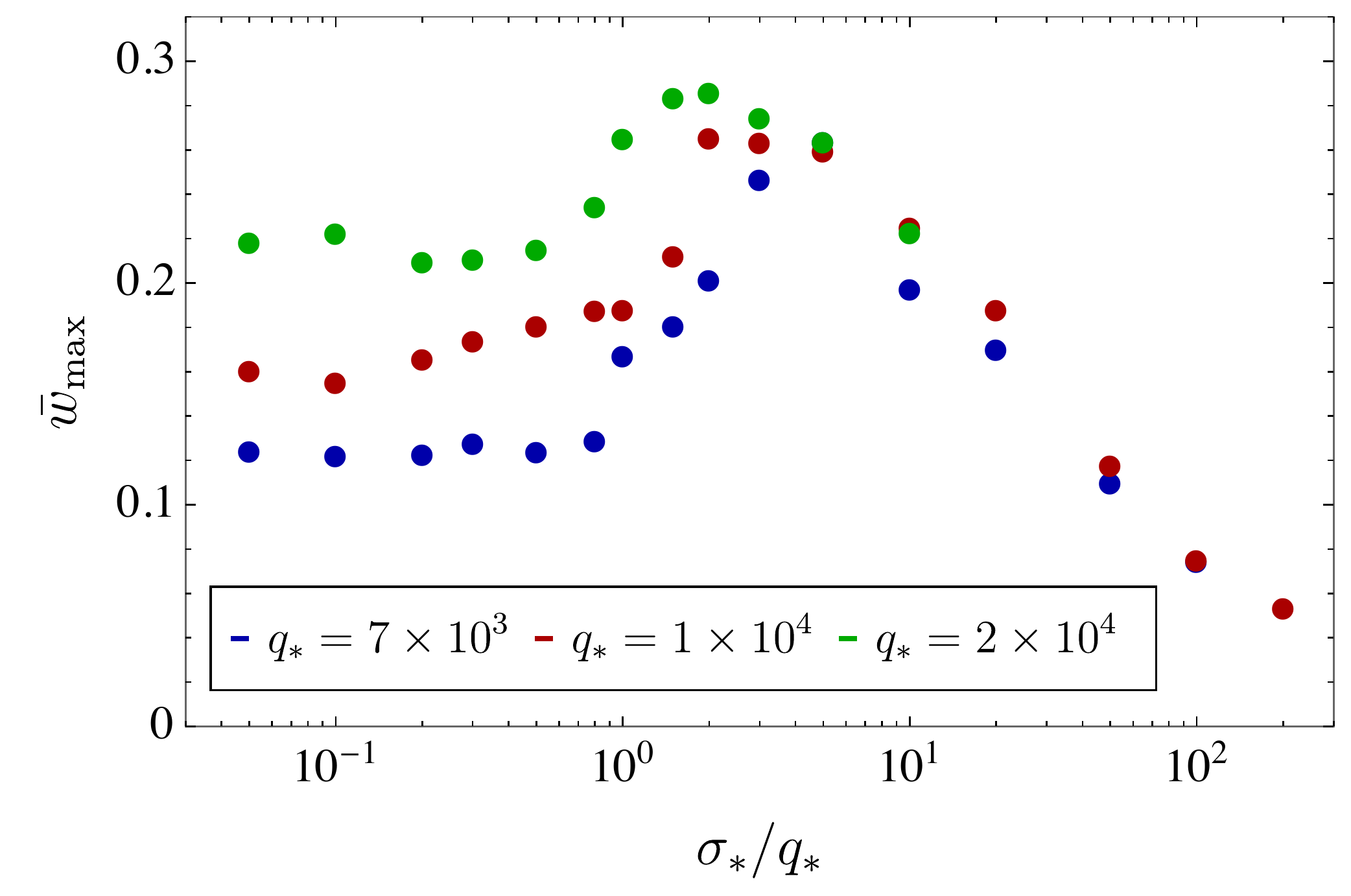}
    \caption{[Potential (\ref{eq:Pot3}), $p=2$] Top panels: Evolution of energy ratios for $q_*=10^4$ and two different self-interaction strengths: $\sigma_{*} /q_*=0.1$ (top-left) and $\sigma_{*} / q_* =2\cdot10^2$ (top-right). The vertical lines indicate when $\tilde{q}=1$ (dash-dotted) and $\tilde{\sigma}=1$ (dashed). Bottom-left panel: Maximum value attained by the energy ratios $\bar \varepsilon_{\rm g}^\varphi $, $\bar \varepsilon_{\rm g}^\chi $, $ \bar \varepsilon_{\rm k}^\chi $, $ \bar \varepsilon_{\rm p}^{\chi^4}$ and $ \bar \varepsilon_{\rm p}^{\varphi^2\chi^2}$, extracted from simulations with $q_*=10^{4}$ and different values of $\sigma_{*}$. Bottom-right: Maximal value attained by the effective equation of state  as a function of the ratio $\sigma_{*}/q_*$, for $p=2$ and three different choices of $q_*$.} 
    \label{fig:p2eos-selfint}
\end{figure*}

\textbf{a)} $\boldsymbol{p=2}$: In Fig.~\ref{fig:p2eos-selfint} we show the evolution of the energy distribution for $q_*=10^4$ and two choices of $\sigma_*$: $\sigma_*/q_*=0.1 ({<}1)$ (top-left) and $\sigma_*/q_*=2\cdot10^2 ({>}1)$ (top-right). In the case $\sigma_*/q_*=0.1$, the effect of the self-interaction in the post-inflationary dynamics is negligible, so the energy distribution evolves in a very similar way as the $\sigma_* = 0$ case discussed in Sect.~\ref{Sec:Nd_1_lambda_0}. More specifically, the daughter field gets excited through broad parametric resonance, which triggers the decay of the inflaton homogeneous mode through backreaction effects at the time $u \sim 10^2$. On the other hand, for $\sigma_* /q_* = 2 \cdot 10^2$ we observe that, although the energy of the daughter field grows initially through broad resonance as well, this growth is slowed down due to the effective mass $(m_\chi^{\rm eff})^2 \sim\tilde{\sigma}\langle \chi^2\rangle$ and never reaches a relevant magnitude before the resonance parameter falls below $\tilde{q} = 1$ (indicated by the vertical dash-dotted line).

\begin{figure}
    \centering
    \includegraphics[width=7.5cm]{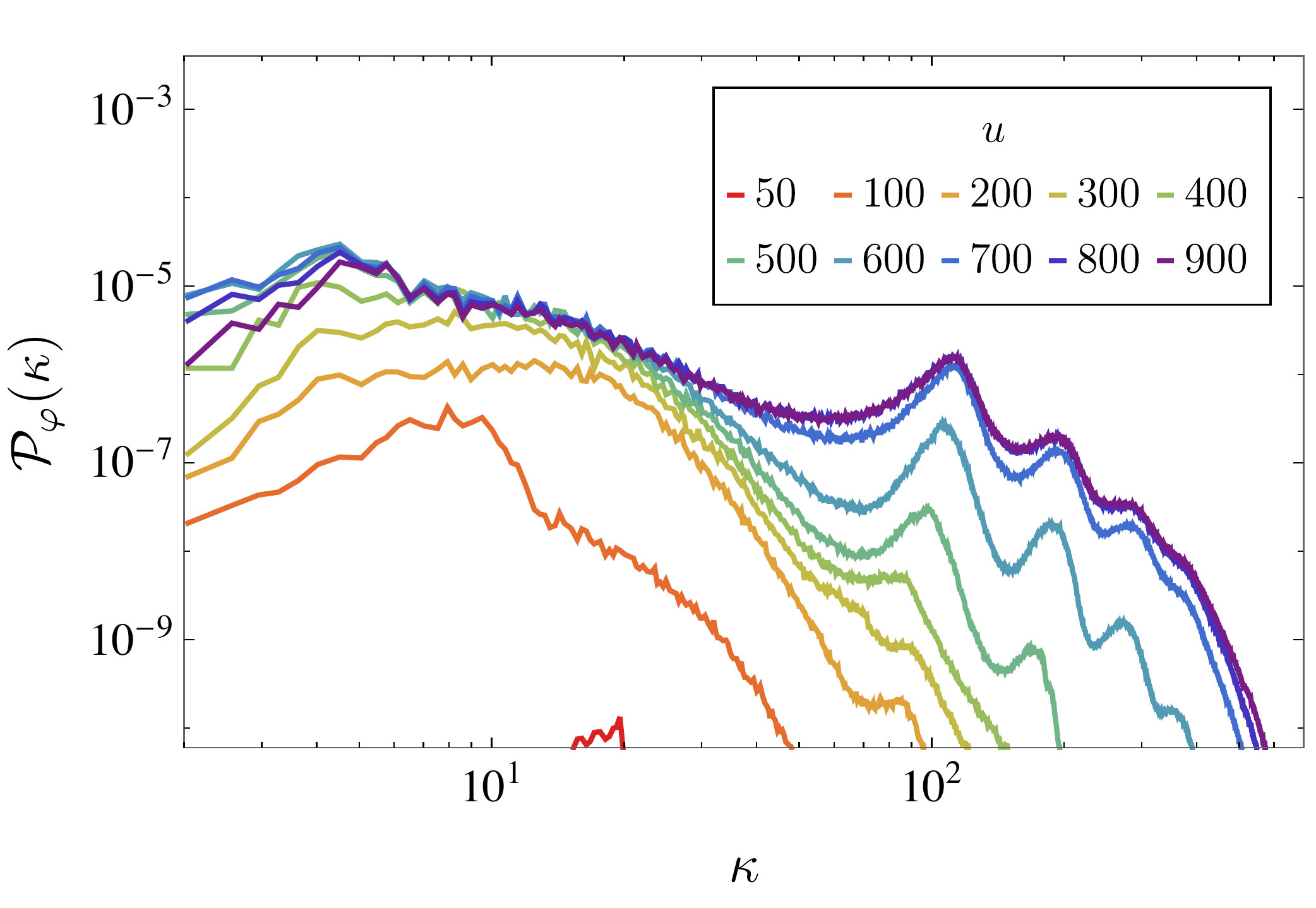}
    \includegraphics[width=7.5cm]{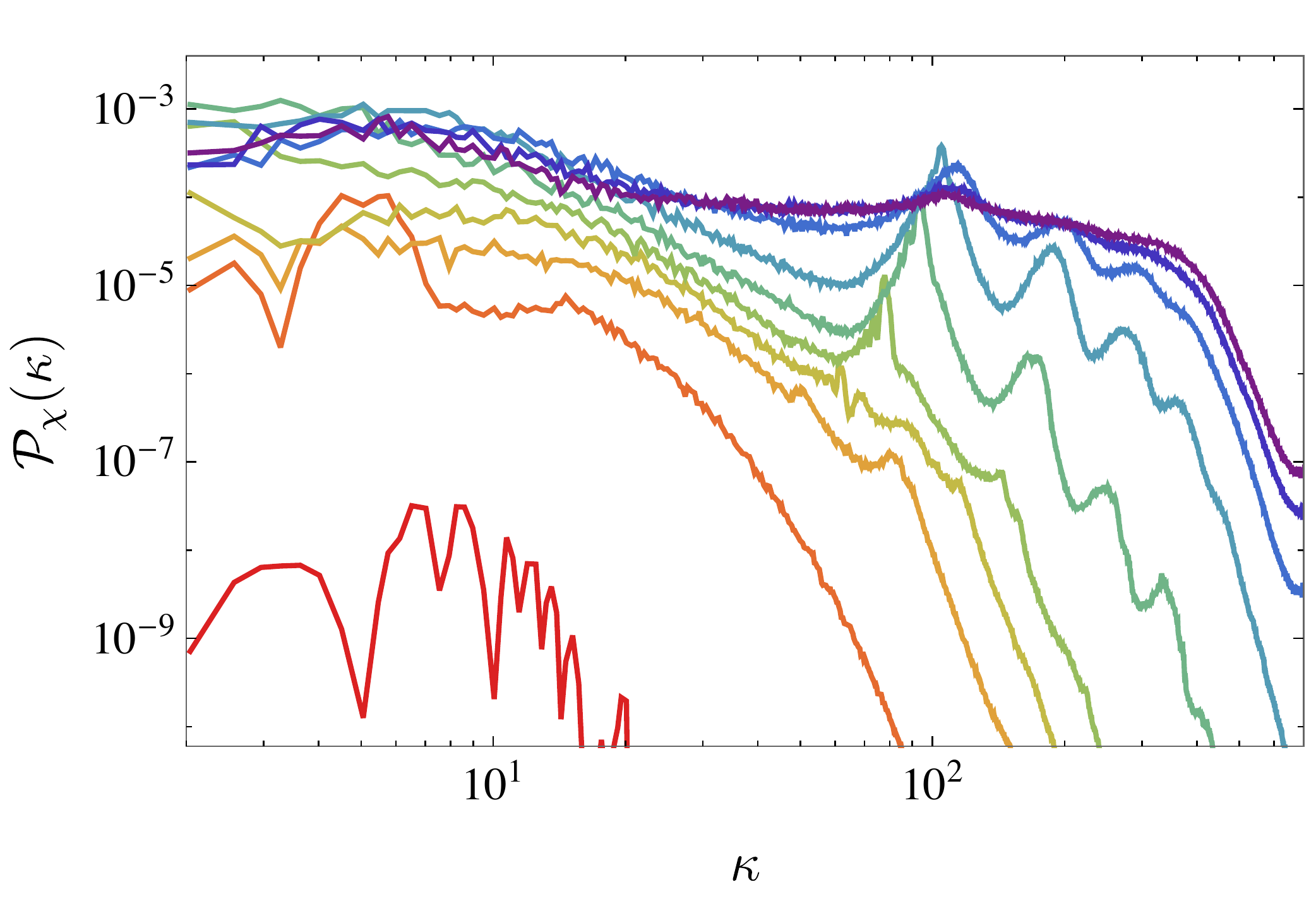}
    \caption{[Potential (\ref{eq:Pot3}), $p=2$] Evolution of the inflaton (left) and daughter field (right) spectra for $q_*=10^4$ and  $\sigma_{*}=2\cdot 10^6$, for times $u \sim 0- 10^3$.} \vspace{0.4cm}
    \label{fig:p2specs-selfint}
    \includegraphics[width=7.5cm]{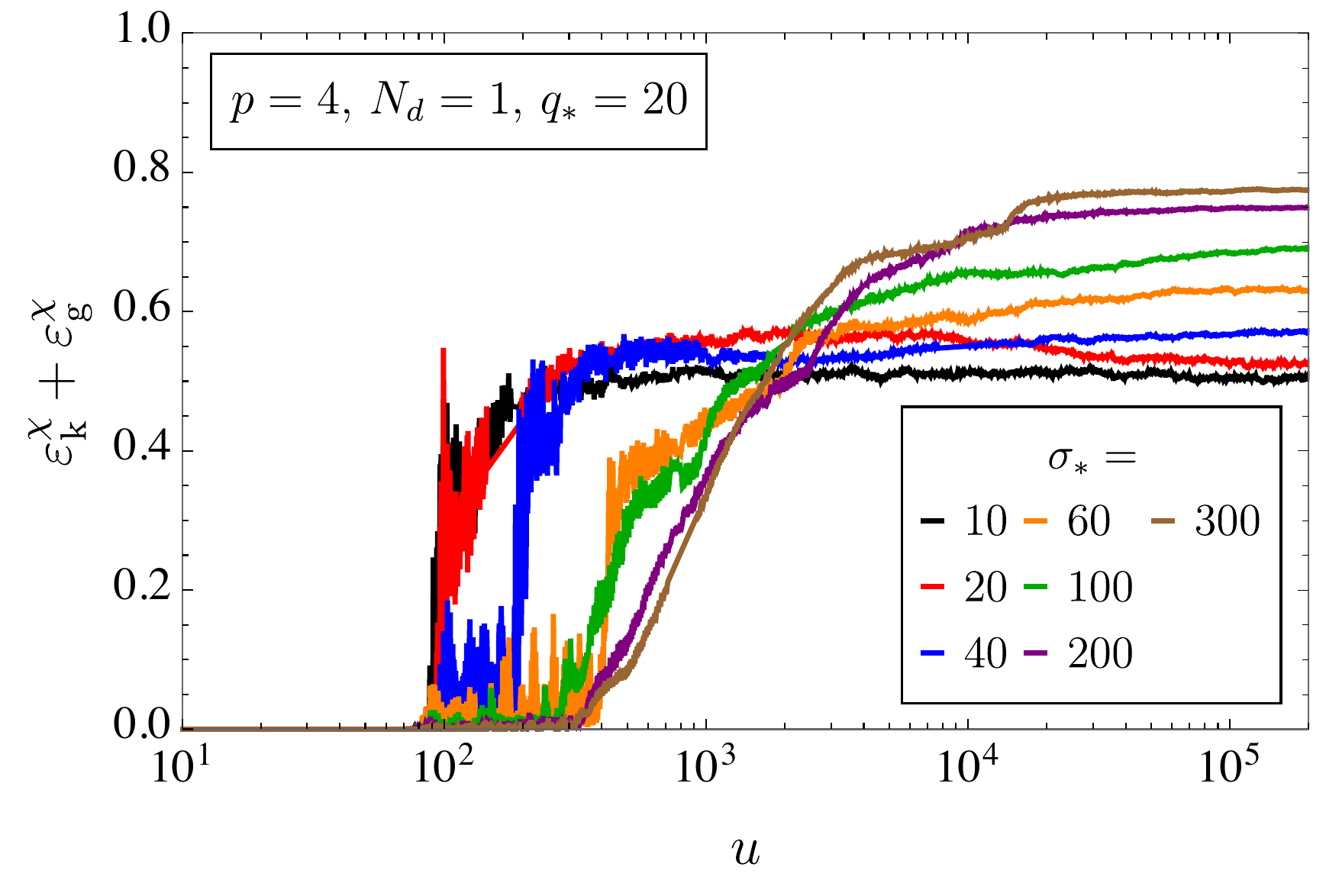}
    \includegraphics[width=7.5cm]{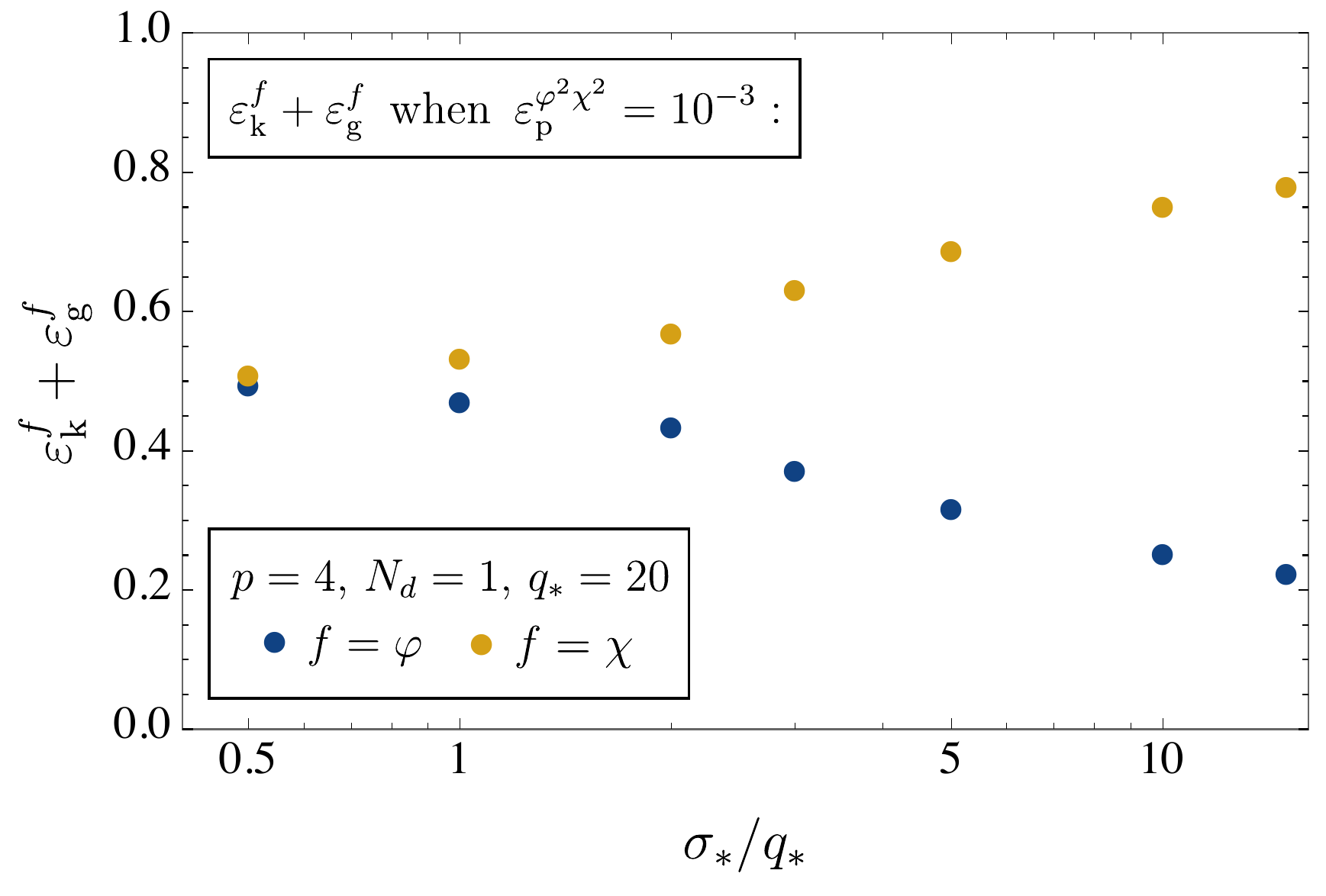}
    \caption{[Potential (\ref{eq:Pot3}), $p=4$] Left: Evolution of the fraction of energy stored in the daughter field for $q_* = 20$ and different choices of $\sigma_*$. More specifically, we depict the sum $ \varepsilon_k^{\chi} + \varepsilon_g^{\chi}$. Right: Fraction of energy stored by the daughter field at very late times, for different choices of $\sigma_* / q_*$ (more specifically, when $\bar\varepsilon_{\rm p}^{\varphi^2 \chi^2} \simeq 10^{-3}$).} 
    \label{fig:p4enrgies-selfint}
\end{figure}

The energy distribution also evolves in different ways during the non-linear regime. In particular, in the case $\sigma_*/q_*=2\cdot10^2$, the daughter field develops a relevant homogeneous mode during the early stage of broad resonance due to its quartic potential. This triggers a late growth of the daughter field fluctuations through a process of self-resonance, analogous to the one experienced by the inflaton for $p > 2$. This leads to an exponential growth of $\bar{\varepsilon}_{\rm g}^{\chi}$ at times $u \sim 4 \cdot 10^2 - 10^3$, which can be observed in the top-right panel of Fig.~\ref{fig:p2eos-selfint}. This effect can also be clearly observed in the spectral evolution of the fields, depicted in Fig.~\ref{fig:p2specs-selfint}. At these times, the spectrum of the daughter field shows a distinct structure of narrow peaks, reminiscent of the narrow bands that appeared in the Floquet diagram of inflaton self-resonance (see left panel of Fig.~\ref{fig:stabilitychart}). These peaks also get imprinted on the inflaton due to their interaction to the daughter field. Remarkably, backreaction effects lead to a wash out of these peaks in the daughter field spectrum, while they stay imprinted in the inflaton spectrum due to the lack of inflaton self-interactions. 

The effective self-coupling parameter $\tilde{\sigma} (a) \equiv  \sigma_* a^{\frac{6 (p-4)}{p+2}} =   \sigma_* a^{-3}$ decreases with time, so the process of self-resonance terminates when $\tilde{\sigma}\lesssim 1$. For $\sigma_*/q_*=2\cdot10^2$ this happens when $u \sim 1.5\cdot10^3$ (gray dashed line), when $\bar{\varepsilon}_{\rm g}^\chi$ reaches its maximum. The inflaton homogeneous mode eventually recovers all the energy of the system, and we get $\bar \varepsilon_{\rm k}^\varphi  \simeq \bar \varepsilon_{\rm p}^{\varphi^2} \rightarrow 0.5$ at late times.

Although the inflaton homogeneous mode recovers 100\% of the energy at late times independently of the strength of the self-interaction, it can strongly affect the field dynamics at intermediate times. In order to illustrate this, in the bottom-left panel of Fig.~\ref{fig:p2eos-selfint} we have depicted the maximum value attained by the energy ratios during the simulation, for different ratios $\sigma_* / q_*$.  The (temporary) energy transfer to the daughter field gets maximized for intermediate ratios $\sigma_*/q_*\sim 1-50$, while it gets strongly suppressed for $\sigma_* / q_* \gg 1$ due to the suppression of the resonance effects by the effective mass. Similarly, the self-interaction affects the evolution of the equation of state. Its qualitative evolution is similar to the $\sigma_* = 0$ case discussed in Sect.~\ref{Sec:Nd_1_lambda_0}: the production of field fluctuations triggers a transitory deviation from $\bar w = 0$ to $\bar w = \bar w_{\rm max} < 1/3$, which then relaxes back to $\bar w=0$ at late times. We illustrate this in the bottom-right panel of Fig.~\ref{fig:p2eos-selfint}, which shows $\bar w_{\rm max}$ for different choices of $\sigma_* /q_*$. The maximum deviation towards radiation-domination takes place again for intermediate values  $\sigma_* / q_* \sim 1 - 50$, while the deviation is minimal for very large ratios $\sigma_* / q_* \gg 1$. \vspace{0.2cm}

\textbf{b)} $\boldsymbol{2<p <4}$: The evolution of the daughter field energies is similar to the $p=2$ case. However, the inflaton now fragments due to self-resonance, so we have $\bar{\varepsilon}_{\rm k}^\varphi\simeq\bar{\varepsilon}_{\rm g}^\varphi \rightarrow 0.5$  at late times (and consequently $\bar{\varepsilon}_{\rm k}^\chi\simeq \bar{\varepsilon}_{\rm g}^\chi \rightarrow 0$). The system will eventually arrive at a radiation dominated state as well. However, for large enough $\sigma_*$ the transition phase can span over several e-folds, while it happens rather fast in the case of negligible self-interaction. \vspace{0.2cm}

\textbf{c)} $\boldsymbol{p\geq 4}$: In the left panel of Fig.~\ref{fig:p4enrgies-selfint} we show the fraction of energy stored in the daughter field, for $p=4$, $q_* = 20$, and different choices of $\sigma_*$. For $\sigma_* / q_* \lesssim 1$, the effect of the daughter field's quartic self-interaction in the post-inflationary dynamics is negligible, so the energy distribution evolves in a similar way as in the $\sigma_* = 0$ case discussed in Sect.~\ref{Sec:Nd_1_lambda_0}. In particular, at late times the energy is equally distributed between the inflaton and the daughter field, according to Eq.~(\ref{eq:EnergyRatios-Ndg1}).

On the other hand, for $\sigma_* / q_* \gtrsim 1$ the daughter field fluctuations experience a self-resonance process during the late-time regime, as described above for the $p=2$ case. Due to this, the energy from the inflaton is transferred much more efficiently and the fraction of energy stored in the daughter field at late times can be larger than 50\%. In fact, as seen in the right panel of Fig.~\ref{fig:p4enrgies-selfint}, the larger the ratio $\sigma_* / q_*$, the larger the amount of transferred energy. We have observed the same behaviour for lattice simulations of the $p=4.5$ case, and we expect it to happen for all values of $p>4$. These results show that even in the case of one daughter field,  a significant amount of energy density can be extracted from the inflaton for power-law coefficients $p\geq 4$, as long as a quartic self-interaction with $\lambda \gg g^2$ is present in the theory.

For $p>4$ the late excitation of fluctuations for large ratios $\sigma_* / q_*$, similar to the case seen in Fig.~\ref{fig:p4enrgies-selfint}, has a relevant influence on the equation of state. While for $\sigma_* \simeq 0$ the transition from the initial homogeneous to radiation dominated averaged equation of state happens rather fast, it appears for $\sigma_* \gg q_*$ as a smooth transition spanning over several e-folds.

\subsection{Multiple daughter fields with (self-)interactions: \texorpdfstring{$N_d \geq1$}{Nd >1}, \texorpdfstring{$\lambda_{nm}\geq 0$}{lmn>0} } \label{Sec:Ndg1-lambdag0} 

Finally, let us discuss the case of multiple daughter fields with interactions of the type $\lambda_{nm} X_n^2 X_m^2$, represented by potential (\ref{eq:Pot4}). Due to the large number of free parameters, a detailed parametric analysis of the post-inflationary dynamics is not possible in this case. However, we will consider three particularizations of potential (\ref{eq:Pot4}), that allow us to learn about the generic features of the post-inflationary energy distribution in this model. We fix the number of daughter fields to two ($N_d = 2$) for simplicity, but our results can be easily generalized to $N_d > 2$.  \vspace{0.2cm}

\textbf{a)} $\boldsymbol{\lambda_{11}> 0, \lambda_{22} > 0, \lambda_{12} = 0}$: We first consider a direct combination of the scenarios discussed in Sects.~\ref{Sec:MultipleDaughterFields} and \ref{Sec:SelfInteract}: an inflaton coupled to two daughter fields with quartic self-interactions,
\be 
V (\phi, X_1, X_2) = V_{\rm t}(\phi) + \frac{1}{2}g_1^2 \phi^2 X_1^2 + \frac{1}{2}g_2^2 \phi^2 X_2^2  + \frac{1}{4} \lambda_1 X_1^4 + \frac{1}{4} \lambda_2 X_2^4 \ .  \label{eq:Pot4-typeA}
\ee
If $\sigma_*^{(1)} / q_*^{(1)} \lesssim 1$ and $\sigma_*^{(2)} / q_*^{(2)} \lesssim 1$, the role of both self-interactions is negligible and we recover the results of Sect.~\ref{Sec:MultipleDaughterFields}: the energy is equally distributed between the three fields at very late times if $p \geq 4$, while the inflaton eventually recovers 100\% of the energy for $2\leq p < 4$.

On the other hand, if $\sigma_*^{(i)} / q_*^{(i)} \gtrsim 1$ for any of the daughter fields $i=1,2$, the energy transferred to that field during the non-linear regime gets enhanced due to the self-resonance induced by the quartic self-interaction (the same effect discussed in Sect.~\ref{Sec:SelfInteract}). If $2\leq p < 4$, the inflaton will still recover 100\% of the energy at very late times, as the effective parameter $\tilde{\sigma}^{(nm)} (a) \equiv  \sigma_*^{(nm)} a^{\frac{6 (p-4)}{p+2}}$ decreases with time. More interesting is the $p \geq 4$ case: in this case the three fields no longer equilibrate with the same energy, and in particular the daughter fields with larger ratio $\sigma_*^{(i)} / q_*^{(i)}$ will get a larger percentage of the energy at late times. The amount of energy that remains in the inflaton depends strongly on the particular ratios, but it is always less than 33\% because more energy gets extracted from it.

\begin{figure*} \centering
    \includegraphics[width=0.48\textwidth]{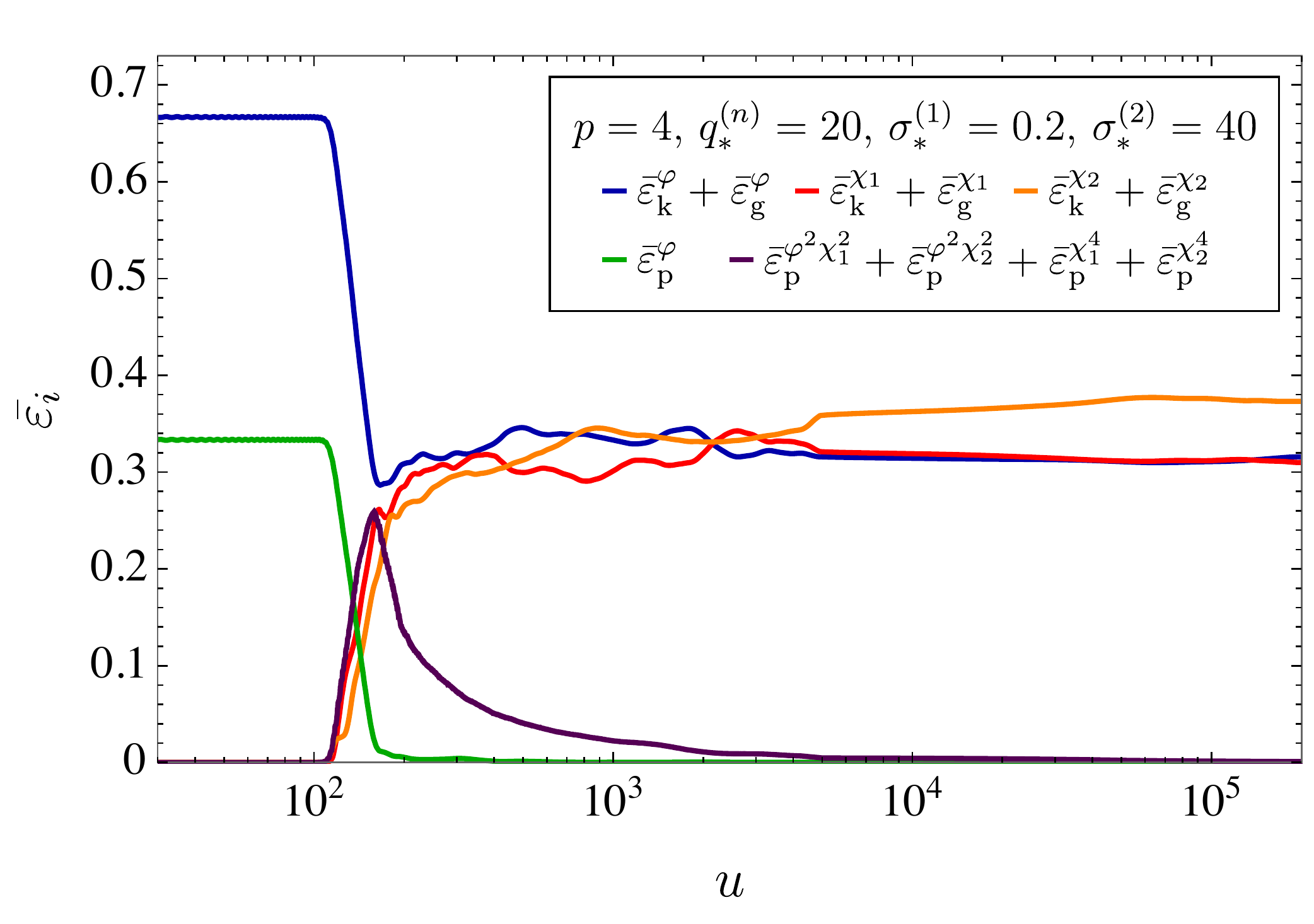} \hspace{0.2cm}
    \includegraphics[width=0.48\textwidth]{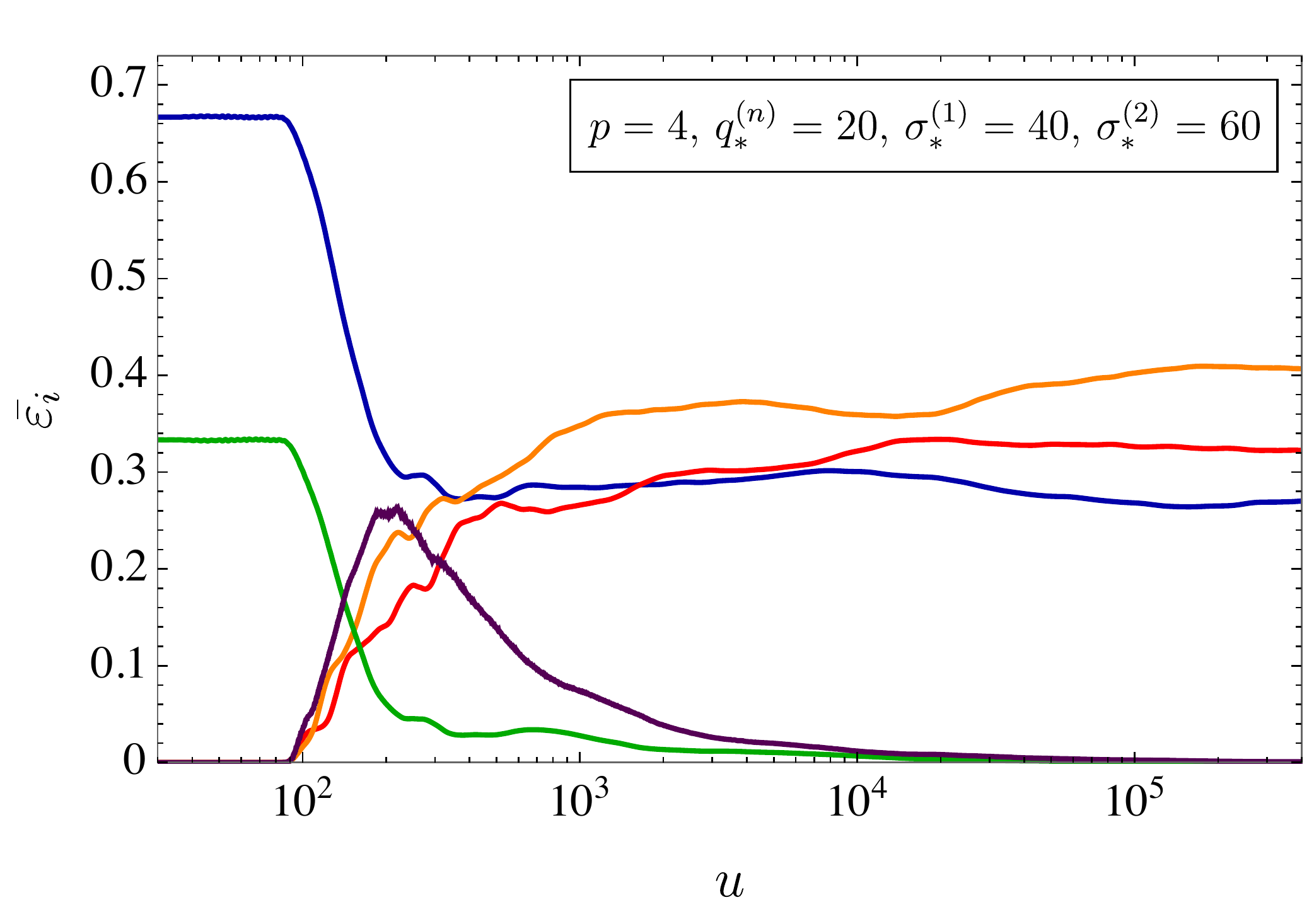}   \\ 
    \caption{[Potential (\ref{eq:Pot4-typeA}), $p=4$] Evolution of the energy ratios for $q_*^{(1)} = q_*^{(2)} = 20$, and different $\sigma_*^{(n)}$ Left: A case with $\sigma_*^{(1)}/q_*^{(1)}=0.01\, (< 1)$ and $\sigma_*^{(2)}/q_*^{(2)}=2\, (>1)$. Right: A case with $\sigma_*^{(1)}/q_*^{(1)}=2\, (>1)$ and  $\sigma_*^{(2)}/q_*^{(2)}=3\, (>1)$.}
    \label{fig:p4Ndsi}
\end{figure*}

We present two illustrative examples for $p=4$ and $N_d = 2$ in Fig.~\ref{fig:p4Ndsi}. The left panel shows a case in which $\sigma_*^{(1)}/q_*^{(1)}=0.01\, (< 1)$ and $\sigma_*^{(2)}/q_*^{(2)}=2\, (>1)$, so only $X_2$ experiences significant self-resonance during the non-linear regime. We  observe that the field $X_2$ gets more than $33\%$ of the energy at very late times (approximately $38\%$), while both $\phi$ and $X_1$ equilibrate with less than that (each one gets $31\%$ of the energy). On the other hand, the right panel shows a case in which $\sigma_*^{(1)}/q_*^{(1)}=2\, (>1)$ and  $\sigma_*^{(2)}/q_*^{(2)}=3\, (>1)$, so both $X_1$ and $X_2$ experience self-resonance: in this case both daughter fields get more energy than $\phi$ at late times, and in fact $X_2$ gets more energy than $X_1$ due to its larger coupling ratio.\vspace{0.2cm}

\textbf{b)} $\boldsymbol{\lambda_{11} = \lambda_{22} = 0, \lambda_{12} > 0}$: Let us now consider a scenario in which both daughter fields are coupled to each other through a quadratic-quadratic interaction, but neither of them have quartic self-interactions. The potential reads as
\be 
V(\phi, X_1, X_2) = V_{\rm t}(\phi) + \frac{1}{2}g_1^2 \phi^2 X_1^2 + \frac{1}{2}g_2^2 \phi^2 X_2^2  + \frac{1}{2} \lambda_{12} X_1^2 X_2^2 \ .  \label{eq:Pot4-typeB}
\ee

\begin{figure}
    \centering
    \includegraphics[width=7.5cm]{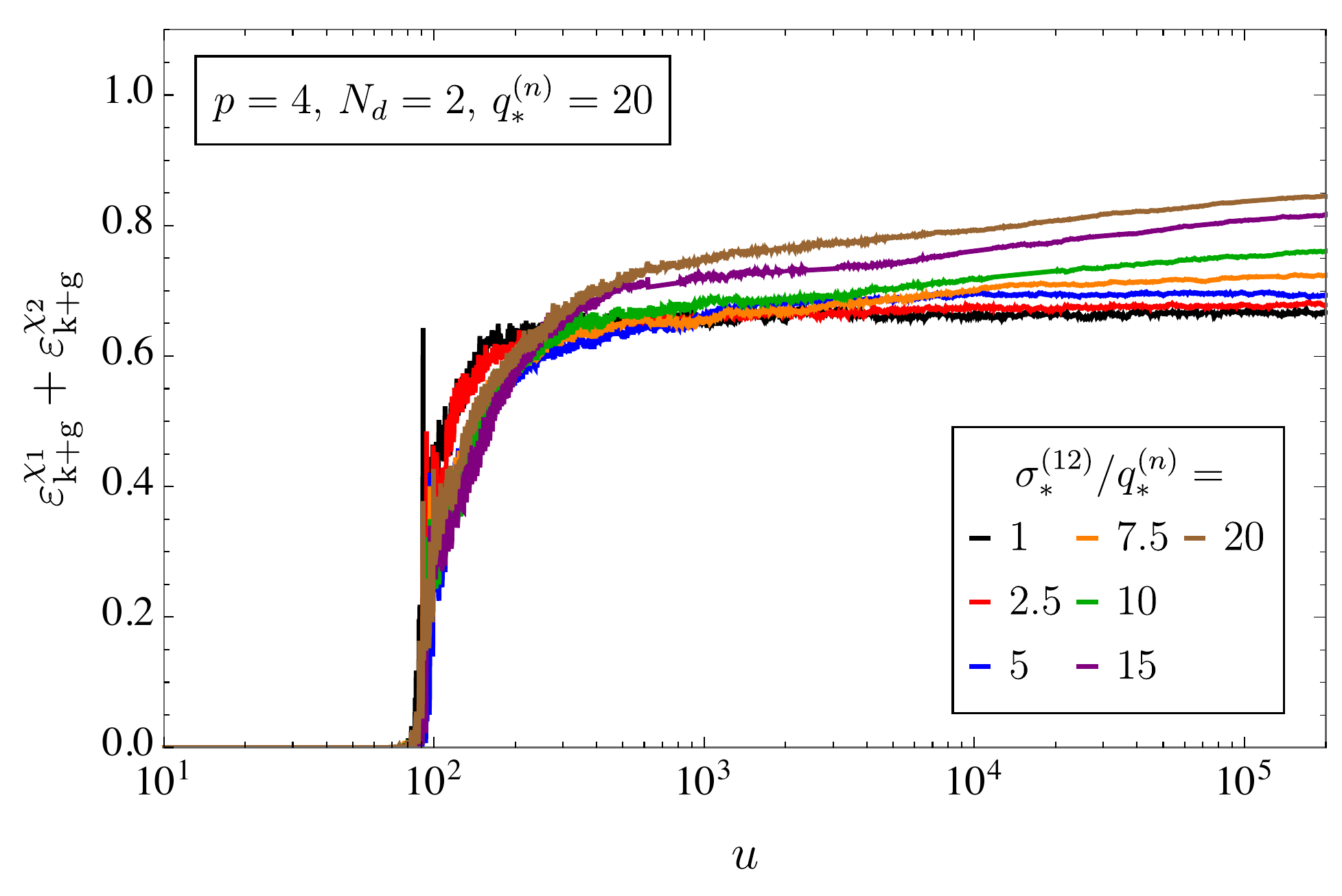}
    \includegraphics[width=7.5cm]{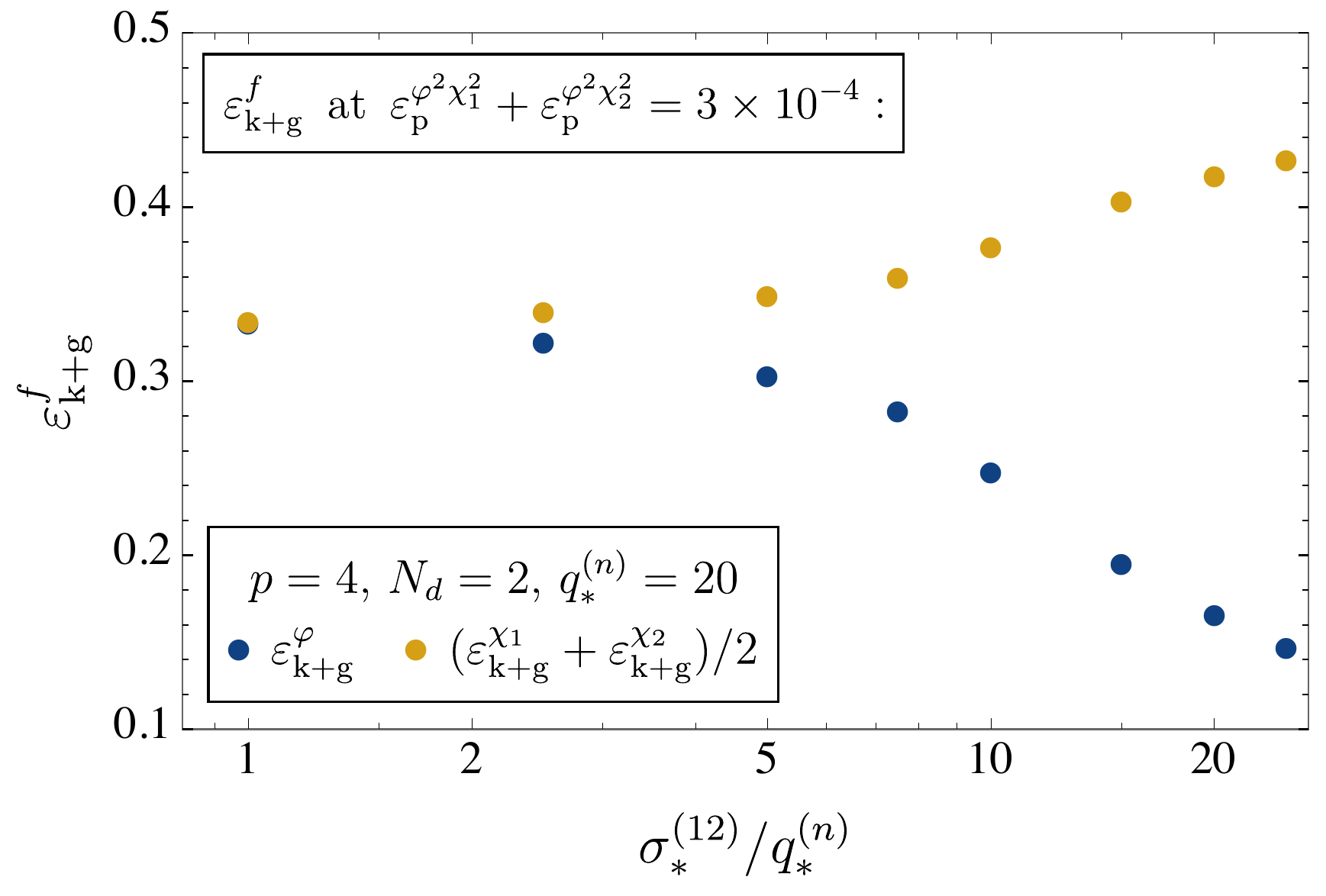}
    \caption{[Potential (\ref{eq:Pot4-typeB}), $p=4$] Left: Evolution of the fraction of energy stored in the daughter fields (i.e.~the sum $\varepsilon_{{\rm k}+{\rm g}}^{\chi_1}+\varepsilon_{{\rm k}+{\rm g}}^{\chi_2}=\varepsilon_{\rm k}^{\chi_1} +  \varepsilon_{\rm g}^{\chi_1}  +  \varepsilon_{\rm k}^{\chi_2}  + \varepsilon_{\rm g}^{\chi_2}$), for $q_* = 20$ and different ratios $\sigma_*^{(12)} / q_*$. Right: Energy stored in the inflaton and daughter fields for different choices of $\sigma_*^{(12)} / q_*$ at late times, when $\varepsilon_{\rm p}^{\varphi^2 \chi_1^2} + \varepsilon_{\rm p}^{\varphi^2 \chi_2^2} \simeq 3\cdot 10^{-4}$.} \vspace{0.3cm} \label{fig:p4enrgies-int}
       
    \includegraphics[width=7.5cm]{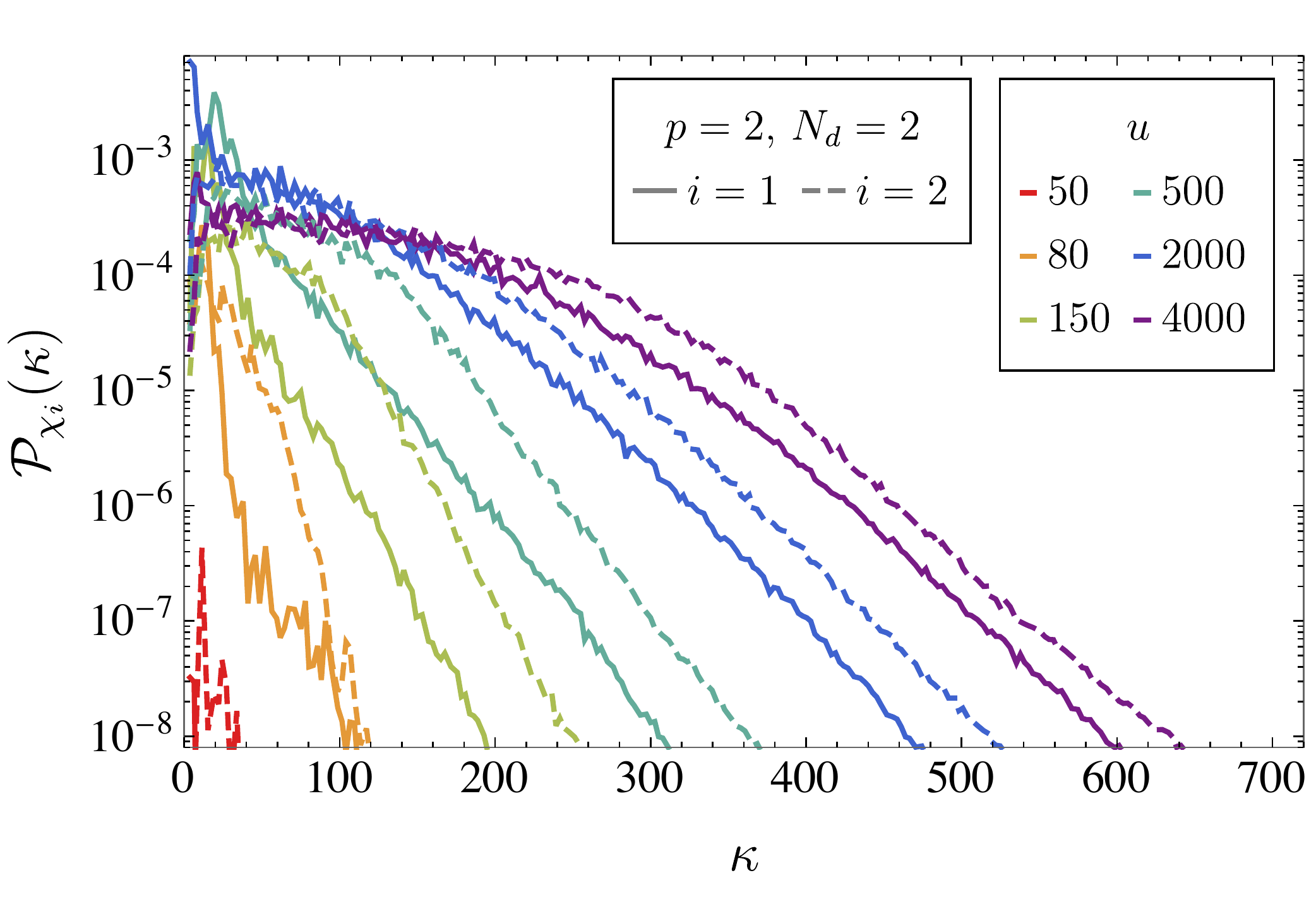}
    \includegraphics[width=7.5cm]{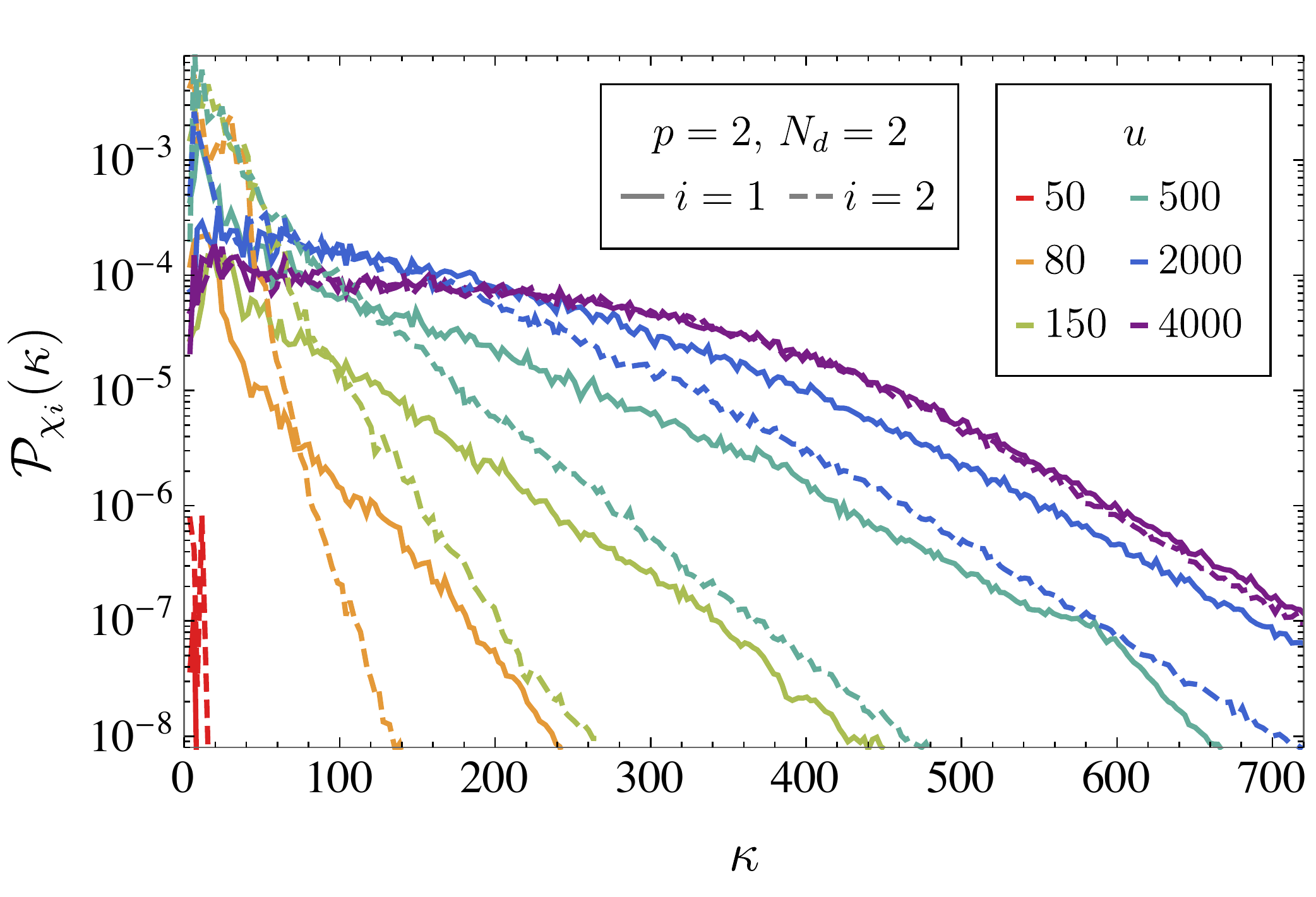}
    \caption{[Potential (\ref{eq:Pot4-typeB}), $p=2$] Comparison of the spectra of fields $\chi_1$ (solid) and $\chi_2$ (dashed) at different times. Left: $q_*^{(1)}=7\cdot10^3$, $q_*^{(2)}=3.5\cdot10^5$ and $\sigma_*^{(12)}=1.4\cdot10^5$. The fields do not equilibrate in momentum space at late times due to the freeze out of the couplings. Right: $q_*^{(1)}=7\cdot10^3$, $q_*^{(2)}=3.5\cdot10^4$ and $\sigma_*^{(12)}=7\cdot10^5$. At late times the fields equilibrate due to the large value of $\sigma_*^{(12)}$ compared to $q_*^{(1),(2)}$, which allows an efficient transfer of energy between each other.} 
    \label{fig:p2specs-int}
\end{figure}

During the linear regime, the strength of the resonance and the momenta excited are characterized, for each of the two fields $i=1,$ 2, by the corresponding resonance parameter $q_{*}^{(i)} \equiv g_i^2 \phi_*^2 / \omega_*^2$.  However, during the later non-linear regime, the quadratic-quadratic interaction $ \lambda_{12} X_1^2 X_2^2$ triggers a resonant process for \textit{both} fields $X_1$ and $X_2$, in a similar way as the quartic term $\lambda_i X_i^4$ did for the field $i$ in the previous example. This allows for an efficient distribution of energy in the daughter field sector.

In order to illustrate this, let us consider first the case $p=4$. Remarkably, in the simulations we have seen that both daughter fields equilibrate very quickly during the non-linear regime (much faster than without such an interaction), and get the same fraction of energy at late times, i.e.~$\bar \varepsilon_{\rm k}^{\chi_1} +  \bar \varepsilon_{\rm g}^{\chi_1} \approx \bar \varepsilon_{\rm k}^{\chi_2} +  \bar \varepsilon_{\rm g}^{\chi_2}$. This happens also when $q_*^{(1)} \neq q_*^{(2)}$, as well as when $q_*^{(1)}, q_*^{(2)}\gg \sigma_*^{(12)}$. Due to this, in the left panel of Fig.~\ref{fig:p4enrgies-int} we have depicted the (at late times) \textit{total} fraction of energy attained by the daughter field sector as a function of time (i.e.~the sum $\bar \varepsilon_{\rm k}^{\chi_1} +  \bar \varepsilon_{\rm g}^{\chi_1}  + \bar \varepsilon_{\rm k}^{\chi_2}  + \bar \varepsilon_{\rm g}^{\chi_2}$), for $q_*^{(1)}=q_*^{(2)}=20$ and different choices of $\sigma_*^{(12)} / q_*^{(1),(2)}$. We observe that for $\sigma_*^{(12)} / q_* \lesssim 1$, the effect of the $\lambda_{12} X_1^2 X_2^2$ interaction is negligible, so the daughter field sector gets 66\% of the energy at late times, in agreement with Eq.~(\ref{eq:EnergyRatios-Ndg1}). However, as  $\sigma_*^{(12)} / q_*$ increases, the transfer of energy to the daughter field sector gets larger. For very large ratios it is difficult to simulate the system long enough to observe the achievement of the stationary regime. Thus, we have depicted in the right panel of Fig.~\ref{fig:p4enrgies-int} the summed energies $\varepsilon_{{\rm k}+{\rm g}}^{\varphi} \equiv \varepsilon_{\rm k}^{\varphi} +  \varepsilon_{\rm g}^{\varphi}$ and $\varepsilon_{{\rm k}+{\rm g}}^{\chi_1}+\varepsilon_{{\rm k}+{\rm g}}^{\chi_2}\equiv(\varepsilon_{\rm k}^{\chi_1} +   \varepsilon_{\rm g}^{\chi_1}+ \varepsilon_{\rm k}^{\chi_2} +  \varepsilon_{\rm g}^{\chi_2})/2$ when $\varepsilon_{\rm p}^{\varphi^2 \chi_1^2} + \varepsilon_{\rm p}^{\varphi^2 \chi_2^2} \simeq 3\cdot 10^{-4}$.  A similar behavior has been observed in lattice simulations of scenarios with $p>4$.

For $p<4$, the inflaton recovers all the energy at late times as expected, while the energy stored in the daughter field sector becomes subdominant (for $p=2$ we get $\bar{\varepsilon}_{\rm k}^{\varphi}$, $\bar{\varepsilon}_{\rm p}^{\varphi^2} \rightarrow 1/2$, and for $2 < p < 4$ we get $\bar{\varepsilon}_{\rm k}^{\varphi}$, $\bar{\varepsilon}_{\rm g}^{\varphi} \rightarrow 1/2$). This happens because the effective self-coupling parameter decreases with time as $\tilde{\sigma}^{(12)} (a) \equiv  \sigma_*^{(12)} a^{\frac{6 (p-4)}{p+2}}$, so the self-resonance triggered by it becomes eventually too weak. However, an interesting effect  takes place in the daughter field sector for these power-law coefficients if $q_*\ll\sigma^{(12)}_*$: the daughter fields tend to equilibrate at late times, in contrast to the case of very small $\sigma^{(12)}_*$. In Fig.~\ref{fig:p2specs-int} we show the evolution of the spectra of both daughter fields for the cases: $q_*^{(1)}=7\cdot10^3$, $q_*^{(2)}=3.5\cdot10^5$ and $\sigma^{(12)}_*=1.4\cdot10^5$ (left panel),  and $q_*^{(1)}=7\cdot10^3$, $q_*^{(2)}=3.5\cdot10^4$ and $\sigma^{(12)}_*=7\cdot10^5$ (right panel). We can see that in the first case (for which $q_*^{(2)}>\sigma^{(12)}$) the two spectra do not equilibrate at late times (similar to the case discussed in Fig.~\ref{fig:p2specs-selfint}), while in the second case the two daughter fields equilibrate, i.e.\ both spectra match at late times.\vspace{0.2cm}

\begin{figure*} \centering
    \includegraphics[width=0.48\textwidth]{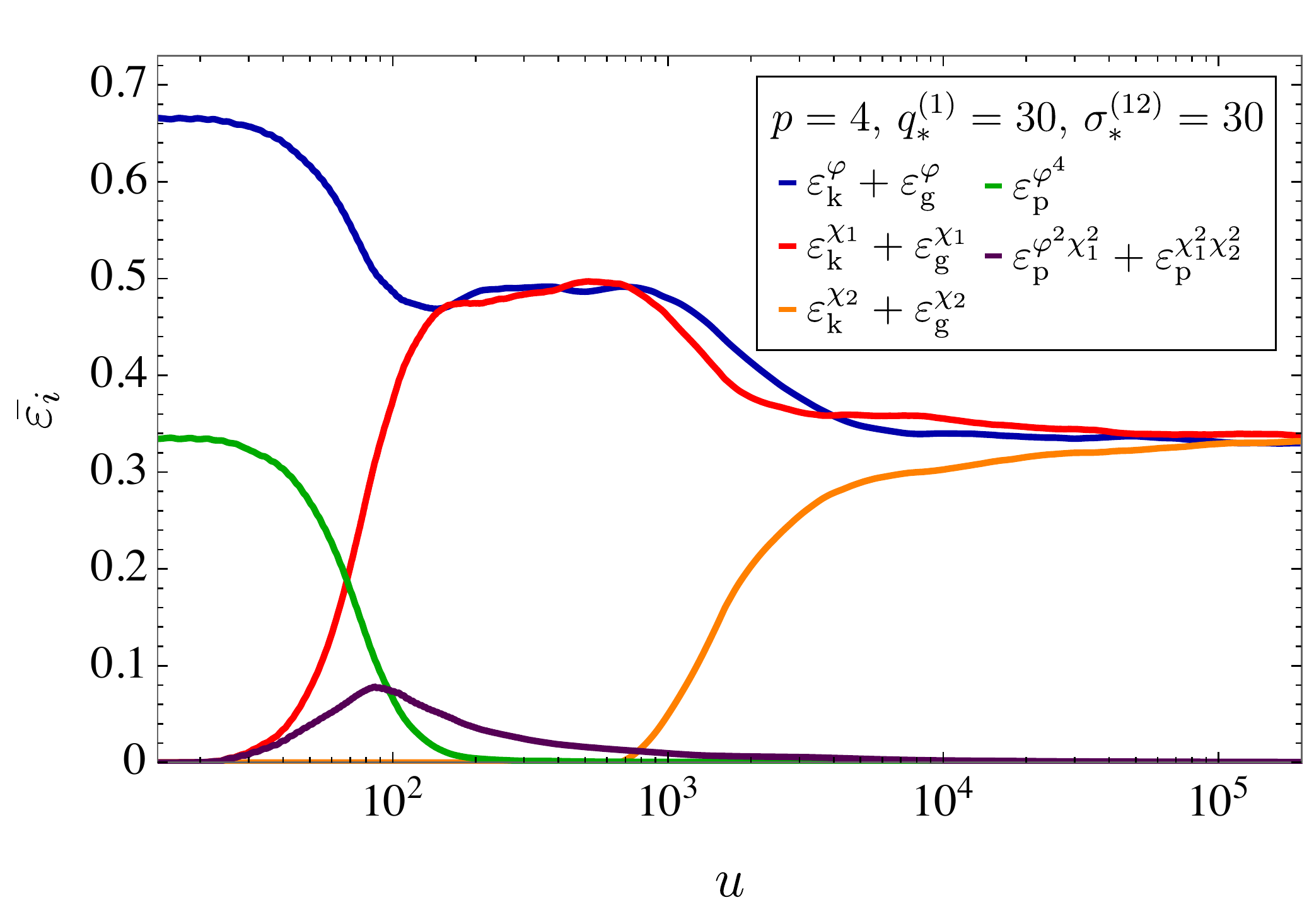} \hspace{0.2cm}
    \includegraphics[width=0.48\textwidth]{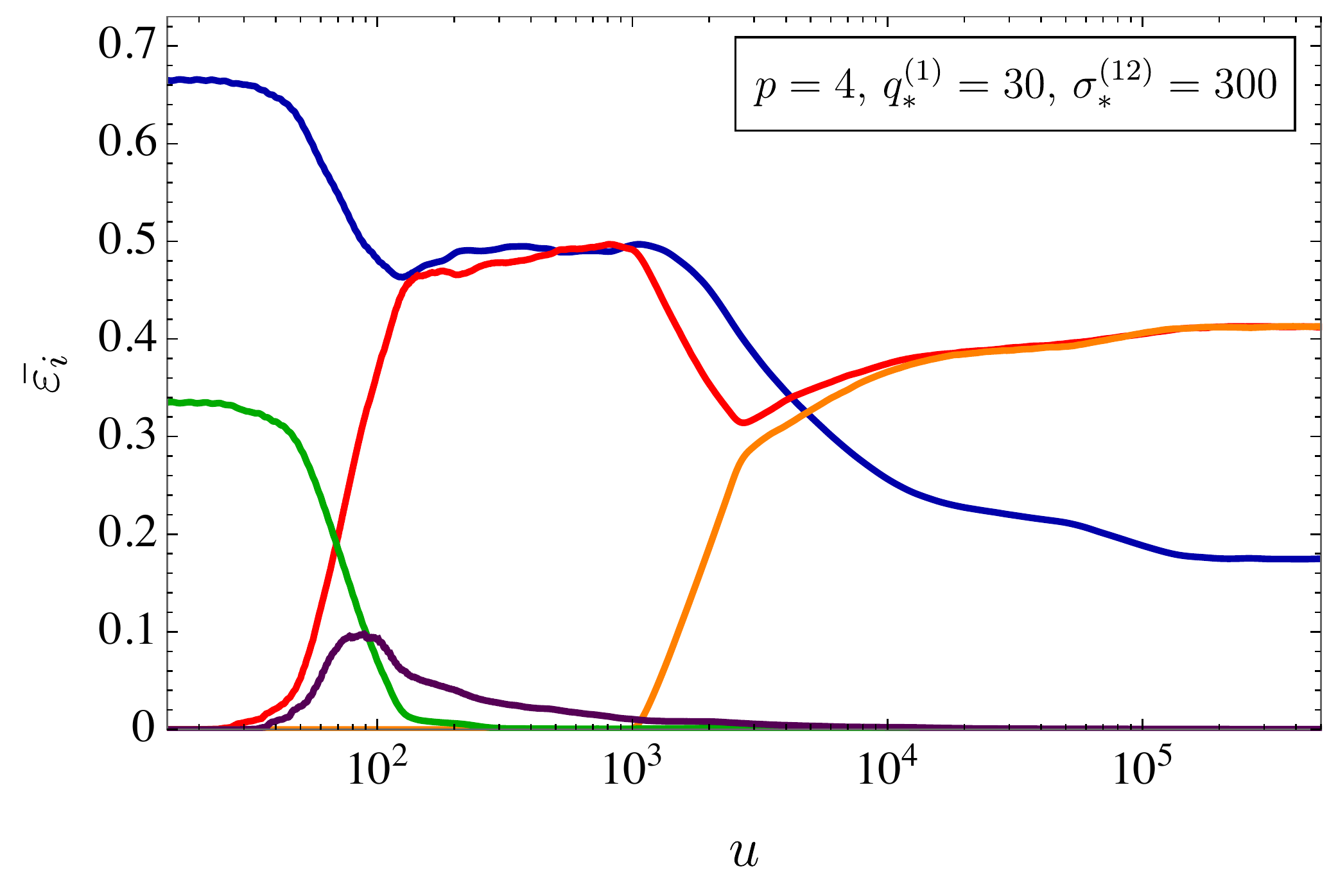}
    \hspace{0.2cm}
    \caption{[Potential (\ref{eq:PotChain}), $p=4$] Evolution of the energy ratios in a `chain' scenario with two daughter fields. We fix $q_*^{(1)} = 30$ and $q_*^{(2)} = 0$ in both panels, and consider the values $\sigma_*^{(12)} / q_*^{(1)} = 1$ (left panel) and $\sigma_*^{(12)} / q_*^{(1)} = 10$ (right panel).} 
    \label{fig:p4enrgies-chain}
\end{figure*}

\textbf{c)} $\boldsymbol{g_1^2 > 0, g_2^2=0, \lambda_{12} > 0, \lambda_{11} = \lambda_{22} = 0}$: Finally, let us consider a `chain' scenario in which only one daughter field ($X_1$) is coupled directly to the inflaton. The second daughter field ($X_2$) is coupled to $X_1$ through a quadratic-quadratic interaction, which allows a transfer of energy from the inflaton to $X_2$ in a two-step process. The potential reads as
\be  V (\phi, X_1, X_2) = V_{\rm t}(\phi) + \frac{1}{2}g_1^2 \phi^2 X_1^2 + \frac{1}{2} \lambda_{12} X_1^2 X_2^2 \ , \label{eq:PotChain} \ee
There are two unspecified parameters to fix: $q_*^{(1)}$ and $\sigma_*^{(12)}$ (or equivalently, $g_1^2$ and $\lambda_{12}$). In Fig.~\ref{fig:p4enrgies-chain} we depict two illustrative examples for $p=4$: we have fixed $q_*^{(1)} = 30$, and plotted the cases $\sigma_*^{(12)}/q_*^{(1)} = 1$ (left panel) and $\sigma_*^{(12)}/q_*^{(1)}=10$ (right panel). As expected, $X_1$ is excited during the initial linear stage through a process of broad parametric resonance in both cases. Later, we get a transitory equilibration phase during the early non-linear regime, in which the inflaton and $X_1$ each hold approximately 50\% of the total energy (this happens for time scales $u \sim 10^2 - 10^3$ in both cases). At these times, the energy of $X_2$ remains subdominant because it is not coupled directly to the inflaton, and hence it is not excited via parametric resonance. However, in a later stage $X_1$ starts to transfer energy into $X_2$ through the $\lambda_{12} X_1^2 X_2^2$ interaction. This is reminiscent of the case studied in Ref.~\cite{Antusch:2015vna}, where parametric resonance was induced by an inhomogeneous field. Eventually, we end in a situation in which $X_1$ and $X_2$ have equilibrated with the same energy, despite $X_2$ not being coupled directly to the inflaton. This happens for both ratios of $\sigma_*^{(12)} / q_*^{(1)}$ considered. However, the amount of energy depleted from the inflaton does indeed increase for larger ratios of $\sigma_*^{(12)} / q_*^{(1)}$. If $\sigma_*^{(12)} \lesssim q_*^{(1)}$ we find that each of the three fields get 33\% of the energy at late times, but if $\sigma_*^{(12)} > q_*^{(1)}$ much more energy is eventually transferred to the daughter field sector. In the depicted case of $\sigma_*^{(12)}/q_*^{(1)}=10$ only $\sim 18\%$ of the energy remains in the inflaton.

We observe the same qualitative behavior for values of $p > 4$. On the other hand, for $p<4$ the energy ratios of both daughter fields attain a maximum when $\tilde{q}^{(1)}\sim1$ and then start decreasing again. In the later case, the second daughter field usually stays subdominant (except for very large values of $q_{*}^{(1)}$ and $\sigma_*^{(12)}$), as there is not enough time to enhance it significantly before $\tilde{\sigma}_{12}$ falls below unity.

\section{Summary and discussion} \label{sec:Summary}

The aim of this work has been to study the post-inflationary dynamics of the universe when the inflaton is coupled to multiple interacting daughter fields. As a set-up we have considered an inflaton that oscillates after inflation in a potential that is monomial around the minimum $V(\phi) \propto |\phi|^p$, and is coupled to daughter fields through scale-free, quadratic-quadratic interactions $g_n^2 \phi^2 X_n^2$. If $q_*^{(n)} \equiv g^2_n \phi_{*}^2 / \omega_{*}^2 \gtrsim 1$, the daughter fields get excited during the linear regime through a process of broad parametric resonance. We have also included additional scale-free interactions of the type $\lambda_{nm} X_n^2 X_m^2$, including also quartic self-interactions when $n=m$. We have first studied the excitation of the field fluctuations during the linear regime by means of a Hartree approximation, see Sect.~\ref{Sec:Linear}. This analysis has allowed us to obtain a prediction for the maximum variance attained by the daughter field as a function of the particle couplings, see e.g.~Fig.~\ref{fig:hartree}. 

We have then simulated the post-inflationary dynamics of the system with (2+1)-dimensional lattice simulations, see Sect.~\ref{sec:LatSims}, which have allowed to capture the full dynamics from the end of inflation until the achievement of a stationary regime for a large set of model parameters. For clarity purposes, we have divided our results in four subsections \ref{Sec:Nd_1_lambda_0} - \ref{Sec:Ndg1-lambdag0}, which cover different particularizations of the most generic potential, see Eqs.~(\ref{eq:Pot1})-(\ref{eq:Pot4}). As a reference, we have used the results for one daughter field without quartic self-interaction. This case was considered extensively in our \textit{Letter} \cite{Antusch:2020iyq} and in \PartI, and the results have been summarized in Sect.~\ref{Sec:Nd_1_lambda_0}. There we observed that the amount of energy transferred to the daughter field after equilibration depends on $p$ in the following way: i) for $p=2$, 100\% of the energy ends up in the homogeneous mode of the inflaton, ii) for $2<p<4$, 100\% of the energy ends up in a fragmented inflaton, and iii) for $p \geq 4$, the inflaton and the daughter field end up fragmented with each field getting 50\% of the energy. In this paper, we have investigated how these results change when two or more daughter fields are included in the theory, as well as scale-free $\lambda_{nm} X_n^2 X_m^2$ and $\lambda_n X_n^4$ interactions:

\begin{itemize}
    \item In Sec.~\ref{Sec:MultipleDaughterFields} we have considered the case of multiple daughter fields without $X_m^2 X_n^2$ interactions. For $p \leq 4$, the results for the energy distribution at very late times remain unchanged with respect to the single daughter field case; however, a larger transfer of energy to the daughter field sector takes place at intermediate times. This has noticeable effects on the equation of state for the $p=2$ case, see Fig.~\ref{fig:En-Ndp2B}, and therefore on the predictions of the CMB observables $n_s$ and $r$. On the other hand, for $p \geq 4$ the energy ends up equally distributed between all fields at very late times, see Eq.~(\ref{eq:EnergyRatios-Ndg1}) and Fig.~\ref{fig:En-MultipleDaughter0}. This is true even if the daughter fields are coupled to the inflaton with different strengths, as long as they are excited in broad resonance, see Fig.~\ref{fig:En-MultipleDaughter}. Remarkably, this means that in systems with many daughter fields, a significant amount of energy can be depleted from the inflaton, without relying on perturbative decay channels or other elements to the theory.
    
    \item In Sec.~\ref{Sec:SelfInteract} we have explored how the results for one daughter field change when a self-interaction $\lambda X^4$ is included in the theory. Results are basically unchanged if $\lambda /g^2 \ll 1$. However, if  $\lambda /g^2 \gtrsim 1$, the self-interaction does significantly affect the evolution of the energy distribution after inflation. First, during the linear regime it appears as an effective mass to the daughter field fluctuations, which suppresses the excitation process. However, during the non-linear regime it triggers a self-resonance process, which enhances the amount of energy transferred to the daughter field. Due to this, for $p \geq 4$ the daughter field gets more than 50\% of the energy at late times, see Fig.~\ref{fig:p4enrgies-selfint}. In fact, the larger the ratio $\lambda /g^2$, the more energy is transferred. Correspondingly, the inflaton retains less than 50\% of the total energy density. For $p < 4$, the inflaton recovers 100\% of the energy at very late times as in the $\lambda = 0$ case, but a larger amount of energy gets temporally transferred at intermediate times during the early non-linear stage, see Fig.~\ref{fig:p2eos-selfint}. For all power-law coefficients (except $p=4$), a large enough ratio $\lambda /g^2$ significantly affects the evolution of the equation of state, see e.g.~the bottom right panel of Fig.~\ref{fig:p2eos-selfint} for $p=2$. For $2<p<4$ and $p>4$, the large ratio also gives rise to a prolonged transition to radiation domination. These changes also affect the predictions for the CMB observables $n_s$ and $r$.
    
    \item Finally, in Sec.~\ref{Sec:Ndg1-lambdag0} we have considered different scenarios involving multiple daughter fields and scale-free interactions $\lambda_{nm} X_n^2 X_m^2$ and $\lambda_n X_n^4$. We have observed that a quadratic-quadratic interaction between two daughter fields, such as e.g.~$\lambda_{12}X_1^2 X_2^2$, allows for an efficient exchange of energy between the $X_1$ and $X_2$ fields, and if $\lambda_{12} \gtrsim g_1^2$, $g_2^2$, they can indeed achieve an equilibration regime with the same energy at late times, see Fig.~\ref{fig:p2specs-int}. Moreover, this term can also induce a resonance during the non-linear regime to both fields $X_1$ and $X_2$, similar to the one induced by the $\lambda_1 X_1^4$ interaction to $X_1$. For $p \geq 4$, this interaction can increase the amount of energy transferred from the inflaton to the daughter fields at late times, see Fig.~\ref{fig:p4enrgies-int}. In fact, even when a given daughter field (say $X_2$) does not have a direct interaction to the inflaton, the resonance induced by the $\lambda_{12}X_1^2 X_2^2$ interaction does indeed allow for an efficient transfer of energy to $X_2$ by means of a two-step process: first there is an energy transfer from $\phi$ to $X_1$ through the $\phi^2 X_1^2$ interaction, and later on from $X_1$ to $X_2$ through the $X_1^2 X_2^2$ interaction. An example of this is shown in  Fig.~\ref{fig:p4enrgies-chain}.
\end{itemize}

Furthermore we like to note that in our set-up, a significant amount of energy remains on the inflaton at late times (one exception is the hypothetical scenario $p\geq4$ and $N_d\rightarrow \infty$, where all energy is depleted from the inflaton). The energy must be transferred to light fields via some other mechanisms, such as perturbative decay channels. A similar mechanism is necessary to achieve a radiation-dominated stage in the $p=2$ case. Small extensions of our scenario are required for this purpose, such as additional couplings that only become relevant at a later stage. However, let us emphasize that for $p>2$, the transition towards a radiation-dominated state is completed after few e-folds (which we can capture with lattice simulations), and the equation of state will not change anymore until BBN.

In this work we have assumed that during the post-inflationary stage of inflaton oscillations, the inflaton potential can be approximated by a simple monomial function $V(\phi) = |\phi|^p$ with $p \geq 2$ around the minimum, and that only quadratic-quadratic interactions exist between the different fields. However, it would be very interesting to explore multi-daughter field theories beyond these two assumptions. For example, one could consider lower-scale inflaton potentials in which the minimum of the potential can be expanded as $V(\phi) = |\phi|^p$ around the minimum, but the inflaton does indeed oscillate over flatter-than-quadratic regions (e.g.~the case of potential (\ref{eq:inflaton-potential}) with $M \lesssim m_{\rm pl}$). Furthermore, studying the effect of trilinear interactions between the different fields in the post-inflationary energy distribution and equation of state would also be very interesting. One could incorporate terms of the type $\phi X_n^2$ like in Ref.~\cite{Dufaux:2006ee}, which induce a stage of tachyonic resonance during the linear regime. One could also incorporate trilinear interactions between the different daughter fields such as $X_m^2 X_n$ or $X_m X_n X_p$. However, unlike quadratic-quadratic interactions, these terms generate additional mass scales into the theory which could be difficult to properly capture on the lattice.

Another interesting extension of our work could be to study the evolution of the metric perturbations in set-ups with multiple daughter fields. It has been shown that the post-inflationary oscillations of the homogeneous inflaton can trigger a resonant growth of metric perturbations at sub-Hubble scales, a process known as `metric preheating' \cite{Nambu:1996gf,Hamazaki:1996ir,Bassett:1997az,Bassett:1998wg,Bassett:1999mt,Jedamzik:2010dq}. This may lead to interesting phenomenology such as production of  black holes or gravitational waves. Metric preheating has been recently studied in the presence of interactions of the inflaton to cosmological fluids in the form of perturbative decays \cite{Martin:2020fgl} (see also \cite{Martin:2021frd} for an study in multi-field inflation). In the scenario considered in our work, the inflaton is coupled to (one or multiple) daughter fields through quadratic-quadratic interactions, which cause its fragmentation few e-folds after the end of inflation (for $p>2$, inflaton fragmentation happens via self-resonance even in the absence of such interactions). Therefore, in order to fully understand of the fate of metric preheating in our set-up, one would need to study the dynamics of the system with numerical relativity simulations, which allow for metric perturbations.

\acknowledgments

We thank Daniel G.~Figueroa for our collaboration in \PartI~of this project. We also acknowledge partial support from the grant 200020/175502 of the Swiss National Science Foundation.

\appendix

\section{Expressions for energy density ratios and equation of state} \label{App:EnergiesEos}

In this Appendix we provide expressions for the energy density ratios of the multi-daughter field system under consideration, as well as for the resulting equation of state. We also present the equipartition identities that govern the evolution of the system.

The energy and pressure density can be written in terms of natural field and spacetime variables (\ref{eq:newvars2})-(\ref{eq:newvars1}) as
\be \rho =  \frac{\omega_*^2 \phi_*^2}{a^{\frac{6p}{2+p}} }  \left( E_{\rm k} + E_{\rm g} + E_{\rm p} \right) \ , \hspace{0.5cm}
p = \frac{\omega_{*}^2 \phi_{*}^2}{a^{\frac{6p}{2+p}} }   \left( E_{\rm k} - \frac{1}{3} E_{\rm g} - E_{\rm p} \right)  \ , \ee
where the subindices `k', `g' and `p' denote the (total) kinetic, gradient, and potential energy contributions. The kinetic and gradient terms can be decomposed as a sum of field contributions as ($f=\varphi,\chi_n$),
\begin{align}
    E_{\rm k} &= \sum_f  E_{\rm k}^f \ , \hspace{0.5cm}  E_{\rm k}^f  =  \frac{1}{2} \left( f' - \frac{6}{p+2} \frac{a'}{a} f \right)^2   \ , \label{eq:Enk} \\
    E_{\rm g} &= \sum_f  E_{\rm g}^f  \ , \hspace{0.5cm}  E_{\rm g}^f \equiv  \frac{1}{2}  a^{\frac{4p-16}{p+2}} |\nabla_{\vec{y}} f |^2 \ . \label{eq:Eng}
\end{align}
Similarly, let us write the potential as $V (\phi,\{X_n\}) =  \sum_t V^{(t)} (\phi,\{X_n\})$, where $t$ labels the different terms. The potential energy contribution can then be written in natural variables as
\be E_{\rm p} = \sum_{ t} E_{\rm p}^t \ , \hspace{0.5cm}  E_{\rm p}^t \equiv \frac{a^{\frac{6p}{2+p}}}{\omega_*^2 \phi_*^2} V^{(t)} (a^{\frac{-6}{2+p}}\phi_* \varphi, \{ a^{\frac{-6}{2+p}}\phi_* \chi_n \} )  \ . \label{eq:Enp}
\ee
For potential (\ref{eq:potential}), $E_{\rm p}$ is a sum of the following four terms,
\begin{align*}
E_{\rm p}^{\varphi^p} &\equiv    \frac{1}{p} |\varphi|^p  \ ,   & E_{\rm p}^{\varphi^2 \chi_n^2} &\equiv  \frac{1}{2} a^{\frac{6p-24}{p+2}} q_{*}^{(n)} \varphi^2 \chi_n^2 \ , \\
E_{\rm p}^{\chi_n^4} & \equiv \frac{1}{4} a^{\frac{6p-24}{p+2}} \sigma_{*}^{(n)} \chi_n^4 \ , & E_{\rm p}^{\chi_m^2 \chi_n^2} &\equiv \frac{1}{2} a^{\frac{6p-24}{p+2}}  \sigma_{*}^{(mn)}  \chi_m^2 \chi_n^2 \:.
\end{align*}
We can define \textit{energy ratios} for each energy component as the corresponding fractional contribution to the total energy density, i.e.~$\varepsilon_i \equiv \langle E_i \rangle / \langle \sum_{j} E_j \rangle$, where $\langle \dots \rangle$ denotes a volume average and $i$ labels each contribution. The ratios obey $\sum_i \varepsilon_i\equiv1$ by construction. The (instantaneous) equation of state of the system, i.e. the ratio of pressure and energy density $w\equiv p/\rho$, is simply
\be w =  \varepsilon_{\rm k} - \frac{1}{3} \varepsilon_{\rm g} -  \varepsilon_{\rm p} \  . \ee
Similarly, we can define the (oscillation-averaged) energy density ratios as $ \bar \varepsilon_i \equiv \langle E_i \rangle_{T_ {\bar \varphi}} / \langle \sum_{j} E_j \rangle_{T_ {\bar \varphi}}$, which source the effective (i.e.~oscillation-averaged) equation of state $\bar{w}$.

Lattice studies have shown that this kind of field systems virialize very quickly \cite{Boyanovsky:2003tc,Lozanov:2016hid,Figueroa:2016wxr,Lozanov:2017hjm}, with the fields obeying the following `equipartition' identities when averaged over both volume and oscillations,
\be \langle \dot{f}^2 \rangle_{T_ {\bar \varphi}} = \langle |\nabla f|^2 \rangle_{T_ {\bar \varphi}} + \left\langle f \frac{\partial V}{\partial f} \right\rangle_{T_ {\bar \varphi}} \ , \hspace{0.3cm} (f=\varphi,\chi_n) \ . \ee

These identities can be evaluated for the multi field model under consideration, described by potential (\ref{eq:potential}). If we take the monomial approximation (\ref{eq:powlaw-pot}) for the inflaton potential, these can be expressed in terms of energy contributions as follows,
\begin{align}
\langle E_{\rm k}^{\varphi} \rangle_{T_ {\bar \varphi}} & \simeq   \langle E_{\rm g}^{\varphi}  \rangle_{T_ {\bar \varphi}} + \frac{p}{2} \langle E_{\rm p}^{\varphi^p} \rangle_{T_ {\bar \varphi}} + \sum_{n=1}^{N_d}\langle E_{\rm p}^{\varphi^2\chi_n^2} \label{eq:Virial1}  \rangle_{T_ {\bar \varphi}} \  , \\
\langle E_{\rm k}^{\chi_n}  \rangle_{T_ {\bar \varphi}} & \simeq  \langle E_{\rm g}^{\chi_n}  \rangle_{T_ {\bar \varphi}} + 2 \langle E_{\rm p}^{\chi_n^4} \rangle_{T_ {\bar \varphi}} + \langle E_{\rm p}^{\varphi^2\chi_n^2} \rangle_{T_ {\bar \varphi}} + \sum_{\substack{m=1\\ n\neq m}}^{N_d} \langle E_{\rm p}^{\chi_m^2\chi_n^2}  \rangle_{T_ {\bar \varphi}} \ . \label{eq:Virial2} 
\end{align}
Typically, the sum $E_{\rm tot}\equiv\sum_{i}E_i$ does not change significantly over one oscillation, thus one can also express these identities in terms of the (oscillation-averaged) energy density ratios by replacing $\langle E_i\rangle_{T_ {\bar{\varphi}}}\rightarrow \bar{\varepsilon}_i$.

  \bibliography{References.bib}
  \bibliographystyle{utphys}

\end{document}